\documentclass[10pt,letterpaper]{article}
\usepackage{amsmath}
\usepackage{amsfonts}
\usepackage{multicol}
\usepackage{empheq}
\usepackage{caption}
\usepackage{float}
\usepackage{hyperref}
\usepackage{amsrefs} 
\usepackage{amssymb}
\usepackage{amsthm}
\usepackage{array}
\usepackage{bm}
\usepackage{datetime}
\usepackage{amsmath}
\usepackage{fancyhdr}
\usepackage{fontenc}
\usepackage{ifthen}
\usepackage{accents}
\usepackage{tabularx}

\newboolean{proofmode}
\newcommand{\StTag}[1]{ \label{st:#1}
\ifthenelse{\boolean{proofmode}}{\ \marginpar{\quad\scriptsize st:#1} }{}      }
\newcommand{\EqTag}[1]{
\ifthenelse{\boolean{proofmode}}
{ {\label{eq:#1}}
  \stepcounter{equation}
  \tag{\theequation \rlap{\kern 23 pt{\scriptsize eq:#1}}}
}
{\label{eq:#1}}
 }

\numberwithin{equation}{section}
\numberwithin{table}{section}

\theoremstyle{definition}

\newtheorem{Theorem*}{Theorem}
\newfont{\deffont}{cmbxti10}
\newfont{\german}{eufm10}
\newfont{\mymath}{cmr12}

\newcommand{\bfalpha}{{\boldsymbol\alpha}}
\newcommand{\bfbeta}{{\boldsymbol\beta}}
\newcommand{\bfgamma}{{\boldsymbol\gamma}}
\newcommand{\gmodz}{\mathfrak g/\liez}

\newcommand{\fluidparm}{z}
\newcommand{\boostparm}{w}
\newcommand\bfk{\boldsymbol{k}}
\newcommand\bfell{\boldsymbol{\ell}}
\newcommand\bfm{\boldsymbol{m}}

\newcommand\lieg{\mathfrak g}
\newcommand\lier{\mathfrak r}
\newcommand\lies{\mathfrak s}

\newcommand\lieso{\mathfrak s \mathfrak o}
\newcommand\liez{\mathfrak z}

\newcommand\barM{
	\hbox{\kern 2.3 true pt 
	\vbox{\hrule width 7  true pt height .3 true pt \kern .9 true pt
	\hbox{\kern -2.3 true pt $M$}}}}

\newcommand\barMs{
    \hbox{\kern 1 true pt
    \vbox{\hrule width 5  true pt height .2 true pt \kern .9 true pt
    \hbox{\kern -1.8 true pt \hbox{$\scriptstyle M$}}}}}

\newcommand\barZ{
	\hbox{\kern 2.3 true pt 
	\vbox{\hrule width 6.5  true pt height .3 true pt \kern .9 true pt
	\hbox{\kern -1.3 true pt $Z$}}}}

\newcounter{SpTcounter}

\newcounter{Solncounter}
\newcommand\SolnNo{\stepcounter{Solncounter}\theSolncounter}

\newcounter{Petrovcounter}
\newcommand\PetrovNo{\stepcounter{Petrovcounter}\thePetrovcounter}

\newcounter{ConEincounter}
\newcommand\ConEinNo{\stepcounter{ConEincounter}\theConEincounter}

\usepackage{graphicx}
\graphicspath{%
    {converted_graphics/}
    {/}
}

\newcommand\heisRS{{\bf heisRS}} 
\newcommand\heisLT{{\bf heisLT}} 
\newcommand\heisLS{{\bf heisLS}} 
\newcommand\heisLN{{\bf heisLN}} 
\newcommand\heisNS{{\bf heisNS}}
\newcommand\heisNN{{\bf heisNN}}

\newcommand\abelthreeRS{{\bf abel3RS}}
\newcommand\abelthreeRZ{{\bf abel3RZ}}
\newcommand\abelthreeLT{{\bf abel3LT}} 
\newcommand\abelthreeLS{{\bf abel3LS}} 
\newcommand\abelthreeLN{{\bf abel3LN}}
\newcommand\abelthreeLZ{{\bf abel3LZ}}
\newcommand\abelthreeNS{{\bf abel3NS}}
\newcommand\abelthreeNZ{{\bf abel3NZ}}
\newcommand\abelthreeNN{{\bf abel3NN}}

\newcommand\LevitwoS{{\bf simpCS}} 
\newcommand\LevitwoT{{\bf simpCT}} 
\newcommand\LevitwoN{{\bf simpCN}} 

\newcommand\abeltwoRone{{\bf abel2R1}}
\newcommand\abeltwoRtwo{{\bf abel2R2}}
\newcommand\abeltwoRthree{{\bf abel2R3}}
\newcommand\abeltwoLone{{\bf abel2L1}}
\newcommand\abeltwoLtwo{{\bf abel2L2}}
\newcommand\abeltwoNone{{\bf abel2N1}}
\newcommand\abeltwoNtwo{{\bf abel2N2}}
\newcommand\abeltwoNthree{{\bf abel2N3}}
\newcommand\abeltwoNfour{{\bf abel2N4}}
\newcommand\abeltwoNfive{{\bf abel2N5}}

\begin{document}

\setlength{\baselineskip}{12pt}
 
\newtheorem{SpaceTime}{}

\title{\bf Spacetime Groups
} 

\author{ 
Ian Anderson 
\\ Department of Mathematics and Statistics
\\ Utah State University  
\\ Logan, Utah 
\\USA 84322
\and 
Charles Torre 
\\Department of Physics
\\ Utah State University
\\Logan, Utah 
\\USA 84322
}
\date{April 10, 2020}
\maketitle

\bigskip\bigskip\bigskip
\begin{abstract}

	A  spacetime group is a connected 4-dimensional Lie group $G$ endowed with a 
	left invariant Lorentz metric $h$ 
	and such that the connected component   of  the isometry group of $h$ is $G$ itself.  The Newman-Penrose formalism 
	is used to give an algebraic classification of spacetime groups, that is, we determine a complete list of  inequivalent spacetime Lie algebras, 
	which are pairs $(\mathfrak g, \eta)$, with $\mathfrak g$ being a 4-dimensional Lie algebra and $ \eta$ being a  Lorentzian inner product on 
	$\mathfrak g$.  
	A full analysis of the equivalence 
	problem for spacetime Lie algebras is given which leads to a completely algorithmic solution to the problem 
	of determining when two spacetime Lie algebras  are isomorphic.  
	
	The utility of our classification is 
	demonstrated by a number of applications. The results of a detailed study of the Einstein field equations for various matter fields on spacetime groups are given, which resolve a number of open cases in the literature. The possible Petrov types  of spacetime groups that, generically,  are algebraically special are completely characterized.  Several examples of conformally Einstein spacetime groups are exhibited.
	
	Finally, we describe some novel features of a software 	package 
	created to support the computations and applications of this paper.
\end{abstract}
\thispagestyle{empty} 

\newpage 
\tableofcontents
\thispagestyle{empty} 
\newcommand{\TestFilename}{SpaceTimeLieGroupEquationInfo.mm}
\immediate\openout5=\TestFilename
\immediate\write5{#####################################}
\immediate\write5{#}
\immediate\write5{#  This file is generated during the LaTeX compilation of the paper}
\immediate\write5{#  It is read during the compilation of the package SpaceTimeGroups}
\immediate\write5{#}
\immediate\write5{####################################}

\newcommand\DGlabel[1]{\label{#1}%
 \immediate\write5{"\getrefnumber{#1}" = "#1",
}}

\newcommand\SpecialDGlabel[1]{\label{#1}%
 \immediate\write5{"2.\getrefnumber{#1}" = "#1",
}}

\vfill\eject
\parskip = 9pt

 \setcounter{page}{1} 
\section{Overview}{}

\subsection{Introduction}

	A  spacetime group is a connected 4-dimensional Lie group $G$ endowed with a 
	left invariant Lorentz metric $h$ 
	and such that the connected component  $\text{Iso}_0(h)$ of  the isometry group of $h$ is $G$ itself. 
	Alternatively, one may say that a spacetime group is a connected 4-dimensional Lorentzian manifold
	$(M, h)$ for which $\text{Iso}_0(h)$ is {\it simply transitive}.
	Two spacetime groups $(G_1, h_1)$ and $(G_2, h_2)$ are  equivalent if there 
	is a  smooth  Lie group isomorphism $\varphi : G_1 \to G_2$  
	such that $\varphi^*(h_2) = h_1$.   
	
	Spacetime groups play a distinguished role in the 
	study of exact solutions of the Einstein field equations.  As homogeneous spaces, the
	field equations for spacetime groups are reduced to purely algebraic equations. However, 
	for spacetime groups there are no isotropy constraints on the space of $G$-invariant metrics
	and, therefore, these algebraic equations are especially complicated. Indeed, for certain 
	matter fields and/or field equations with cosmological term, 
	the properly homogeneous (multiply transitive) solutions to the Einstein equations are entirely 
	known  but the spacetime group solutions are not. The present work is motivated, in part, 
	by the desire to resolve these long-standing open cases. The present article 
	is also part of a larger program to systematically review 
	Petrov's classification of spacetimes with symmetries \cite{Petrov:1961},
	and to provide new computationally effective tools for addressing the equivalence problem in general relativity.

	The accomplishments of this paper are four-fold. First, 
	we  shall determine a complete list of  inequivalent {\it spacetime Lie algebras}, 
	that is, 4-dimensional Lie algebra, Lorentzian inner product
        pairs $(\mathfrak g, \eta)$.  We thus classify all simply connected spacetime groups by enumerating the corresponding Lie algebras and inner products.
	Our classification scheme, 
	based upon the Newman-Penrose (NP) formalism and standard Lie algebraic invariants,
	leads to an initial list of 25 distinct families of spacetime Lie algebras, 
	each class depending on a number of freely variable 
	NP spin coefficients. 
	No two spacetime Lie algebras belonging to different families of our initial classification
	are equivalent.
	Various partial results along these lines can be found, {\it e.g.,} in
	\cite{Farnsworth-Kerr:1966a}, \cite{Hiromoto-Ozsvath:1978} but, somewhat surprisingly, 
	the work presented here  appears to be the first complete, 
	comprehensive solution to this fundamental classification problem using the NP formalism. 
	
	The second accomplishment of this paper is a full analysis of the equivalence 
	problem for spacetime Lie algebras.  This analysis consists of 3 steps.
	First, for each one of our  25 families of spacetime Lie algebras, 
	we determine the residual freedom in 
	the choice of null tetrad, that is, the subgroup of the Lorentz group whose action on the null tetrad
	preserves the form of the NP structure equations for the given family. We take 
	special care to identify the discrete components of the residual group. Second, 
	we explicitly show for each family  how the connected component of the residual group 
	may be reduced to 1-parameter Lorentz boosts and/or Euclidean rotations by gauge fixing 
	certain NP spin coefficients. 
	This requires a refinement of our initial  classification.
	For many of the families this reduction is a straightforward matter; for others, 
	we must rely upon the classification of normal forms  for $3\times 3$ symmetric matrices under conjugation 
	by (subgroups of) $SO(2, 1)$ \cite{Hall-Morgan-Perjes:1987a}. The following theorem summarizes the main result of this paper.

\par
\medskip
\noindent
{\bf Theorem.}{\it\quad
	A complete list of spacetime Lie algebras consists of 42 distinct families, each family
	depending on 2 to 7 Newman-Penrose spin coefficients.
	
\noindent
	For each family in this list the residual group is generated by 
	either a  1-parameter group of Euclidean rotations or Lorentz boosts  
	(or both) and a finite group of discrete Lorentz transformations.

\noindent
	The residual Euclidean rotations and/or Lorentz boosts can
	always be used either to  normalize a non-zero complex spin coefficient
	to a positive real number or to normalize a real nonzero spin coefficient to $\pm 1$. 
	This done, two spacetime Lie algebras in the same family are equivalent 
	if and only if their spin coefficients are related by a given list of discrete 
	Lorentz transformations.
}

	Our third accomplishment consists of a detailed study of the geometric and physical properties of each of the spacetime Lie algebras according to 	our 		classification.   First, we obtain a list of all spacetime groups which are solutions to
	the Einstein equations for a variety of  matter sources. This reproduces known solutions  
	in the literature, it provides new solutions for matter fields not previously considered, and it
	resolve some long-standing open problems in the exact solution literature. 
	Second, we relate our classification of spacetime Lie algebras to one of the standard 
	classifications of 4-dimensional Lie algebras.  Next, we characterize the various Petrov types 
	 of spacetime Lie groups which are generically
	algebraically special. Finally,  
	we provide examples of conformally Einstein spacetime Lie groups. Altogether, these results 
	underscore the utility of our classification.

	Finally, we have created a Maple software package {\sc SpacetimeGroups}
	to support the computations and results of 
	this paper. This package contains two particularly novel features. 
	First, it contains a database of the structure equations for every
	spacetime Lie algebra  in this paper. For example, the command
	{\tt SpaceTimeLieAlgebra("2.1")} initializes 
	the Lie algebra (2.1) in  our classification at which point
	one can perform a wide range  of tensorial and Lie theoretic 
	computations with this algebra.  
	Furthermore, this database allows for the symbolic, step by step,  verification 
	of our classification proof.
	Secondly, we have created commands which provide for a 
	complete implementation of the equivalence 
	problem for spacetime Lie algebras. 
	The command {\tt ClassifySpacetimeLieAlgebra}  will classify
	any given 4-dimensional spacetime Lie algebra according to the 
	classification scheme summarized in Section 2 of the paper; 
	the command {\tt STLAAdaptedNullTetrad} creates an 
	adapted null tetrad which aligns the structure equations 
	with those of Section 2; and, finally,  the command  {\tt MatchNPSpinCoefficients}
	can be used to find explicit Lie algebra spacetime isomorphisms.
	The software, supporting documentation, and 
	worksheets are available at \cite{Anderson-Torre}.

	We emphasize that our definition of a spacetime group requires that {\it  the isometry 
	algebra of the spacetime metric coincides with the underlying Lie algebra.}
	It is therefore essential  that we are able to calculate the full isometry algebra  
	directly from the structure equations of the spacetime Lie algebra.
	In Appendix A we  review  how this can be done without introducing local coordinates in 
	the spacetime  group and without explicitly solving the Killing equations.

	Our paper is organized as follows. In Sections \ref{ClassificationMethod} and \ref{NPApproach}
	we provide a general discussion
	of the classification problem for spacetime Lie algebras and introduce the basic framework 
	and nomenclature that we shall use  in our solution to this classification problem. 
	A general discussion of the equivalence problem for spacetime Lie algebras is given in  
	Section \ref{EquivalenceMethod}. The solution to our 
	classification of spacetime Lie algebras  is summarized in Section \ref{Summary}.  In Section \ref{GenRel}
	we summarize our results on spacetime Lie group solutions to the Einstein field equations. 
	Additional applications and examples are given in Section \ref{Apps}.
	The explicit proofs of our classification and our solution to the equivalence problem 
	are presented in Sections \ref{Classification} and \ref{Equivalence}.
	Illustrations of the  {\sc SpacetimeGroups} software package are given in Section \ref{software}. 
	There are two appendices to this paper. Appendix A describes our algebraic method for 
	calculating the  isometry algebra of a spacetime Lie group.  
	Appendix B gives a complete list of all 4-dimensional Lie algebras.  Appendix C provides a list of symbols and notation.

\subsection{The Classification of Spacetime Lie algebras}\label{ClassificationMethod}

	To frame the general context of our work,  it is useful to discuss
	the  distinct approaches used in the past  to obtain a  classification of spacetime Lie algebras.
	
	The first classification appears to be due to Petrov; see pages 233--240 of
	\cite{Petrov:1961}. Petrov's approach is an inductive one in local coordinates, starting 
	with an enumeration of all 3-dimensional Lie algebras, given as vector field systems. Petrov  assumes  that the spacetime metric is given in geodesic 		normal coordinates, adapted to each 	3-dimensional 		vector 	field 	system. 
	For reasons we cannot fully understand, Petrov fails to include 
	those spacetime Lie algebras for which, from our viewpoint,  the metric is degenerate on 
	the derived algebra. This implies, for example, that  in  case (32.47) from \cite{Petrov:1961}, 
	where the algebra is $so(3) \oplus {\mathbb R}$,  Petrov does not consider the possibility that
	 the center $\mathbb R$ may be a null subspace.  Also unclear is the extent to which the
	parameter values in the metric components of Petrov's formulas all define inequivalent spacetimes. 
	Finally, the coordinate formulas for the metrics do not 
	seem to be in a form that is advantageous for subsequent analysis and applications.

	A second approach, and perhaps the most natural one,
	is to begin by classifying all 4-dimensional Lie algebras. Specifically, the 
	problem  here is to find all bilinear maps $[\cdot, \cdot] : V \times V \to V$, defined on a	
	real 4-dimensional vector space $V$,  which are skew-symmetric and which satisfy the Jacobi identity.
	Let $\text{GL}(V)$ be the group of invertible linear transformations on $V$.
	Two  Lie brackets  $[\cdot, \cdot]_1$ and $[\cdot, \cdot]_2$ on $V$ define isomorphic Lie algebras if 
	\begin{equation}
	\psi([ x ,\, y ]_1) =  [  \psi(x),\, \psi(y)]_2 
	\quad
	\text {for all $x, y \in V$ and some $\psi \in \text{GL}(V)$.}
	\DGlabel{Lie}
	\end{equation} 
	The problem at hand is to classify all possible Lie brackets  $[\cdot, \cdot]$ on $V$, up to isomorphism.
	This classification problem 
	has been solved by many different authors, dating back to an initial classification
	by Kruchkovich \cite{Kruchkovich:1954}. A
	detailed review of this literature  is given by  MacCallum \cite{ MacCallum:1991}, 
	so only a few simple remarks need to be made here.
 
	In the mathematical physics literature, one finds various classifications of 4-dimensional Lie algebras based  
	 upon  simple tensorial invariants of the structure constants (see, {\it e.g.,} \cite{ MacCallum:1991})
	and/or upon the fact that every 4-dimensional Lie algebra contains a 3-dimensional Lie sub-algebra (see, {\it e.g.,} \cite{Calvaruso}). 
	In the Lie theory literature, 
	the classifications  \cite{Patera-Sharp-Winternitz-Zassenhaus:1976}, \cite{Snobol-Winternitz} are 
	based upon  the Levi canonical decomposition  and  flags of distinguished ideals such as the 
	derived series. It is these latter techniques which will be most useful to us.
	It should be emphasized  that the structure constants of the Lie algebras under consideration 
	often depend upon parameters, in which case  a full and proper classification of 
	such Lie algebras should 
	specify those parameter domains for which the Lie algebras are inequivalent. As noted in
	\cite{ MacCallum:1991}, the classification results in
	\cite{Patera-Sharp-Winternitz-Zassenhaus:1976} (with amendments given in \cite{Snobol-Winternitz}) are the most thorough in this regard.

	Thus, in this second approach to the study of spacetime Lie algebras, 
	one can start with any of the known classifications of 4-dimensional Lie algebras 
	and immediately turn to the classification of inner products on a given 
	Lie algebra taken from that  classification. So, fix a 4-dimensional Lie algebra $\lieg$ and
	let $\text{Aut}(\lieg)$ be the matrix group of automorphisms of $\lieg$.  
	In this context, two Lorentzian inner products, 	$\eta_1$ and $\eta_2$, on  $\lieg$  
	define equivalent spacetime Lie algebras
 if and only if 
\begin{equation*}
	\eta_1( x ,y)=  \eta_2(  \chi(x),\, \chi(y)) 
	\quad
	\text {for all $x, y \in \lieg$ and some $\chi \in \text{Aut}(\lieg)$.}
\end{equation*} 
	An essential issue now arises in that the structure constants of $\lieg$ 
	may be functions of auxiliary parameters and  $\text{Aut}(\lieg)$ may change
	at isolated parameter values.  
	Accordingly, in order to proceed 
	one needs a classification of the 4-dimensional Lie algebras which is refined to the 
	degree that $\text{Aut}(\lieg)$ is unchanged  throughout the specified parameter domains.
	For completeness, we provide such a classification in Appendix B.

	Granted this, one can say that the second approach to the classification
	of spacetime Lie algebras reduces to 
	the enumeration of congruence classes of inner products under a variety of 
	different matrix groups.  Finding a useful enumeration of the possible normal forms for quadratic forms
	with respect to congruence by a variety of different matrix groups is a 
	challenging problem. In \cite{Fee:1979},  G. Fee solves this problem in an ingenious manner. Fee begins 
	with a classification of 4-dimensional  Lie algebras 
	into  a list of 16 families of algebras. He then shows that a generic Lorentz metric 
	on a 4-dimensional vector space can be transformed into 
	1 of 10 normal forms by conjugation with an upper triangular matrix. 
	From each of these normal forms, Fee then constructs parametrized families of
	inequivalent inner products on each Lie algebra. 
	The result is a list of 160 families of spacetime Lie algebras 
	(although a number of these families define spacetimes whose isometry groups are of  dimension greater 4). This work certainly 
	deserves to be better appreciated in the literature; we would only remark that 
	Fee's method  does not yield a classification which is easily related to 
	geometric properties of the spacetime.

	We now describe a third  approach which we believe is better adapted to 
	applications (see,  {\it e.g.,}
	\cite{Farnsworth-Kerr:1966a}, \cite{Hiromoto-Ozsvath:1978}).  Since all inner products are equivalent under 
	conjugation by $\text{GL}(V)$, one first fixes the Lorentzian inner product
	$\eta$ on the vector space $V$. Let $\text{O}(\eta)$ be the group of Lorentz 
	transformations on $V$ which preserve the given inner product. Two  Lie brackets  
	$[\cdot, \cdot]_1$ and $[\cdot, \cdot]_2$ on the inner product space 
	$(V, \eta) $ define    equivalent spacetime Lie algebras in this approach if and only if
	\begin{equation}
	\varphi([ x ,\, y ]_1) =  [  \varphi(x),\, \varphi(y)]_2 
	\quad
	\text {for all $x, y \in V$ and some $\varphi \in \text{O}(\eta)$.}
	\DGlabel{Cartan}
	\end{equation} 
	The classification problem defined by  \eqref{Cartan} is very different, and more difficult,	 
	from that defined by \eqref{Lie}, but it avoids the challenge of classifying quadratic forms as described above.
	To illustrate the difference, suppose one wishes to classify 
	4-dimensional Lie algebras $\lieg$ with a 1-dimensional center $\mathfrak z$. In \eqref{Lie}, 
	one is free to pick a basis $\{e_1, e_2, e_3, e_4\}$ for $\lieg$ such that $\mathfrak z$ is spanned 
	by $e_1$.  In \eqref{Cartan}, one has to consider three cases according to whether  $\mathfrak z$ is 
	a 1-dimensional time-like, space-like, or null subspace. The first goal of this paper is to solve 
	the classification problem posed by 
	\eqref{Cartan}.
	We  turn to a more detailed explanation of this approach.

\subsection{An Approach  to Classification  Using the Newman-Penrose Formalism}\label{NPApproach}

	Let $(G, h)$  be a 4-dimensional spacetime Lie group.  	From the set of left invariant vector fields 
	on $G$, we construct
	 a null tetrad $\{K, L, M, \barM\}$. These  vector fields 
	satisfy
\begin{equation}
	h( K , L ) = -1, \quad h( M , \barM ) = 1,
\DGlabel{NT}
\end{equation}
	with all other inner products vanishing.
	The vector fields $K$ and $L$ are real. The vector field   $M$ is a complex vector field with complex conjugate $\barM$. In accordance with 
	the Newman-Penrose formalism (see, for example, 
	\cites{ Newman-Penrose:1962a, Stephani, Stewart:1991}) we write the commutator formulas for the
	null tetrad as
\begin{equation}
\begin{alignedat}{1}
	[\, K,\, L\,] & = -(\gamma+\bar\gamma)\, K -(\epsilon +\bar\epsilon)\, L+(\pi +\bar\tau)\, M
	+(\bar\pi +\tau)\, \barM
\\[.5\jot]
	[\, K,\, M \, ] & = -(\bar\alpha+\beta-\bar\pi)\, K-\kappa\, L+(\epsilon-\bar\epsilon + \bar\rho)\, M+\sigma\, \barM
\\[.5\jot]
[\, L,\, M] & = \bar\nu\, K+(\bar\alpha+\beta -\tau)\, L+(\gamma-\bar\gamma -\mu)\, M-\bar\lambda\, \barM
\\[.5\jot]
[\, M,\, \barM\,] & = (\mu-\bar\mu)\, K+(\rho-\bar\rho)\, L -(\alpha-\bar\beta)\, M+(\bar\alpha-\beta)\, \barM.
\end{alignedat}
\DGlabel{masterNP}
\end{equation}
	In general, the twelve complex  Newman-Penrose spin coefficients
\begin{equation}
	\mathcal S = \{ \alpha,\beta, \gamma, \epsilon, \kappa, \lambda,
	\mu, \nu, \pi, \rho, \sigma,\tau\} 
\end{equation}
	are functions on the spacetime manifold, but now, in our Lie group setting with a left invariant null tetrad, the spin coefficients
	are constants. They are the structure constants for the Lie algebra of the spacetime group.
	In what follows we shall denote the real and imaginary parts of the spin coefficients with 
	the subscript 0 and 1 respectively; for example,  $\alpha = \alpha_0 + i \alpha_1$.

	It is to be emphasized that in this approach to spacetime groups the Killing vectors for the metric are not explicitly
	determined. Indeed, since we declared the null tetrad to consist of 
	left invariant vector fields on the spacetime, 
	the Killing vectors will be  right invariant vector fields. Of course, the Killing vectors will 
	be expressible in terms of the null tetrad, but only as linear combinations with coefficients which are functions on $G$.	
	
	The first main result of this paper is
	{\it an explicit determination of all possible values of the spin coefficients 
	for which  \eqref{masterNP} are indeed the structure equations of a  (real)
	4-dimensional Lie algebra.} Put differently,  we find  all
	values of the spin coefficients  for which the Ricci and Bianchi identities in the Newman-Penrose 
	formalism are satisfied.
	
	To describe our classification methodology, we recall that every finite-dimensional Lie algebra $\lieg$ 
	admits a semi-direct sum decomposition $\lieg = \mathfrak s  + \lier$, 
	where  $\lies$ is a semi-simple Lie algebra, $\lier$ 
	is a solvable Lie algebra,  and $[\lies, \lier]  \subset  \lier$.  
	This is the Levi decomposition.
	There are no 4-dimensional semi-simple Lie algebras; 
	exactly two 4-dimensional Lie algebras with non-trivial Levi decomposition;
	and all other 4-dimensional Lie algebras are solvable.  
	These simple observations provide  the basis for our classification scheme. For details see \cite{Snobol-Winternitz}.
	
	The two 4-dimensional Lie algebras $\lieg$ with non-trivial Levi decomposition are actually
	the direct sum of a 3-dimensional simple Lie algebra and a 1-dimensional center 
	$\liez$, that is, $\lieg = \lieso(2,1)  \oplus \liez$ or $\lieg = \lieso(3)  \oplus \liez$.
	We emphasize
	that these direct sums are generally {\it not}  orthogonal direct sums with respect to the given 
	Lorentz inner product \eqref{NT}  on $\lieg$. 
	For these Lie algebras our classification is  based upon the spacetime signature of the center. 
	
	For  solvable Lie algebras $\lieg$, the derived algebra $\lieg' = [\lieg, \lieg]$ 
	is a nilpotent algebra properly contained in  $\lieg$. The derived algebra is 
	the primary invariant we use for the classification of solvable spacetime algebras.
	If the derived algebra is 3-dimensional, then it is 
	either the 3-dimensional Heisenberg algebra or the 3-dimensional   
	abelian algebra. The classification then proceeds
	according to the signature of the inner product on the derived algebra. 
	To complete our classification in this case we shall identify a  privileged 1-dimensional subspace
	in the derived algebra and then further case split according to its spacetime character.  When $\lieg'$ is the Heisenberg
	algebra the privileged subspace is $\lieg''$, which is the center of $\lieg'$.  
	When $\lieg'$ is 3-dimensional and
	abelian the privileged subspace is defined as follows.
	
	Let $v$ be any vector complementary to the 3-dimensional derived subalgebra
	$\lieg'$.  Since the derived algebra is now assumed abelian, the  
	adjoint matrix  $A^a_b= [\text{ad}(v)]^a_b$ is independent of the direction of $v$. 
	Next, let $N$ be any vector  which is orthogonal to $\lieg'$. We then define
\begin{equation}
	\zeta^a = \epsilon^{abcd} A_{cd} N_b,
	\quad\text{where  $A_{cd} = \eta_{ca} A^a_d$.}
\DGlabel{defzeta}
\end{equation} 	
	We are free to scale both $v$ and $N$ so that $\zeta$ defines a 1-dimensional subspace only. 
	The vector $\zeta$ is orthogonal to $N$ and hence $\zeta \in \lieg'$.  If the metric restricted to the derived algebra 
	is non-degenerate, then we may take $v = N$. In this case,
	the vector $\zeta =0$  if and only if the  skew-symmetric part of $A$ 
	with respect to the inner product  vanishes, in other words,
	{\it $\text{ad}(v)$ is self-adjoint if and only if $\zeta$ vanishes}.  We will refer to the subspace defined by $\zeta$ as the {\it skew-adjoint line}.
	
	If the derived algebra $\lieg'$ is 2-dimensional, then  it is necessarily abelian and, 
	again, we proceed according to the signature of the metric restricted to $\lieg'$.
	Finally, it will be a simple matter to show there are no spacetime Lie algebras 
	which are solvable and have a 1-dimensional derived algebra (since the isometry algebra will 
	necessarily have dimension greater than 4).

	\newcommand\Strut{\rule[-6pt]{0pt}{18pt}}

	The possibilities and the nomenclature  we use   are summarized in the following tables.  The structure equations for all the numbered Lie algebras 		are given 	in Section 2.
	
\begin{table}[h]
\caption {\sc Solvable Spacetime Lie Algebras with 3-Dimensional Heisenberg Derived Algebra} \label{hiesTable}
\medskip
\centering
\begin{tabular}{|l|l|l|l|l|l|}
\hline
\Strut 
 \ref{HeisRS}. & \heisRS & $\lieg$ solvable, $\lieg'$  = Heisenberg   & 
 \ref{HeisLN}. & \heisLN & $\lieg$ solvable, $\lieg'$  = Heisenberg
\\[-1\jot]
\Strut
   &&  $\lieg'$ Riemannian,   $\lieg''$ space-like 
   &&&  $\lieg'$ Lorentzian,   $\lieg''$ null
\\
\hline
\Strut	
\ref{HeisLT}. & \heisLT & $\lieg$ solvable, $\lieg'$  = Heisenberg  &
\ref{HeisNS}. &  \heisNS & $\lieg$ solvable, $\lieg'$  = Heisenberg  
\\[-1\jot]
\Strut
   &&  $\lieg'$ Lorentzian,   $\lieg''$ time-like
   &&&  $\lieg'$ null,   $\lieg''$ space-like \\
\hline	
\Strut
 \ref{HeisLS}. & \heisLS &  $\lieg$ solvable, $\lieg'$  = Heisenberg &
 & \heisNN &  $\lieg$ solvable, $\lieg'$  = Heisenberg  
\\[-1\jot]
\Strut
   &&  $\lieg'$ Lorentzian,   $\lieg''$ space-like
&&&  $\lieg'$ null,   $\lieg''$ null \\
&&&&& {\sl Isometry group is 6-d}
\\
\hline
\end{tabular}
\end{table}

\begin{table}[h]
\caption{\sc Solvable Spacetime Lie algebras with 3-Dimensional Abelian Derived Algebra} \label{abel3Table}
\medskip
\centering
\begin{tabular}{|l|l|l|l|l|l|}
\hline
\Strut 
\ref{IIIdAbelianRS}. & \abelthreeRS & $\lieg$ solvable, $\lieg'$  =  3-d~abelian~& 
\ref{IIIdAbelianLN}. & \abelthreeLN & $\lieg$ solvable, $\lieg'$  =  3-d~abelian
\\[-1\jot]
\Strut
   &&  $\lieg'$ Riemannian,   $\zeta$ space-like 
   &&&  $\lieg'$ Lorentzian,    $\zeta$ null
\\
\hline
\Strut	
\ref{IIIdAbelianRZ}. &  \abelthreeRZ & $\lieg$ solvable, $\lieg'$  = 3-d~abelian  &
\ref{IIIdAbelianLZ}. & \abelthreeLZ & $\lieg$ solvable, $\lieg'$  = 3-d~abelian  
\\[-1\jot]
\Strut
   &&  $\lieg'$ Riemannian,     $\zeta =0 $ 
   &&&  $\lieg'$ Lorentzian,  $\zeta = 0$\\
\hline	
\Strut
\ref{IIIdAbelianLT}. & \abelthreeLT &  $\lieg$ solvable, $\lieg'$  = 3-d~abelian &
\ref{IIIdAbelianNS}. & \abelthreeNS & $\lieg$ solvable, $\lieg'$  = 3-d~abelian  
\\[-1\jot]
\Strut
   &&  $\lieg'$ Lorentzian,  $\zeta$ time-like
&&&  $\lieg'$ null,  $\zeta$ space-like  \\
\hline	
\Strut
\ref{IIIdAbelianLS}. & \abelthreeLS &  $\lieg$ solvable, $\lieg'$  = 3-d~abelian &
& \abelthreeNN & $\lieg$ solvable, $\lieg'$  = 3-d~abelian
\\[-1\jot]
\Strut
   &&  $\lieg'$ Lorentzian,   $\zeta$ space-like 
&&{\bf abel3NZ}&  $\lieg'$ null,   $\zeta$ null or $\zeta=0$
\\
&&&&& {\sl Isometry group is 6-d.}
\\[1\jot]
\hline
\end{tabular}
\end{table}

\begin{table}[h]
\caption{\sc  Spacetime Lie Algebras with 3-Dimensional Simple Derived Algebra}\label{LeviTable}
\medskip
\centering
\begin{tabular}{|l|l|l|l|l|l|}
\hline
\Strut 
\ref{simpT}. & \LevitwoT & $\lieg = \lieg' \oplus \liez $, $\lieg'$  = simple  & 
\ref{simpN}. &  \LevitwoN &  $\lieg = \lieg' \oplus  \liez $, $\lieg'$  = simple
\\[-1\jot]
\Strut
   &&   $\liez$ time-like  
   &&&  $\liez$ null  
\\
\hline
\Strut	
\ref{simpS}. & \LevitwoS &  $\lieg = \lieg' \oplus \liez $, $\lieg'$  = simple  & &&
\\[-1\jot]
\Strut
   &&   $\liez$ spacelike  
   &&&   
\\ 
\hline	
\end{tabular}
\end{table}

\begin{table}[H]
\medskip
\caption{\sc Solvable Spacetime Lie Algebras with 2-Dimensional Abelian Derived Algebra}\label{abel2Table}
\medskip
\centering
\begin{tabular}{|l|l|l|l|l|l|}
\hline
\Strut 
\ref{IIAbelR1} --\ref{IIAbelR3}. & \abeltwoRone--{\bf 3} & $\lieg$ solvable,  $\lieg'$ = 2-d~abelian&
\ref{IIAbelN1}--\ref{IIAbelN5}. & \abeltwoNone--{\bf 5} & $\lieg$ solvable,   $\lieg'$ = 2-d~abelian
\\[-1\jot]
\Strut
   &&   $\lieg'$  Riemannian
   &&&  $\lieg'$ null  
\\
\hline
\Strut 
\ref{IIAbelL1} -- \ref{IIAbelL2}. & \abeltwoLone--{\bf 2} & $\lieg$ solvable,   $\lieg'$ = 2-d~abelian&&&
\\[-1\jot]
\Strut
   &&   $\lieg'$    Lorentzian
   &&& 
\\
\hline
\end{tabular}
\end{table}

\subsection{The Equivalence Problem for Spacetime Lie Algebras}\label{EquivalenceMethod}
	
	In  relativity, the equivalence problem -- determining whether two 
	given Lorentzian manifolds are locally isometric -- can be addressed  in general by
	the Cartan-Karlhede algorithm \cite{Karlhede}, \cite{Stephani}.
	In the special case  of spacetime groups, the equivalence 
	problem reduces to determining whether two spacetime 
	Lie algebras are isomorphic by a Lorentz 	transformation (see (\ref{Cartan})).
	Since the  25 classes of spacetime Lie algebras (as enumerated in tables 1--4) are characterized by Lie algebraic and tensorial 
	invariants, {\it no two different classes can contain equivalent spacetime Lie algebras}. Thus, in order 
	to completely solve the equivalence problem for spacetime groups, it remains to address the 
	issue of equivalence of spacetime Lie algebras within the same class.

	By definition, the {\it residual group}  of a given class is the subgroup of the Lorentz group which preserves its defining properties as listed
	in Tables 1--4 above.  For example, the residual group for the spacetime Lie algebras in Table 1 will preserve the derived and second derived algebras, while the residual 	group for the algebras in Table 3 will preserve the 1-dimensional center.  See Table 5 (in Section 5) for an enumeration of all the residual groups in terms of adapted null tetrads.  The 
	residual group will have a well-defined action on the free (independent) Newman-Penrose coefficients defined by the adapted tetrad for any given class. In this way the residual group transforms a given spacetime Lie algebra into another algebra within the same class.

	In the parlance of the general theory of equivalence problems, the residual group is 
	called the reduced structure group (the initial structure group being the full Lorentz group). Within 
	the classical field theory community, the residual group could be called the gauge group
	of the given spacetime Lie algebra class. The central problem is now to determine whether two spacetime Lie algebras in a given class are isomorphic by an element of the residual group.
		The strategy is  to use transformation properties of the 
	spin coefficients under the residual group
	to {\it normalize} the spin coefficients, that is, put them into a standard form.   These normalizations further reduce the residual group.  In the physics literature 
	one might call this normalization procedure ``gauge fixing'',  while differential geometers would say this process is picking 
	a cross-section to the gauge group.  Our goal is to reduce the residual group in each case to a discrete group.  We do this in two steps (see Section 6 for details).
	
	First, we systematically reduce the residual group to the group generated by the standard NP boosts and/or rotations and/or a finite group of discrete Lorentz transformations.  In all but 5 of these cases, this simply involves gauge fixing the null rotations when they appear in the residual group.   There are 4 cases, namely \abelthreeRZ, \abelthreeLZ, \LevitwoT, and  \LevitwoS,
	where the residual  group is a 3-dimensional simple group and  1 case, namely  \LevitwoN,
	where the residual  group is 4-dimensional.  In each of these latter 5 cases there is a symmetric tensor naturally defined on the derived algebra.  The residual group	
	can  be used to cast  this symmetric tensor into a  canonical form.   All together, these normalizations lead to case splitting of the 25 spacetime Lie algebra classes to a total of 42 inequivalent classes of algebras.
	
	Second we show how to gauge fix the remaining rotations and/or boosts.  The only way this two step normalization procedure could fail is  if in a given case there is an inadequate number of independent spin coefficients to normalize, {\it e.g.}, too 
	many of the spin coefficients are zero or invariant, so that the normalization cannot be fully performed.
        In Section 6 we  explicitly show how to perform the normalizations in each case, and therefore this issue does not arise.
	
%
%

In the end, we use  gauge fixing of the  spin coefficients to reduce the 
	residual group for each class to a finite discrete group. One could now use the remaining discrete residual group to further normalize the spin coefficients,  but it is 		more convenient to leave this discrete group intact.  In each case it is a simple matter to determine whether two spacetime Lie algebras are related by 	the 		action of the discrete residual group (see Section 7.7).  The equivalence problem is now solved: 
	 {\it  two spacetime Lie algebras are	
	equivalent if and only if they have the same class and, after normalization, the
	remaining spin coefficients are equal up to an explicitly given finite discrete group.}

\newcommand\mybigskip{\vspace{5pt}}
\newpage
\newpage
\section{Classification of Spacetime Lie Algebras: Summary of Results}\label{Summary}

In this section we summarize our classification of spacetime Lie algebras.
For each algebra in our classification, we give the  structure 
equations in terms of the Newman-Penrose spin coefficients as well as  information regarding the relevant subalgebras used for the classification. To define these subalgebras, we use the notation $\langle A, B, C, \dots\rangle$ to denote the vector space spanned by $A, B, C,\dots\,$.
The generators of the residual group, that is, the subgroup of the Lorentz group which preserves the form of the 
structure equations, are given. For definitions of the various subgroups, see Section \ref{Lorentz}. 
Where appropriate, we provide gauge fixing conditions which reduce the residual group  to no more than that generated by a one parameter family of rotations, and/or one parameter families of boosts, and/or a finite set of discrete transformations.   

We also give particular values of the structure constants for which the dimension of the isometry algebra jumps to values greater than 4.  These prohibited values are not exhaustive, but their omission guarantees the ability to  normalize the spin coefficients.
For all other values of the structure constants (which keep the dimension of the isometry group at 4) the residual group can be reduced by normalization to at most a discrete group.

Algebras 1--5 are those for which the derived algebra is the 3-dimensional Heisenberg algebra. In this case we are able to explicitly provide  open conditions on the spin coefficients  which ensure that the 
derived algebra is neither lower-dimensional nor abelian.  

Algebras 6--12 are those spacetime Lie algebras for which the derived algebra is the 3-dimensional abelian algebra.  Our classification here  depends on the skew-adjoint line defined by the vector $\zeta$ in  (\ref{defzeta}), which we  exhibit (up to an overall factor).
If $\{E_1, E_2, E_3,E_4\}$ is a basis adapted to the derived algebra (with $\{E_1, E_2, E_3\}$ being a basis for $\lieg'$ and $E_4$ complementary to $\lieg'$), 
then we require that  $[E_1, E_4]$, $[E_2, E_4]$, $[E_3, E_4]$ are linearly independent, that is
\begin{equation}
	[E_1, E_4] \wedge  [E_2, E_4]\wedge [E_3, E_4]  \neq 0,
\end{equation} 
so that $\lieg'$ is 3-dimensional.
These conditions  are too unwieldy to display explicitly.

For algebras 4, 5, 10, 12 we provide gauge conditions  (``Null Rotation Gauge") which eliminate the null rotation group from the residual group.  For algebras 7 and 11 the gauge conditions (``Gauge'') put the adjoint matrix of the complement of the derived algebra $\lieg'$, which is symmetric with respect to the induced inner product on $\lieg'$, into normal form.  See Section 6 for details.

Algebras 13--15 are those spacetime Lie algebras for which the derived algebra is a simple Lie algebra. Such algebras always admit a 1-dimensional center $\liez$ and the spacetime character of the center is the basis for our classification.  The derived algebra is (semi-) simple if and only if its Killing form is non-degenerate, which implies the Killing form of $\mathfrak g/\liez$  is non-degenerate.  These non-degeneracy conditions are  given in terms of quantities $c_1,\dots, c_6$, which are explicitly defined in Section 6. For these algebras, the residual group is gauge-fixed by putting  the Killing form of $\mathfrak g/\liez$ into standard form. The relevant conditions are denoted by ``Gauge''.  See Section 6 for details.

The spacetime Lie algebras 16--25 all have 2-dimensional derived algebras  $\lieg'$. If $\{E_1, E_2, E_3,E_4\}$ is a basis adapted to the derived algebra (with $\{E_1, E_2\}$ being a basis for $\lieg'$ and $\{E_3, E_4\}$ complementary  to $\lieg'$), the conditions that 
$[E_1, E_3]$,   $[E_1, E_4]$, $[E_2, E_3]$,   $[E_3, E_4]$ span $\lieg'$, that is, the $E_1$, $E_2$ plane, are tacitly assumed. 
Again, our classification is based upon the spacetime signature of the derived algebra but now, in addition,
some case-splitting analysis is needed to impose the Jacobi identities. The conditions listed here reflect that analysis.

Throughout, we refer to the algebras either by their number in the table or by an extension of nomenclature introduced earlier. For example, 2.10.3 and {\bf abel3LN3} refer to the same  Lie algebra.  

\newpage
\begin{center}
 \fbox{\rule[-5pt]{0pt}{20pt} \enspace \large
	\bf  Spacetime Groups with Heisenberg Derived Algebra\enspace} 
\end{center}

\newcommand\LR{\mathcal R}
\newcommand\LS{\mathcal S}
\newcommand\LT{\mathcal T}

\columnsep =24pt
\medskip
\noindent
\begin{multicols*}{2}
\setlength{\abovedisplayskip}{2pt}
\setlength{\belowdisplayskip}{2pt}
\setlength{\baselineskip}{8pt}
\parindent = 0pt
\columnseprule = .5pt
\parskip = 4pt
\noindent
\small
\SpaceTime{\bf \normalsize \heisRS}\label{HeisRS}
\begin{flalign*}
	[\, K,\, L\,] & = 2\,\mu_0\,( K-  L) &
\\[.5\jot]
	[\, K,\, M\,]  &= \kappa(K-L)+(i\epsilon_1+i\gamma_1-\mu_0)\, M+\sigma\, \barM 
\\[.5\jot]
	[\, L,\, M\,] & = \kappa(K-L)+(i\epsilon_1+i\gamma_1-\mu_0)\, M+\sigma\, \barM
\\[.5\jot]
	[\, M,\, \barM\,]  &= 2\,i\,(\gamma_1 - \epsilon_1)\, (K- L)
\end{flalign*}
	Derived Series\,:  $\lieg' = \langle K-L,\  M,\   \barM\rangle$, $\lieg'' = \langle K-L \rangle$
\par		
	3 Dim. Derived\,:  $(\epsilon_1 +\gamma_1)^2
		+\mu_0^2-\bar\sigma\,\sigma\neq 0$ 
\par
	Heisenberg: $\epsilon_1 - \gamma_1 \neq 0$
\par
	Isometry Jump\,: $\kappa = \sigma  = 0$
\par
	Residual\,: $\{ R_{M \barMs},\ \mathcal T,\ \mathcal Y,\ \mathcal Z \}$
\SpecialDGlabel{HeisRS}
\mybigskip
\SpaceTime{\bf \normalsize \heisLT}\label{HeisLT}
\SpecialDGlabel{HeisLT} 
\vspace{-2pt}
\begin{flalign*}
	[\, K,\, L\,] & = 2\,\mu_0(K + L)&
\\[1\jot]
	[\, K,\, M\,] & = -\kappa(K + L)-(i\gamma_1 -i\epsilon_1 -\mu_0)\, M+\sigma\, \barM
\\[1\jot]
	[\, L,\, M\,] & = \kappa\, (K+ L) +( i\gamma_1 - i\epsilon_1 - \mu_0)\, M-\sigma\, \barM
\\[.5\jot]
	[\, M,\, \barM\,] & = 2\,i\,(\gamma_1+ \epsilon_1)\,(K+ L)
\end{flalign*}
	Derived Series\,:  $\lieg' = \langle K + L,\  M,\   \barM\rangle$,  $\lieg'' = \langle K + L \rangle$
\par	
	3 Dim. Derived\,: $(\epsilon_1 -\gamma_1)^2
		+\mu_0^2-\bar\sigma\,\sigma\neq 0$ 
\par
	Heisenberg\,: $\gamma_1  +\epsilon_1 \neq 0$
\par
	Isometry Jump\,: $\kappa = \sigma  = 0$
\par
	Residual\,:  $\{ R_{M \barMs},\ \mathcal T,\ \mathcal Y,\ \mathcal Z \}$
\mybigskip
\noindent
\SpaceTime{\bf  \normalsize\heisLS}\label{HeisLS}
\SpecialDGlabel{HeisLS}
\vspace{-2pt}
\begin{flalign*}
	[\, K,\, L\,] & = 2\,\pi_0\, (M+ \barM)&
\\[.5\jot]
	[\, K,\, M\,] & = -i\,(2\,\beta_1 +3\,\tau_1)\, K-i\kappa_1\, L+2\,i\epsilon_1(M+ \barM)
\\[.5\jot]
	[\, L,\, M\,] & = -i\nu_1\, K+ i\,(2\beta_1 + \tau_1)\, L+2\,i\gamma_1 (M+ \barM)
\\[.5\jot]
	[\, M,\, \barM\,] & = 2\,i\tau_1(M + \barM)
\end{flalign*} 
	Derived Series\,:  $\lieg' = \langle K,\ L,\  M +   \barM\rangle$,  $\lieg'' = \langle M + \barM \rangle$
\par			
	3 Dim. Derived\,:  $  \nu_1\,\kappa_1+3\,{\tau_1}^{2}+8\,\tau_1\,\beta_1+4\,{\beta_1}^{2} \neq 0$
\par
	Heisenberg\,: $\pi_0  \neq 0$     
\par
	Isometry Jump\,: $\epsilon_1 = \gamma_1= \kappa_1 = \nu_1=0$
\par
	Residual\,:   $\{ B_{KL}^*,\ \mathcal R,\ \mathcal Y,\ \mathcal Z \}$
\begin{flalign*}
\\[0pt]
\end{flalign*}            
\columnbreak
\mybigskip
\noindent
\SpaceTime{\bf \normalsize \heisLN}\DGlabel{HeisLN}
\SpecialDGlabel{HeisLN}
\immediate\write5{"2.4.1" = "HeisLN1",}
\immediate\write5{"2.4.2" = "HeisLN2",}
\begin{flalign*}
	[\, K,\, L\,] & = 0 &
\\[.5\jot]
	[\, K,\, M\,] & = \,i\,(2\,\beta_1 - \tau_1)\, K
\\[.5\jot]
	[\, L,\, M\,] & = \bar\nu\, K  -i\,(2\beta_1 + \tau_1)\, L + i\,(2\,\gamma_1-\mu_1)\,(M + \barM)
\\[.5\jot]
	[\, M,\, \barM\,] & = 2\,i\mu_1 K+4\,i\epsilon_1 L-4\,i\beta_1( M + \barM)
\end{flalign*}
	Derived Series\,:  $\lieg' = \langle K,\ L,\  M +  \barM \rangle$, $\lieg'' = \langle K \rangle$ 
\par
	3 Dim. Derived\,:
	$ \tau_1\,\beta_1+2\,\beta_1^2-2\,\epsilon_1\,\gamma_1+\epsilon_1\,\mu_1 \neq 0$
\par
	Heisenberg\,: $\nu_0 \neq 0 $
\par
	Isometry Jump\,: $\epsilon_1 = \gamma_1= \mu_1 = \nu =0$
\par
	Residual\,: $\{ B_{KL}^*,\  N_{K,u},\ \mathcal R, \ \mathcal Y \}$
\par
	{\bf 4.1.} Null Rotation Gauge\,: If $\epsilon_1 \neq 0$,  set $\tau_1 = 0$
\par
	{\bf 4.2.} Null Rotation Gauge\,: If  $\epsilon_1 = 0$ and $\tau_1 + 2 \beta_1 \neq 0$, set $\mu_1 = 0$
\mybigskip
\noindent
\SpaceTime{\bf \normalsize \heisNS }\label{HeisNS}
\SpecialDGlabel{HeisNS}
\immediate\write5{"2.5.1" = "HeisNS1",}
\begin{flalign*}
	[\, K,\, L\,] & = -2\,\gamma_0 K+2\,\bar\tau\, M+2\,\tau\, \barM &
\\[.5\jot]
	[\, K,\, M\,] & = 2\,i\epsilon_1( M +  \barM)
\\[.5\jot]
	[\, L,\, M\,] & = -i\nu_1 K + (2\,i\gamma_1 - \mu_0)\, M+(\gamma_0 + 2\,i\,\gamma_1)\, \barM
\\[.5\jot]
	[\, M,\, \barM\,] & = 0
\end{flalign*}
	Derived Series\,: $\lieg' =  \langle K,\ M,\ \barM \rangle$, $\lieg'' =  \langle M + \barM \rangle$
\par
	3 Dim. Derived\,: $ \nu_1\,\tau-\nu_1\,\bar\tau+i{\gamma_0}^{2}+i\mu_0\gamma_0
	\neq 0$
\par
	Heisenberg\,: $\epsilon_1 \neq 0$
\par
	Residual\,: $\{ B_{KL}^*,\  N_{K,iv},\ \mathcal R, \ \mathcal Y \}$
\par
	{\bf 5.1.}  Null Rotation Gauge\,: Set $\tau_0 =0$.  
\medskip	

\begin{flalign*}
\\[100pt]
\end{flalign*}

\end{multicols*}
\newpage
%
%
\setlength{\topmargin}{.1in}
\setlength{\headheight}{-.1in}
\setlength{\headsep}{0in}
\setlength{\textwidth}{7in}
\begin{center} \fbox{\rule[-5pt]{0pt}{20pt} \enspace \large \bf  Spacetime Groups with 3 Dimensional Abelian Derived Algebra \enspace}
\end{center}
\smallskip
\begin{multicols}{2}
\small
\setlength{\abovedisplayskip}{3pt}
\setlength{\belowdisplayskip}{2pt}
\setlength{\baselineskip}{9pt}
\parindent = 0pt
\columnseprule = .1pt
\parskip = 3pt

\noindent
\SpaceTime{\bf \normalsize \abelthreeRS}\label{IIIdAbelianRS}
\SpecialDGlabel{IIIdAbelianRS}
\begin{flalign*}
	[\, K,\, L\,] & = 2\,\epsilon_0( K  - L )-2\,\bar\kappa\, M-2\,\kappa\, \barM &
\\[.5\jot] 
	[\, K,\, M\,] & = \kappa\, (K- L)+(\rho_0 + 2\,i\epsilon_1)\, M+\sigma\, \barM
\\[.5\jot]
	[\, L,\, M\,] & = \kappa\,(K-\, L)+(\rho_0 + 2\,i\epsilon_1)\, M+\sigma\, \barM
\\[.5 \jot]
	[\, M,\, \barM\,] & = 0
\end{flalign*}    
	Derived Algebra\,: $\lieg' = \langle\, K - L,\, M,\, \barM \,\rangle$ 
\par                
	Skew-Adjoint Line\,:  $\zeta = \epsilon_1(K-L) \neq 0$
\par	
	Isometry Jump\,:  $\kappa = \sigma = 0$
\par
	Residual\,: $\{ R_{M \barMs},\ \mathcal T,\ \mathcal Y,\ \mathcal Z \}$
\mybigskip
\noindent
\SpaceTime{\bf  \normalsize \abelthreeRZ}\label{IIIdAbelianRZ}
\SpecialDGlabel{IIIdAbelianRZ}
\immediate\write5{"2.7.1" = "IIIdAbelianRZ1",}
\begin{flalign*}
	[\, K,\, L\,] & = 2\,\epsilon_0(K- L)-2\,\bar\kappa\, M-2\,\kappa\, \barM &
\\[.5\jot]
	[\, K,\, M\,] & = \kappa\,(K- L)+ \rho_0\, M+\sigma\, \barM
\\[.5\jot]
	[\, L,\, M\,] & = \kappa\,(K- L) + \rho_0\, M+\sigma\, \barM
\\[.5\jot]
	[\, M,\, \barM\,] & = 0
\end{flalign*}
	Derived Algebra\,: $\lieg' = \langle\, K - L,\ M,\ \barM\, \rangle$                          
\par
	Skew-Adjoint Line\,:  $\zeta = 0$
\par
	Isometry Jump\,: $\rho_0 = \pm\,\sigma_0 -2\epsilon_0$ or $\sigma_0= 0$
\par
	Residual\,: Rotation group  $O(3)$ on $\mathfrak g'$, $\mathcal T$
\par
       {\bf 7.1} Gauge\,: $ \kappa = 0$, $\sigma_1 = 0$;
\par
	\quad\enskip\  Residual : 
	$\{ \mathcal T,\ \mathcal U,\  \mathcal V,\ \mathcal Y,\ \mathcal Z\}$
\mybigskip
\noindent
\SpaceTime{\bf  \normalsize \abelthreeLT}\label{IIIdAbelianLT}
\SpecialDGlabel{IIIdAbelianLT}
\begin{flalign*}
	[\, K,\, L\,] & = -2\,\epsilon_0( K +  L)  + 2\,\bar\kappa\, M+2\,\kappa\, \barM &
\\[.5\jot]
	[\, K,\, M\,] & = -\kappa\, (K + L)+(\rho_0-2\,i\gamma_1)\, M+\sigma\, \barM
\\[.5\jot]
	[\, L,\, M\,] & = \kappa\, (K+ L)-(\rho_0-2\,i\gamma_1)\, M-\sigma\, \barM
\\[.5\jot]
	[\, M,\, \barM\,] & = 0
\end{flalign*}
	Derived Algebra\,:   $\lieg' = \langle \,K +L,\, M,\, \barM \,\rangle$
\par
	Skew-Adjoint Line\,:  $\zeta = \gamma_1\,(K +L) \neq 0 $
\par
	Isometry Jump\,: $\kappa= \sigma = 0$
\par
	Residual\,: $\{R_{M \barMs},\ \mathcal T,\ \mathcal Y,\ \mathcal Z\}$
\mybigskip
\SpaceTime{\bf \normalsize \abelthreeLS}\label{IIIdAbelianLS}
\SpecialDGlabel{IIIdAbelianLS}
\begin{flalign*}
	[\, K,\, L\,] & = 0 &
\\[.5\jot]
	[\, K,\, M\,] & = i\,(\alpha_1-\beta_1-\tau_1)\, K-i\,\kappa_1 L+i\,\epsilon_1\,(M + \barM)
\\[.5\jot]
	[\, L,\, M\,] & = -i\nu_1K -i(\alpha_1 -\beta_1 + \tau_1)\, L+i\mu_1\,(M + \barM)
\\[.5\jot]
	[\, M,\, \barM\,] & = 2\,i\mu_1 K+2\,i\epsilon_1 L -i\,(\alpha_1 + \beta_1)\,\,(M + \barM)
\end{flalign*}
	Derived Algebra\,: $ \lieg' =  \langle\, M + \barM,\ K,\ L \rangle$
\par
	Skew-Adjoint Line\,:  $\zeta = (\alpha_1 - \beta_1)\,(M + \barM) \neq 0$
\par
	Isometry Jump\,: $\epsilon_1 = \kappa_1= \mu_1 = \nu_1 = 0$ 
\par	Residual\,: $\{ B_{KL}^*,\ \mathcal R,\ \mathcal Y,\ \mathcal Z \}$
\vfill
\columnbreak
\small
\setlength{\abovedisplayskip}{2pt}
\setlength{\belowdisplayskip}{2pt}
\setlength{\baselineskip}{10pt}
\parskip = 4pt
\parindent = 0pt
\noindent
\SpaceTime{\bf \normalsize \abelthreeLN}\label{IIIdAbelianLN}
\SpecialDGlabel{IIIdAbelianLN}
\immediate\write5{"2.10.1" = "IIIdAbelianLN1",}
\immediate\write5{"2.10.2" = "IIIdAbelianLN2",}
\immediate\write5{"2.10.3" = "IIIdAbelianLN3",}
\immediate\write5{"2.10.4" = "IIIdAbelianLN4",}
\begin{flalign*}
	[\, K,\, L\,] & = 0&
\\[.5\jot]
	[\, K,\, M\,] & = -i\tau_1  K-i\kappa_1 L+i\epsilon_1\,(M + \barM)
\\[.5\jot]
	[\, L,\, M\,] & = -i\nu_1 K-i\tau_1 L +i\,(2\,i\gamma_1 -\mu_1)\,(M + \barM)
\\[.5\jot]
	[\, M,\, \barM\,] & = 2\,i\mu_1 K+2\,i\epsilon_1 L-2\,i\alpha_1\,(M + \barM)
\end{flalign*}
	Derived Algebra\,: $\lieg' =  \langle K,\ L,\ M + \barM \rangle$  
\par
	Skew-Adjoint Line\,:  $\zeta = (\mu_1 - \gamma_1)\,K \neq  0$
\par
	Isometry Jump\,: $\{\epsilon_1 = \gamma_1 =  \kappa_1 = \mu_1 = \nu_1 =0\}$ or
\par
	\quad $\{\epsilon_1 =\gamma_1 = \kappa_1 =0, \ \tau_1 = -2\alpha_1\}$
\par
	Residual\,: $\{B_{KL}^*,\  N_{K,u},\ \mathcal R, \ \mathcal Y \}$
\par
	{\bf 10.1} Null Rotation  Gauge\,: If $\kappa_1 \neq 0$, then  $\epsilon_1 =0$
\par
	{\bf 10.2} Null Rotation Gauge\,: If $\kappa_1  = 0,\ \epsilon_1 \neq 0$, then  $\tau_1 + 2\alpha_1 =0$
\par
	{\bf 10.3} Null Rotation Gauge\,: If $\kappa_1  = 0,\ \epsilon_1 =0$ and  $\tau_1 + 2\alpha_1 \neq 0 $, then  
	$\gamma_1 =0$
\par
	{\bf 10.4} Null  Rotation Gauge\,: If $\kappa_1  = 0,\ \epsilon_1 =0,\ \tau_1 + 2\alpha_1= 0$, $\gamma_1 \neq 0$,
	 then  $\nu_1 =0$
\mybigskip
\immediate\write5{"2.11.1" = "IIIdAbelianLZ1",}
\immediate\write5{"2.11.2" = "IIIdAbelianLZ2",}
\immediate\write5{"2.11.3" = "IIIdAbelianLZ3",}
\immediate\write5{"2.11.4" = "IIIdAbelianLZ4",}
\SpaceTime{\bf \normalsize \abelthreeLZ}\label{IIIdAbelianLZ}
\SpecialDGlabel{IIIdAbelianLZ}
\begin{flalign*}
	[\, K,\, L\,] & = 0 &
\\[.5\jot]
	[\, K,\, M\,] & = -i\,\tau_1 K-i\kappa_1 L+i\,\epsilon_1( M+  \barM)
\\[.5\jot]
	[\, L,\, M\,] & = -i\,\nu_1 K-i\,\tau_1 L+ i\,\gamma_1( M+ \barM)
\\[.5\jot]
	[\, M,\, \barM\,] & = 2\,i\gamma_1 K+2\,i\epsilon_1 L-2\,i\alpha_1 (M + \barM)
\end{flalign*}
	Derived Algebra : $\lieg' = \langle K,\, L,\, M + \barM \rangle$
\par
	Skew-Adjoint Line\,:  $\zeta = 0 $
\par
	Residual\,: Lorentz group $O(2,1)$ on $\lieg'$, $\mathcal Y$
\par	
	{\bf 11.1} Gauge I\,: $\nu_1 = \kappa_1,\ \epsilon_1 =0,\ \gamma_1=0$
\par
	Residual: $\{\mathcal R,\  \mathcal T,\ \mathcal V,\ \mathcal Y,\ \mathcal Z\}$
\par	
	Isometry Jump: $\alpha_1 = -(\tau_1 \pm \kappa_1)/2$ or $\kappa_1 = 0$ 
\par
	{\bf 11.2} Gauge II\,:  $\nu_1 = -\kappa_1,\ \epsilon_1 =0,\ \gamma_1=0$
\par
	Residual: $\{\mathcal R,\ \mathcal T,\ \mathcal Y,\  \mathcal Z\}$; 
	Isometry Jump: NA
\par
	{\bf 11.3} Gauge III\,: $ \epsilon_1 =0, \gamma_1=0, \kappa_1 =0$, 
\par
	Residual: $\{B_{KL}^*,\ \mathcal R,\ \mathcal Y \}$

	Isometry Jump\,: $\nu_1 = 0$ or $\tau_1 = -\alpha_1$
\par
	{\bf 11.4} Gauge IV\,: $\epsilon_1=0,\ \kappa_1 =0,\ \nu_1 =0,\ \tau_1 = -2\alpha_1$, 
	Residual\,: $\{B_{KL}^*,\ \mathcal R,\ \mathcal Y \}$;
	Isometry Jump\,: $\gamma_1 = 0$

\mybigskip
\SpaceTime{\bf \normalsize \abelthreeNS }\label{IIIdAbelianNS}
\SpecialDGlabel{IIIdAbelianNS}
\immediate\write5{"2.12.1" = "IIIdAbelianNS1",}
\begin{flalign*}
	[\, K,\, L\,] & = -2\,\gamma_0 K -4\,i\,\beta_1 (M -  \barM), \quad [\, K,\, M\,]  = 0  &
\\[.5\jot]
	[\, L,\, M\,] & = \bar\nu\, K -\mu_0 M-\bar\lambda\, \barM, \quad [\, M,\, \barM\,]  = 0
\end{flalign*}
	Derived Algebra\;:  $\lieg'= \langle K,\ M,\ \barM \rangle$  

	Skew-Adjoint Line: $\zeta =  \beta_1(M + \barM) \neq 0$

	Isometry Jump: $\nu =0,\ \lambda = 0,\ \mu_0 = 0$

Residual: $\{ B_{KL}^*,\ N_{K,iv},\ \mathcal R, \  \mathcal Y\}$

{\bf 12.1} Null Rotation Gauge: $\gamma_0 = 0$

\medskip
\end{multicols}
\vfill
\newpage
\begin{center}
\fbox{\rule[-5pt]{0pt}{20pt} \enspace \large \bf Spacetime Groups with Simple  Derived Algebra\enspace}
\end{center}
\setlength{\abovedisplayskip}{2pt}
\setlength{\belowdisplayskip}{2pt}
\setlength{\baselineskip}{10pt}
\parindent = 0pt
\columnseprule = .5pt
\parskip = 3pt
\noindent
\SpaceTime{\bf \LevitwoT}\label{simpT}
\SpecialDGlabel{simpT}
\immediate\write5{"2.13" = "simpT",}
\immediate\write5{"2.13.1" = "simpT1",}
\immediate\write5{"2.13.2" = "simpT2",}
\setlength{\abovedisplayskip}{3pt}
\setlength{\belowdisplayskip}{3pt}
\setlength{\parskip}{4pt}
\begin{flalign*}
[\, K,\, L\,] & = 0, \quad
[\, K,\, M\,]  = -\bar\nu\, K+ (4\,\beta-3\,\bar\nu)\, L - i\,(2\,\gamma_1- \mu_1)\, M+\bar\lambda\, \barM
\\[.5\jot]
[\, L,\, M\,] & = \bar\nu\, K+(-4\,\beta+3\,\bar\nu)\, L +i\,(2\,\gamma_1 -\mu_1)\, M-\bar\lambda\, \barM &
\\[.5\jot]
[\, M,\, \barM\,] & = 2\,i\mu_1\, K+ 2\,i\,(2\epsilon_1 + 2\gamma_1 - \mu_1)\, L+ 2\,(2\,\bar\beta-\nu)\, M -2\,(2\,\beta -\bar\nu)\, \barM
\end{flalign*}
	Center\,: $\liez = \langle K + L \rangle$.  
\enskip Simple\,: $c_1\,c_2\, c_3 \neq 0$. 
\enskip	Residual\,: $O(3)$ acting on $\gmodz$, $\mathcal T$
\par	
	Gauge: $\beta_0 = \nu_0/2,\ \lambda_0=0,\ \nu_1=-2\, \beta_1$. \enskip  Isometry Jump\,: $ \beta= 0 $ or $\{\lambda_1=0,\ \mu_1 = 2\epsilon_1\}$

\par
\qquad {\bf \ref{simpT}.1} Distinct Eigenvalues. 
	Residual\,: $\{\mathcal T,\ \mathcal U,\ \mathcal V,\ \mathcal Y,\ \mathcal Z \}$
\par
\qquad {\bf \ref{simpT}.2} Repeated Eigenvalues: $\lambda_1 = 0$. 
	Residual\,:  $\{ R_{M \barMs},\  \mathcal T ,\ \mathcal Y,\ \mathcal Z\}$
\par
\mybigskip
\noindent
\SpaceTime{\bf \LevitwoS}\label{simpS}
\SpecialDGlabel{simpS}
\immediate\write5{"2.14" = "simpS",}
\immediate\write5{"2.14.1" = "simpS1",}
\immediate\write5{"2.14.2" = "simpS2",}
\immediate\write5{"2.14.3" = "simpS3",}
\immediate\write5{"2.14.4" = "simpS4",}
\immediate\write5{"2.14.5" = "simpS5",}
\immediate\write5{"2.14.6" = "simpS6",}
\immediate\write5{"2.14.7" = "simpS7",}
\immediate\write5{"2.14.8" = "simpS8",}
\setlength{\abovedisplayskip}{3pt}
\setlength{\belowdisplayskip}{3pt}
\setlength{\parskip}{4pt}
\begin{flalign*}
[\, K,\, L\,] & = 0, \quad [\, K,\, M\,]  = \bar\nu\, K -(4\,\beta +3\,\bar\nu)\, L +i\,(2\,\gamma_1 - \mu_1)\, M-\bar\lambda\, \barM &
\\[.5\jot]
[\, L,\, M\,] & = 
\bar\nu\, K -(4\,\beta + 3\,\bar\nu)\, L +i\,(2\,\gamma_1 - \mu_1)\, M-\bar\lambda\, \barM
\\[.5\jot]
[\, M,\, \barM] & = 2\,i\mu_1\, K+ 2\,i\,(2\,\epsilon_1-2\,\gamma_1+ \,\mu_1)\, L+ 2\,(2\,\bar\beta+\,\nu)\, M -2\,(2\,\beta +\,\bar\nu)\, \barM
\end{flalign*}
	Center\,: $\liez= \langle K -L \rangle$.	
	\enskip Simple\,: $c_3\,\neq 0$. 
	\enskip Residual\,: O(2,1) acting on $\gmodz,\ \mathcal Z$.
\par
 Gauge {\bf I}\,: $\nu_0 = -2\beta_0, \enskip \nu_1= 2\beta_1, \enskip  \lambda_0 = 0$.
\par	
	
\qquad {\bf \ref{simpS}.1} Distinct Eigenvalues.
	Residual\,: $\{ \mathcal T,\ \mathcal U,\ \mathcal Y,\ \mathcal Z \}$. \enskip Isometry Jump\,:  $\{\lambda_1 =0,\ \mu_1 = -2\epsilon_1\}$.
\par
\qquad {\bf \ref{simpS}.2} 
	Two Equal Eigenvalues\,: $\lambda_1 = -\mu_1 - 2\epsilon_1$.\enskip   
	Residual\,: $\{B^*_{K+L, \ M + \barMs},\ \mathcal R,\ \mathcal Y,\ \mathcal Z\}$ 
\par
\qquad \qquad \ Isometry Jump\,: $\{ \beta_1 =0,\ \gamma_1 = \epsilon_1\}$
\par
\qquad {\bf \ref{simpS}.3} 
	Two Equal Eigenvalues: $\lambda_1 =0$. \ 
	Residual:  $\{ R_{M, \barMs},\ \mathcal T,\  \mathcal Y,\  \mathcal Z\}$. 	
\enskip   Isometry Jump\,:  $\beta =0$.

\par	
	 Gauge {\bf II}\,: $\nu_0 = -2\beta_0, \enskip  \lambda_0 = 0, \enskip 
	\lambda_1 = -2\epsilon_1 - \mu_1,\ c_6 \neq 0$.	 
\quad	 Simple\,: $c_3\,(c_4^2 + c_6^2) \neq 0$.
\par
\qquad {\bf \ref{simpS}.4} Residual $\{ \mathcal R,\ \mathcal T,\ \mathcal Y,\ \mathcal Z\}$
\par
	Gauge {\bf III}\,: $\nu_0 = -2\beta_0, \enskip  \lambda_0 = 0, \enskip \lambda_1=  -8\,\beta_1 -2\,\epsilon_1 - \mu_1 +4\,\nu_1, \enskip c_6 \neq 0 $.
\quad Simple\,: $c_3c_4  \neq 0$. 

\qquad {\bf \ref{simpS}.5} Two Equal Eigenvalues.\enskip  
	Residual\,: $\{B^*_{K+L,M + \barMs},\  \mathcal R \circ \mathcal T,\  \mathcal Y,\  \mathcal Z\}$.
\par
\qquad\qquad \	Isometry Jump\,: $\{\beta_1 =0,\ \nu_1 =0,\ \epsilon_1 = \gamma_1\}$
\par
\qquad {\bf \ref{simpS}.6} Three Equal Eigenvalues\,: $\beta_0=0$, $\nu_1=2\,\beta_1+ 2/3\,\epsilon_1 + 1/3\,\mu_1$. 

\qquad\qquad Residual: $\{B^*_{K+L,M + \barMs},\  \mathcal R \circ \mathcal T,\  \mathcal Y,\  \mathcal Z \}$.
 \ Isometry Jump\,: $\{   \beta_1 + \epsilon_1/6+  \gamma_1/2 +
	\mu_1/3 = 0\}$  

\par
\qquad {\bf \ref{simpS}.7} Three Equal Eigenvalues\,: $\beta_1 =0,\  \epsilon_1 = - 3 \gamma_1 -
	2\mu_1$, $\nu_1=-2\gamma_1 - \mu_1$. 

\qquad\qquad Residual: $\{\mathcal R \circ \mathcal T,\  \mathcal Y\}$.
 \ Isometry Jump\,: $\{  \beta_0=0\}$  
	
\par
	Gauge {\bf IV}\,: 
	$\nu_0=-2\,\beta_0 +2\gamma_1/3 -\mu_1/3,\
	\nu_1= 2\,\beta_1 +4\gamma_1/3-2\mu_1/3,\
	\epsilon_1 = 2\gamma_1 - 3\mu_1/2,$
	
	\quad\qquad\qquad$\lambda_0=\lambda_1=4\gamma_1/3-2\mu_1/3$
\par
	\qquad Simple\,: $c_4 c_6 \neq 0$
\par
\qquad	{\bf \ref{simpS}.8} Residual $\{\mathcal R \circ \mathcal T,\   \mathcal Z\}$. 
\enskip 	Isometry Jump\,: $\{\lambda_0 = \lambda_1 = 0\}$	
\noindent
\SpaceTime{\bf \LevitwoN}\label{simpN}
\SpecialDGlabel{simpN}
\immediate\write5{"2.15.1" = "simpN1",}
\immediate\write5{"2.15.2" = "simpN2",}
\immediate\write5{"2.15.3" = "simpN3",}
\setlength{\abovedisplayskip}{3pt}
\setlength{\belowdisplayskip}{3pt}
\setlength{\parskip}{4pt}
\begin{flalign*}
	[\, K,\, L\,] & = 0,  \quad [\, K,\, M\,] = 0, \quad 
	[\, L,\, M\,]  = \bar\nu\, K-4\,\beta\, L + i\,( 2\gamma_1 -\mu_1)\, M-\bar\lambda\, \barM &
\\[.5\jot]
[\, M,\, \barM\,] & = 2\,i\mu_1 K+4\,i\epsilon_1 L+4\,\bar\beta\, M-4\,\beta\, \barM
\end{flalign*}
	Center\,: $\liez = \langle\, K \,\rangle$.  
\enskip
	Residual\,: $B^*_{KL}$,  $R_{M\barMs}$,  $N_K$, $\mathcal Y$
\par
 Gauge {\bf V}\,:  $ \beta_0 = \beta_1 = \lambda_0 =0$.\enskip 
	Simple\,: $c_1 c_2  \neq 0$
	
	\qquad {\bf \ref{simpN}.1} Distinct Eigenvalues. Residual\,: $\{ B_{K,L}^*,\ \mathcal U,\ \mathcal Y \}$
\par
	\qquad {\bf \ref{simpN}.2} Repeated Eigenvalues: $\lambda_1 = 0.$ \enskip
	Residual\,: $\{ B_{K,L}^*,\ R_{M\barMs},\ \mathcal Y \}$.
	\enskip Isometry Jump\,: $\lambda_1 = \nu = 0$
\par
Gauge {\bf VI}\,:  $\beta_1 =0, \enskip \epsilon_1 =0, \enskip \lambda_0 =0, \enskip 
	\lambda_1 = 2\, \gamma_1 - \mu_1$.
	\quad Simple\,: $\beta_0\,(2\gamma_1 - \mu_1) \neq0 $.
\par
	\qquad {\bf \ref{simpN}.3} 
	Residual\,: $\{B^*_{KL},\ \mathcal R,\ \mathcal Y\}$
\noindent
\vfill
\newpage
\begin{center}
 \fbox{\rule[-5pt]{0pt}{20pt} \enspace \large \bf Spacetime Groups with 2-Dimensional Derived Algebra\enspace}
\end{center}
\par
\medskip
\begin{multicols}{2}
\small
\setlength{\abovedisplayskip}{4pt}
\setlength{\belowdisplayskip}{4pt}
\setlength{\baselineskip}{10pt}
\parindent = 0pt
\columnseprule = .5pt
\parskip = 4pt

{\bf  Riemannian 2D Abelian Derived Algebra}

Derived Algebra\,: $\lieg' = \langle \, M,\,  \barM \rangle, \quad [\,M, \barM\,] = 0$

\SpaceTime{\abeltwoRone}\label{IIdAbelR1}
\SpecialDGlabel{IIAbelR1}
\begin{flalign*}
	[\, K,\, L\,] & = 4\,\bar\beta\, M+4\,\beta\, \barM &
\\[.5\jot]
	[\, K,\, M\,] & = (2\,i\,\epsilon_1  + \rho_0)\, M+\sigma_0 \barM
\\[.5\jot]
	[\, L,\, M] & = -(\mu_0 + 2\,iq\epsilon_1)\, M-\sigma_0q\, \barM	
\end{flalign*}
	Conditions\,:  $\sigma_0 \neq 0$, $q^2 = 1$

	Residual Group\,: $\{\mathcal T,\ \mathcal U,\ \mathcal Y,\ \mathcal Z \}$
\mybigskip
\noindent
\SpaceTime{\abeltwoRtwo}\label{IIdAbelR2}
\SpecialDGlabel{IIAbelR2}
\begin{flalign*}
	[\, K,\, L\,] & = 4\,\bar\beta\, M+4\,\beta\, \barM &
\\[.5\jot]
	[\, K,\, M\,] & = (2\,i\,\epsilon_1 + \rho_0)\, M+\sigma\, \barM
\\[.5\jot]
	[\, L,\, M\,] & = -\mu_0 M
\end{flalign*} 
	Conditions\,: $\sigma \neq 0$
\par
	Residual Group\,: $\{R_{M \barMs},\enskip B_{KL}^*,\ \mathcal Y\}$

\mybigskip
\noindent
\SpaceTime{\abeltwoRthree}\label{IIdAbelR3}  
\SpecialDGlabel{IIAbelR3}  
\begin{flalign*}
	[\, K,\, L\,] & = 4\,\bar\beta\, M+4\,\beta\, \barM &
\\[.5\jot]
	[\, K,\, M\,] & = (2\,i\,\epsilon_1 + \rho_0)\, M
\\[.5\jot]
	[\, L,\, M\,] & = (2\,i\,\gamma_1 -\mu_0)\, M.
\end{flalign*}
	Conditions\,: $\sigma=\lambda=0$
\par	
	Isometry Jump\,:  $\beta = 0$ or $\{\epsilon_1= \gamma_1 = \mu_0 = \rho_0 = 0 \}$.
\par
	Residual Group\,: $\{R_{M \barMs},\enskip B_{KL}^*, \ \mathcal Y,\ \mathcal Z\}$
\par
\mybigskip
\noindent
\hbox{\bf   Lorentzian 2D Abelian Derived Algebra}

Derived Algebra\,: $\lieg' = \langle\, K, L \, \rangle, \quad [\, K, L\,] = 0 $.

\SpaceTime{\abeltwoLone}\label{IIdAbelL1} 
\SpecialDGlabel{IIAbelL1}  
\begin{flalign*}
[\, K,\, M\,] & = -(2\,\beta_0+\tau)\, K-\kappa_0 L &
\\[.5\jot]
[\, L,\, M\,] & = \nu_0  K+(2\,\beta_0 -\tau)\, L
\\[.5\jot]
[\, M,\, \barM\,] & = 4\,i\,\gamma_1  K+4\,i\,\epsilon_1 L
\end{flalign*}
	Conditions\,: $\beta_0^2 + \kappa_0^2 + \nu_0^2 \neq 0$
\par	
	Isometry Jump\,: $\epsilon_1= \gamma_1 = \kappa_0 = \nu_0 = 0$ 
\par
	Residual  Group\,: $\{ B_{KL}^*, \ \mathcal R,\ \mathcal Y,\ \mathcal Z\}$
\par
\mybigskip
\noindent
\SpaceTime{\abeltwoLtwo}\label{IIdAbelL2} 
\SpecialDGlabel{IIAbelL2}
\begin{flalign*}
	[\, K,\, M\,] & = -\tau\, K,  \quad 
	[\, L,\, M\,]  = -\tau\, L &
\\[.5\jot]
	[\, M,\, \barM\,] & = 4\,i\,\gamma_1 K+4\,i\,\epsilon_1 L
\end{flalign*}
	Conditions\,: $\kappa_0 = \nu_0 = \beta_0 =0$
\par
	Isometry Jump\,: $\tau=0$ or $\{\epsilon_1 = \gamma_1= 0\}$
\par
	Residual Group\,: $\{ R_{M \barMs}, \enskip B_{KL}^*, \ \mathcal Y,\ \mathcal Z\}$
\vfill
\columnbreak

\noindent
{\bf  Null 2D Abelian Derived Algebra}

Derived Algebra\,: $\lieg' = \langle K,\, M + \barM \rangle $.

\SpaceTime{\abeltwoNone}\label{IIAbelN1}  
\SpecialDGlabel{IIAbelN1}   
\begin{flalign*}
[\, K,\, L\,] & = 2\,\mu_0 K, \quad [\, K,\, M]  = 2\,i\,\epsilon_1 (M + \barM) &
\\[.5\jot]
[\, L,\, M\,] & = -i\,\nu_1 K+(2\,i\,\gamma_1 -\mu_0 - i\, \mu_1)\,( M + \barM)
\\[.5\jot]
[\, M,\, \barM\,] & = 2\,i\,\mu_1 K-2\,i\,\beta_1 (M +\barM)
\end{flalign*}
Conditions\,: $\epsilon_1 \neq 0$

Residual Group\,: $\{ B_{KL}^*,\ \mathcal R, \  \mathcal Y \}$

\mybigskip
\noindent
\SpaceTime{\abeltwoNtwo}\label{IIAbelN2}
\SpecialDGlabel{IIAbelN2}    
\begin{flalign*}
	[\, K,\, L\,] & = -2\,\gamma_0 K+2\,\tau_0 (M+ \barM) &
\\[.5\jot]
	[\, K,\, M\,] & = 4\,i\,\beta_1 K , \quad [\, L,\, M]  = \bar\nu\, K 
\\[.5\jot]
	[\, M,\, \barM\,] & = -4\,i\,\beta_1 (M + \barM)
\end{flalign*}
	Conditions\,: $\tau_0 \neq 0$
\par
	Isometry Jump\,: $\nu = \gamma_0 = 0$
\par
	Residual Group\,:$\{ B_{KL}^*,\ \mathcal R, \  \mathcal Y \}$

\mybigskip
\noindent
\SpaceTime{\abeltwoNthree}\label{IIAbelN3} 
\SpecialDGlabel{IIAbelN3}      
\begin{flalign*}
	[\, K,\, L\, ] & = 2\,\mu_0 K , \quad [\, K,\, M]  = -2\,i\,\tau_1 K&	
\\[.5\jot]
	[\, L,\, M\, ] & = -i\,\nu_1 K+(2\,i\,\gamma_1 -\mu_0)\,( M+  \barM)
\\[.5\jot]
	[\, M,\, \barM\, ] & = -i\,(2\,\beta_1 -\tau_1 )\, (M+  \barM)
\end{flalign*}
	Conditions\,: $\tau_1 + 2\beta_1 \neq 0$
\par
	Isometry Jump\,: $\gamma = \mu_0 = \nu_1 =0$
\par
	Residual Group\,: $\{ B_{KL}^*,\  \mathcal R, \  \mathcal Y \}$

\par 
\mybigskip
\noindent
\SpaceTime{\abeltwoNfour}\label{IIAbelN4}   
\SpecialDGlabel{IIAbelN4}  
\begin{flalign*}
[\, K,\, L\, ] & = 0, \quad [\, K,\, M] = 4\,i\,\beta_1 K &
\\[.5\jot]
[\, L,\, M\, ] & = \bar\nu\, K -i\,\mu_1\ (M+ \barM)
\\[.5\jot]
[\, M,\, \barM\, ] & = 2\,i\,\mu_1 K-4\,i\,\beta_1 (M +\barM)
\end{flalign*}
Conditions\,: $\beta_1 \neq 0$
\par
Isometry Jump\,: $\mu_1 = \nu = 0$

Residual Group\,: $\{ B_{KL}^*,\ \mathcal R, \  \mathcal Y \}$

\mybigskip
\noindent
\SpaceTime{\abeltwoNfive}\label{IIAbelN5}    
\SpecialDGlabel{IIAbelN5}
\begin{flalign*}
\kern\the\parindent
[\, K,\, L\,] & = -2\,\gamma_0 K,  \quad [\, K,\, M]  = 4\,i\,\beta_1 K&
\\[.5\jot]
[\, L,\, M\,] & = \bar\nu\, K
\\[.5\jot]
[\, M,\, \barM\,] & = -4\,i\,\beta_1 (M + \barM)
\end{flalign*}
	Conditions\,:  $\beta_1 \neq 0$
\par
	Isometry Jump\,: $\gamma_0 = \nu = 0$
\par
	Residual Group: $\{ B_{KL}^*,\ \mathcal R, \  \mathcal Y \}$

\end{multicols}
\newpage
\section{Spacetime Group Solutions to the Einstein Equations}\label{GenRel}
\parindent = 12pt 
\setlength{\baselineskip}{12pt}

	In this section we provide a list of {\it all} spacetime groups which are solutions to
	the Einstein equations with the following matter sources:

\medskip
\begin{tabular}{lll}	
&$\bullet$ 
 	 Vacuum 
&$\bullet$ 
	Einstein (cosmological term)
\\[2\jot]
&$\bullet$ 
	Pure Radiation with Cosmological Term
&$\bullet$ 
	Perfect Fluids
\\[2\jot]
&	$\bullet$ 
 	Inheriting, Non-Null Maxwell with Cosmological Term
&$\bullet$ 
	Non-Inheriting, Non-Null Maxwell
\\[2\jot]
&$\bullet$ 
	Free, Massless Scalar Field
\end{tabular}

\medskip
	The classifications of the simply transitive vacuum and Einstein solutions are
	due to Petrov \cite{Petrov:1962} and  Kaigorodov \cite{Kaigorodov}; see   \cite{Stephani} for details. The spacetime groups which 
	admit  perfect  fluid solutions  are classified in
	\cites{Farnsworth-Kerr:1966a, Ozsvath:1965c}.
	We have independently verified all these 	
	results and present them here within the context 
	of our general classification of spacetime groups.\footnote{
	In some of the fluid solutions we find the formulas for the 
	 energy density and pressure to be slightly different than that given in 
	 \cite{Stephani}.
	Note that the perfect fluid solutions 
	(12.30--32) given in \cite{Stephani} are 
	contained in one case here  -- the splitting into the three cases of \cite{Stephani}
	is only required for the explicit integration of 
	the Newman-Penrose equations to find the coordinate form of the spacetime metric.  }

	According to  \cite{Stephani}, all homogeneous pure radiation 	solutions with vanishing cosmological term have an isometry dimension greater than 	4, so there are no such pure radiation spacetime groups. The classification given in this section  of pure radiation spacetime groups with a 				cosmological term is, to the best of our  knowledge, new.  
	
	The classification of scalar field solutions includes the possibility of a cosmological term.  This classification is new.
	
	According to  \cite{Stephani} (see also \cite{Komrakov}), there are no non-null inheriting electrovacua (no cosmological term) which are spacetime 	groups. A spacetime group which is an inheriting, non-null Einstein-Maxwell solution with a cosmological term appears in 
	 \cite{Ozsvath:1965a}.  (Our calculations correct the formula for the metric in   that paper.)
	  As the author of  \cite{Ozsvath:1965a} remarked, 
	one branch of his classification was incomplete.   Our independent analysis  implies that that this branch yields 
	no additional solutions, so that the solution exhibited in \cite{Ozsvath:1965a} is in fact the only such solution with a simply transitive 			maximal 4-dimensional  isometry group. 

	A 1-parameter family of non-null, non-inheriting solutions to the Einstein-Maxwell equations (without cosmological term) appears in 					\cite{McLenaghan-Tariq:1975a}; see also \cites{Tupper:1976, Stephani}. This family of solutions was derived from an ansatz based upon
	properties of the principal null directions and
	not upon  symmetry considerations so, until now, it was not known if there are any other spacetime groups with  a non-inheriting electromagnetic 	field.
	We have shown this family of solutions in fact represents 
	{\it all} non-null, non-inheriting solutions to the Einstein-Maxwell equations with a simply transitive maximal 4-dimensional isometry 
	group.\footnote{In addition, we have found that there are no such multiply transitive solutions. Details will appear elsewhere.}
	
	For  pure radiation solutions and for perfect fluid solutions, the matter fields are necessarily invariant under the isometry group $G$ of the spacetime.  Consequently, the field equations can be solved purely in terms of the Lie algebra  -- local coordinates are not introduced. A non-trivial  free scalar field $\phi$ on a spacetime group cannot be $G$-invariant; however, the field equations for a massless free field only depend upon  $\omega = d\phi$ which  {\it is} $G$-invariant.  An electromagnetic field on a spacetime group may be $G$-invariant (inheriting) or may not be $G$-invariant (non-inheriting) \cite{Henneaux:1984}. For a non-inheriting electromagnetic field, it is still possible to formulate the field equations in terms of $G$-invariant data. Indeed, the field equations imply the electromagnetic field is $G$-invariant up to a duality rotation.  The duality rotation angle is a function $\theta\colon G\to {\bf R}$ such that $\omega=d\theta$ is $G$-invariant. This implies the existence of a $G$-invariant  2-form $f$ obeying the  field equations displayed below.  The electromagnetic field is   related to $f$ by a duality rotation by $\theta$.  
	For the scalar field solutions and the non-inheriting electromagnetic solutions we express the $G$-invariant closed form $\omega$ in terms of the dual basis $(\Theta_K, \Theta_L, \Theta_M, \Theta_{\barMs})$ associated to $(K, L, M, \barM)$. In summary, for all matter fields considered in this paper, the  field equations can be formulated in terms of $G$-invariant data and hence can be reduced to purely algebraic equations for the spin coefficients and matter variables.  
	
	Our approach to studying the spacetime group solutions to the 
	Einstein equations is simply to evaluate and solve the field equations for each case of our general classification.
	Again, we emphasize that multiply-transitive solutions are excluded from our analysis.
	Groebner basis techniques \cite{Cox-Little-Shea:2000a}, \cite{Heck:2003a} prove essential
	in solving the various 
	algebraic equations.  Indeed, in almost every case, a Groebner basis
	computation for the polynomial system of equations for 
	the spin coefficients and matter variables arising from the field equations yields
	simple  algebraic consequences which
	make solving the field equations feasible.  Details of these computations will appear elsewhere.


It should be emphasized that local 
	coordinates are not needed or used to find  the solutions in this section. To obtain local coordinate expressions for these solutions, one may invoke Lie's Third Theorem \cites{Flanders:1989a , Spivak:1979}.  This theorem asserts that for any abstract $n$-dimensional Lie algebra
	one can find $n$ point-wise independent vector fields $X_1, X_2$, \ldots, $X_n$,  
	defined locally on a coordinate chart, such that the structure equations for these vector fields coincide with that of the Lie algebra. Moreover, if the 
	Lie algebra is solvable, then these vectors fields may be taken to be globally defined on all of ${\bf R}^n$. The spacetime metric is then easily constructed from the dual basis to the vector fields $X_i$.
	For example, for the scalar field solution (3.12), one finds a coordinate form of the metric to be
\begin{align*}
	h = &2\Big\{
-\,dx\otimes dx+(4\,\alpha_1\,x+4\,\kappa_1\,y)\,dx\odot dw+\,dy\otimes dy+(-4\,\alpha_1\,y+4\,\kappa_1\,x)\,dy\odot dw+\,dz\otimes dz+
\\
&
8\,\alpha_1\,z\,dz\odot dw+
(-4\,\alpha_1^{2}{x}^{2}+4\,\alpha_1^{2}{y}^{2}+16\,\alpha_1^{2}{z}^{2}-16\,\alpha_1\,\kappa_1\,xy+4\,\kappa_1^{2}{x}^{2}-4\,\kappa_1^{2}{y}^{2}+1)\,dw\otimes dw \Big\}.
\end{align*}
	Here $A\odot B = \frac{1}{2}(A\otimes B + B\otimes A)$. It is of  interest to note that this formula gives the metric in terms of
	its  own spin coefficients. 
	In section 7 we show how this result was obtained using the {\sc DifferentialGeometry} software.
	
Our conventions are as follows.  The metric signature is $(-+++)$. The curvature, Ricci tensor,  Ricci scalar, and Einstein tensor are defined in terms of the metric and metric-compatible covariant derivative by
\begin{equation*}
Z^d{}_{;ba} - Z^d{}_{;ab} = R^d{}_{cab} Z^c,\quad R_{bc} = R^d{}_{bdc},\quad R= R^a{}_a, \quad G^{ab} = R^{ab}  - \frac{1}{2} R \\h^{ab}.
\end{equation*}
The Hodge star operation on 2-forms, $F_{ab} = - F_{ba}$, is defined in terms of the Levi-Civita tensor $\epsilon_{abcd}$ by 
\begin{equation*}
{}^*F_{ab} = \frac{1}{2} \epsilon_{ab}{}^{cd} F_{cd}.
\end{equation*}
We give explicit forms for the various field equations at the beginning of each of the following tables of solutions.
\vfill\eject
\newcommand\StrutA{\rule[-8pt]{0pt}{22pt}}
 \newcommand\StrutB{\rule[-8pt]{0pt}{26pt}}
 Table 3.1: Solutions to the Einstein Field Equations\\

\begin{tabular}{|l|l|}
\hline
\multicolumn{2}{|l|}{\Strut \sc \normalsize Vacuum Solutions\quad $G^{ab}=0$}
\\
\hline
\StrutA
{\bf \SolnNo. \bf  abel3LZ2} 
\immediate\write5{"3.\theSolncounter" = "VacAbel3LZ1",}

&
$
\StrutB
	\alpha=\dfrac{\sqrt {3}}{3}i\,\kappa_1,\enskip
	\beta= \dfrac{\sqrt {3}}{3}i\,\kappa_1,\enskip
	\gamma=0,\enskip
	\epsilon=0,\enskip
	\kappa=i\,\kappa_1,\enskip
	\lambda=0,\enskip
	\mu=0,\enskip
	\nu=-i\,\kappa_1,\enskip
$
\\
\StrutA

&
$
	\pi =  \dfrac{\sqrt {3}}{3}i\,\kappa_1, \enskip
	\rho = 0,\enskip
	\sigma =0,\enskip
	\tau =\dfrac{\sqrt{3}}{3}\, i \, \kappa_1.
$
	\quad {\sc Petrov Type: I \quad Ref:  \cite{Stephani} (12.14)} 
\\[2\jot]
\hline
\noalign{\vspace{5\jot}}
\hline
\multicolumn{2}{|l|}{\Strut \sc \normalsize  Einstein Spaces \quad  $G^{ab} + \Lambda h^{ab} = 0,\quad \Lambda= {\rm const.}$}
\\
\hline
\StrutA
{\bf\SolnNo. abel3LS} 
\immediate\write5{"3.\theSolncounter" = "EinAbel3LS1",}

&
$
	\alpha=-\dfrac{1}{2}i\,\beta_1,\enskip
	\beta=i\,\beta_1,\enskip
	\gamma=0,\enskip
	\epsilon=i\,\epsilon_1,\enskip
	\kappa=0,\enskip
	\lambda=0,\enskip
	\mu=0,\enskip
	\nu=0,\enskip
	\pi=-\dfrac12\, i\,\beta_1,\enskip
$
\\
&
$	
	\rho=i\,\epsilon_1,\enskip
	\sigma=i\,\epsilon_1,\enskip
	\tau= -\dfrac12 i\,\beta_1,\enskip  
	\Lambda = -\dfrac32\,\beta_1^2.
$
	\quad {\sc Petrov Type: III \quad Ref: \cite{Stephani} (12.35) }
\\[2\jot]
\hline
\noalign{\vspace{5\jot}}
\hline
\multicolumn{2}{|l|}{\StrutA\sc \normalsize  Pure Radiation With Cosmological Term:
$G^{ab} + \Lambda\, h^{ab}  = \phi^2 N^a N^b,\enskip \Lambda, \phi = {\rm const.} \enskip N^aN_a=0. $
} 
\\[2\jot]
\hline
\StrutB
{\bf \SolnNo. abel3LS} 
\immediate\write5{"3.\theSolncounter" = "RadAbel3LS1",}

&
$	\alpha = i\,\alpha_1,\enskip
	\beta= - \dfrac{1}{2}i\,\alpha_1,\enskip
	\gamma = i\,\mu_1,\enskip
	\epsilon = 0,\enskip
	\kappa = 0,\enskip
	\lambda = i\,\mu_1,\enskip
	\mu = i\mu_1,\enskip
	\nu = i\nu_1,\enskip
$
\\
&
$	
	\pi = - \dfrac{1}{2}i\,\alpha_1,\enskip
	\rho = 0, \enskip
	\sigma = 0, \enskip
	\tau = -\dfrac12 i\, \alpha_1, \enskip
	\Lambda = -\dfrac32\alpha_1^2,\enskip
	\phi^2 = 3 \alpha_1\nu_1,\enskip
	 N = K.  \quad 
$
\\
&
\Strut
	\text{\sc Petrov Type: III} 
\\[2\jot]
\cline{2-2}
\StrutB
{\bf \SolnNo.} 
\immediate\write5{"3.\theSolncounter" = "RadAbel3LS2",}

&
$	
	\alpha = - \dfrac{1}2 i\,\beta_1,\enskip
	\beta =  i\,\beta_1,\enskip
	\gamma = 0,\
	\epsilon = i\,\epsilon_1,\enskip
	\kappa = 0 ,\enskip
	\lambda = 0,\enskip
	\mu = 0,\enskip
	\nu = i\,\nu_1,\enskip
	\pi = - \dfrac{i}2\,\beta_1,

$
\\
&
$
	\rho = i\, \epsilon_1,\enskip	
	\sigma = i\, \epsilon_1,\enskip
	\tau = - \dfrac{1}2 i\,\beta_1, \enskip
	\Lambda = -\dfrac32\,\beta_1^2,\enskip
	\phi^2 = -9 \,\beta_1\nu_1,\enskip
	N = K.
$
\\
&
\Strut
	\text{\sc Petrov Type: I}  \enskip 
\\[2\jot]
\hline
\StrutA
{\bf \SolnNo. simpCT1}
\immediate\write5{"3.\theSolncounter" = "RadSimp3CT1",}

&
$ 
	\alpha = 0,\enskip
	\beta = 0,\enskip
	\gamma = i\,\epsilon_1,\enskip
	\epsilon = i\, \epsilon_1,\enskip
	\kappa = 0,\enskip
	\lambda = 2\,i\,\epsilon_1, \enskip
	\mu = 6\,i\,\epsilon_1, \enskip
	\nu = 0,\enskip
	\pi = 0, \enskip
$	
\\
&
$
	\rho = -2\,i\,\epsilon_1,\enskip
	\sigma = -2\,i\,\epsilon_1,\enskip
	\tau = 0, \enskip
	\Lambda = 16\,ie^2,  \enskip
	\phi^2 = 64\,ie^2,  \enskip
	N = K. 
$
\\
&
\Strut 
	\text{\sc Petrov Type: I}  
\\[2\jot]
\hline

\StrutB
{\bf \SolnNo. simpCS1}
\immediate\write5{"3.\theSolncounter" = "RadSimp3CS1",} 
&
$ 
	\alpha= \dfrac{\sqrt {15}}{21} i\, \mu_1,\enskip
	\beta= \dfrac{\sqrt {15}}{7} i\,\mu_1, \enskip
	\gamma= i\,\mu_1,\enskip
	\epsilon= \dfrac{3}{7}\,i\,\mu_1,\enskip
	\kappa=0,\enskip
	\lambda = \dfrac{1}{7}\, i \,\mu_1,\enskip
	\mu=i\mu_1,\enskip	
$	
\\
&\StrutA
$	
	\nu = \dfrac {4\sqrt {15}}{21}\, i\,\mu_1,\enskip
	\pi = \dfrac {2\sqrt {15}}{21}\, i\,\mu_1,\enskip
	\rho= -\dfrac{1}{7}\, i\,\mu_1,\enskip
	\sigma= \dfrac{1}{7}\,i\,\mu_1,\enskip
	\tau=   \dfrac{2  \sqrt {15}}{21}\, i \,\mu_1,\enskip	
$
\\
&\StrutA
$	\Lambda = \dfrac{16}{49}\,\mu_1^2, \enskip
	\phi^2  =  \dfrac{128}{147}\,\mu_1^2 ,  \enskip
	N = K.
$ 	\quad\sc Petrov Type: I  

\\[2\jot]
\cline{2-2}
{\bf \bf \SolnNo.}
\immediate\write5{"3.\theSolncounter" = "RadSimp3CS2",}

&
\StrutA
$
	\alpha=0,\enskip
	\beta=0,\enskip
	\gamma=i\,\gamma_1,\enskip
	\epsilon=-i\gamma_1,\enskip
	\kappa=0, \enskip
	\lambda=2\,i\,\gamma_1,\enskip
	\mu=6\,i\,\gamma_1,\enskip
	\nu=0, \enskip
	\pi =0,
	
$ 
\\
\Strut
&
$	\rho=2\,i\,\gamma_1,\enskip
	\sigma=2\,i\gamma_1,\enskip
	\tau=0, \enskip
	\Lambda = -16\,\gamma_1^2, \enskip
	\phi^2 = 64\,\gamma_1^2, \enskip
	N = K. \quad 
$
\\
& \Strut 
	\text{\sc Petrov Type: I}  
\\[2\jot]
\hline

\end{tabular}

\newpage
 Table 3.1 (cont.)\\
\\
\begin{tabular}{|l|l|}
\hline
\multicolumn{2}{|l|}{\StrutA\sc \normalsize Perfect Fluid Solutions
\quad
$G^{ab} = \phi^2 U^a U^b + \psi h^{ab},\quad U^a U_a = -1, \quad \phi, \psi = {\rm const.}$}
\\
\hline
\StrutB
{\bf \SolnNo. \abelthreeLS}
\immediate\write5{"3.\theSolncounter" = "FluAbel3LS1",}

&
$ 
	\alpha= \dfrac {i}{4a} 
	\,(s^{3} +\sqrt {2}s^{2} -2\,s  -\sqrt {2}),\enskip
	\beta= \dfrac{i}{4a}\,(-s^3 + \sqrt {2}s^{2} +2\,s  - \sqrt {2}),\enskip
	\gamma=0,\enskip
	\epsilon=0,\enskip			
$
\\
\StrutA
&
$
	\kappa= \dfrac{i}{4a}\,(2\,s - \sqrt {2}),\enskip
	\lambda=0,\enskip
	\mu=0, \enskip
	\nu= -\dfrac{i}{4a}\,( 2\,s+\sqrt {2}),\enskip
	\pi = \dfrac{i}{4a}\sqrt {2}, \enskip
	\rho = 0,\enskip		
$
\\[2\jot]
&
$
	\sigma = 0, \enskip
	\tau= \dfrac{i}{4a}\sqrt {2}, 
	\enskip U = \dfrac{\sqrt{2}}{2}(K +L), \enskip
	\phi^2 = \dfrac{1}{a^2}(-2s^4 + 5s^2 -2), \enskip
	\psi = \dfrac{1}{2 a^2}( -s^2 +2).
$
\\[2\jot]
&
\Strut 

	{\sc Petrov Type: I} \qquad  {\sc Ref:} \cite{Stephani} (12.30--32)
\\
\hline
\StrutB
{\bf \SolnNo. simpCT1}
\immediate\write5{"3.\theSolncounter" = "FluSimCT1",}

&
$
	\alpha=0,\enskip
	\beta=0,\enskip
	\gamma= \dfrac{1}{2}\, i\,\fluidparm\,( v+1) ,\enskip
	\epsilon= \dfrac{1}{2}\, i \,\fluidparm\,( v+1) ,\enskip
	\kappa=0,\enskip
	\lambda=\lambda_0+i\lambda_1,\enskip
$
\\
\Strut
&
$
	
	\mu= -i\,\fluidparm\,(v-3), \enskip
	\nu=0,\enskip
	\pi = 0, \enskip
	\rho=i\,\fluidparm\,( 3\,v-1) ,\enskip
	\sigma= \lambda_0 -i\,\lambda_1,\enskip
	\tau=0,\enskip
$
\\
\StrutB
&
	$
	U = \dfrac{\sqrt{2}}{2}(  \dfrac{1}{\sqrt{v}}\, K  + \sqrt{v}\, L), \enskip
	\phi^2 = 32\, \fluidparm^2 v , \enskip 
	\psi = -4\,\fluidparm^2(v-1)^2, \enskip 
	\fluidparm^2 = \dfrac{ \lambda_0^2  + \lambda_1^2}{v^2 - 6v +1}.
	$
\\
&
\Strut
	{\sc Petrov Type: I} \quad  {\sc Ref:} \cite{Stephani} (12.27)
\\
\hline
\StrutB
{\bf \SolnNo. simpCS1}
\immediate\write5{"3.\theSolncounter" = "FluSimCS1",}

&
$
	\alpha=0,\enskip
	\beta=0,\enskip
	\gamma= -\dfrac{1}{2}\,i\,\fluidparm\,( v-1) ,\enskip
	\epsilon= \dfrac{1}{2}\,i\,\fluidparm\,( v-1) ,\enskip
	\kappa=0,\enskip
	\lambda=\lambda_0+i\,\lambda_1,\enskip
$
\\
\Strut
&
$
	
	\mu = i\,\fluidparm\,(v+3), \enskip
	\nu=0,\enskip
	\pi = 0, \enskip
	\rho=i\fluidparm\,( 3\,v+1) ,\enskip
	\sigma= -\lambda_0  + i\,\lambda_1,\enskip
	\tau=0,\enskip
$
\\
\StrutB
&
	$
	U = \dfrac{\sqrt{2}}{2}(  \dfrac{1}{\sqrt{v}}\, K  +\sqrt{v}\, L), \enskip
	\phi^2 = 32 \,\fluidparm^2v , \enskip 
	\psi = 4\, \fluidparm^2(v+1)^2, \enskip 
	\fluidparm^2 = \dfrac{ \lambda_0^2  + \lambda_1^2}{v^2 + 6v +1}.
	$
\\
&
\Strut
	{\sc Petrov Type: I} \quad  {\sc Ref:} \cite{Stephani} (12.28)
\\
\cline{2-2}
\StrutA
{\bf \SolnNo.}
\immediate\write5{"3.\theSolncounter" = "FluSimCS2",}

&
$ 
	\alpha=0,\enskip
	\beta=0,\enskip
	\gamma=0,\enskip
	\epsilon=0,\enskip
	\kappa=0,\enskip
	\lambda=\lambda_0+i\,\lambda_1,\enskip
	\mu=i\mu_1,\enskip
	\nu=0,\enskip
	\pi = 0, \enskip
	\rho=i\mu_1,\enskip
	
$
\\
&
$
	\sigma= -\lambda_0 + i \,\lambda_1,\enskip
	\tau=0,\enskip
	U = \dfrac{\sqrt{2}}{2}\,(K +L), \enskip
	\phi^2 = -4\,\lambda_0^{2}-4\,\lambda_1^{2}+4\,\mu_1^{2}, \enskip
	
$
\\
&	
\Strut
	$\psi =  -2\,\lambda_0^{2}-2\,\lambda_1^{2}+2\,\mu_1^{2}$. \quad  {\sc Petrov Type: D} \quad  {\sc Ref:} \cite{Stephani} (12.29)
\\[2\jot]
\hline
\end{tabular}

\newpage
Table 3.1 (cont.)\\
\\
\begin{tabular}{|l|l|}
\hline
\multicolumn{2}{|l|}{\StrutA\sc \normalsize Massless Scalar Fields  \quad $G^{ab} + \Lambda h^{ab} = \omega^a \omega^b - \frac{1}{2} \omega_c\omega^c h^{ab},\quad \omega_{[a;b]} = 0,  \quad \omega^a_{;a} =0$}
\\
\hline
\StrutA
{\bf\SolnNo. abel3LZ2}

\immediate\write5{"3.\theSolncounter" = "ScaAbel3LZ1",}

&
$ 
	\alpha=i\,\alpha_1,\enskip
	\beta=i\,\alpha_1,\enskip
	\gamma=0,\enskip
	\epsilon=0,\enskip
	\kappa=i\,\kappa_1,\enskip
	\lambda=0,\enskip
	\mu=0,\enskip
	\nu=-i\,\kappa_1,\enskip
$
\\
&
$
	\pi = i\,\alpha_1,\enskip
	\rho=0,\enskip
	\sigma=0,\enskip
	\tau=i\,\alpha_1,\enskip
	\omega = i\phi\,(\Theta_{\barMs} - \Theta_{\vphantom{\barMs} M}),\enskip 
	\Lambda  = 0,
	
$
\\
&
	$
	\phi^2 = 2\,\kappa_1^2-6\,\alpha_1^2.
	$
	\Strut  \quad {\sc Petrov Type: I} \quad  
\\
\hline
\StrutA
{\bf\SolnNo. abel2L1}
\immediate\write5{"3.\theSolncounter" = "ScaAbel2L11",}
&
$ 
	\alpha = 0,\enskip
	\beta = 0,\enskip
	\gamma = 0,\enskip
	\epsilon = 0,\enskip
	\kappa = \kappa_0,\enskip
	\lambda = 0,\enskip
	\mu = 0,\enskip
	\nu = \nu_0,\enskip
	\pi = i \, \tau_1,\enskip
	\rho = 0,\enskip

$
\\
&
\Strut  $
	\sigma = 0,\enskip
	\tau = i\tau_1,\enskip
	\omega = \phi\left(\Theta_{\vphantom{\barMs}M}  + \Theta_{\barMs}\right), \enskip
	\Lambda = -4\tau_1^2, \enskip
	\phi^2 = 2\,\tau_1^2 + 2 \,\kappa_0\nu_0. 
	$ 
\\
\Strut
& 	
	{\sc Petrov Type: I} \qquad  

\\
\cline{2-2}
\StrutA
{\bf\SolnNo. } 
\immediate\write5{"3.\theSolncounter" = "ScaAbel2L12",}
&
$ 
\StrutB 
	\alpha=0,\enskip
	\beta=0,\enskip
	\gamma= 6 i\,\dfrac {\epsilon_1^3}{\kappa_0^2},\enskip
	\epsilon=i\epsilon_1, \enskip
	\kappa=\kappa_0, \enskip
	\lambda=0, \enskip
	\mu= 12\, \,i\,\dfrac {\epsilon_1^3}{\kappa_0^2},\enskip
	\nu=-36\,\dfrac {\epsilon_1^4}{\kappa_0^3},\enskip
	
$
\\
&
\rule[-8pt]{0pt}{24pt}
$
	\pi = 2 \,\dfrac{\epsilon_1^2}{\kappa_0},\enskip
	\rho=2\,i\epsilon_1,\enskip
	\sigma=0,\enskip
	\tau=-2\,\dfrac {\epsilon_1^2}{\kappa_0},\enskip	
	\omega = i\phi\,(   \Theta_{\barMs}- \Theta_{\vphantom{\barMs}M}), \enskip
	\Lambda = -64\,\dfrac{\epsilon_1^4}{\kappa_0^2}, \enskip
	
$ 
\\
&	
\Strut
$ \phi^2 =  80\, \dfrac{\epsilon_1^4}{\kappa_0^2}.$ \quad	

{\sc Petrov Type: I} \quad  
\\[3\jot]

\cline{2-2}
\StrutA
{\bf\SolnNo. }
\immediate\write5{"3.\theSolncounter" = "ScaAbel2L13",}

&
\StrutB 
$ 
	\alpha = -\dfrac{3}{2}\,\tau_0,\enskip
	\beta = -\dfrac32\,\tau_0,\enskip
	\gamma = 0,\enskip
	\epsilon = i\,\epsilon_1,\enskip
	\kappa = \dfrac {\epsilon_1^2}{\tau_0},\enskip
	\lambda = 0,\enskip
	\mu = 0,\enskip
	\nu = 0,\enskip
	\pi = -\tau_0, \enskip	
$
\\
&
\Strut
$	
	\rho= 2i\,\epsilon_1,\enskip
	\sigma = 0,\enskip
	\tau = \tau_0,\enskip
	\omega =  i\phi\,(\Theta_{\barMs} - \Theta_{\vphantom{\barMs}M}),\enskip
	\Lambda = -4\, \tau_0^2,\enskip
	\phi^2= 2\, \tau_0^2.\quad
	
$ 
\\
&\Strut
	 {\sc Petrov Type: III} 
\\

\cline{2-2}
\StrutA
{\bf\SolnNo. }
\immediate\write5{"3.\theSolncounter" = "ScaAbel2L14",}

&
\StrutB 
$ 
	\alpha=- \dfrac{1}{2} \tau_0,\enskip
	\beta=- \dfrac{1}{2}\tau_0,\enskip
	\gamma=0,\enskip
	\epsilon=0,\enskip
	\kappa=\kappa_0,\enskip
	\lambda=0,\enskip
	\mu=0,\enskip
	\nu=0,\enskip
$
\\
&
\Strut
$	\pi = -\tau_0 + i\, \tau _1,\enskip
	\rho=0,\enskip
	\sigma=0,\enskip
	\tau=\tau_0+i\tau_1, \enskip
	\omega =  \sqrt{2}\left[(\tau_1 - i\tau_0)\,\Theta_{\vphantom{\barMs}M} + 
	(\tau_1 + i\tau_0)\,  \Theta_{\barMs}\right],\enskip	
$ 
\\
&\Strut
	$\Lambda = -4\,(\tau_0^2 + \tau_1^2)$. \quad	
{\sc Petrov Type: N} \quad 
\\
\hline
\end{tabular}

\begin{tabular}{|l|l|}
\hline
\multicolumn{2}{|l|}{\StrutA\sc \normalsize Inheriting,  Non-Null Einstein-Maxwell Solutions With Cosmological Constant}
\\
\multicolumn{2}{|l|}{\StrutA $G^{ab} + \Lambda h^{ab} =  F^{ac}F^b{}_c  - \frac{1}{4} F_{cd}F^{cd} h^{ab},\quad d F = d{}^*F = 0$}

\\
\hline
{\bf\SolnNo. abel3LS}
\immediate\write5{"3.\theSolncounter" = "MaxAbel3LS1",}
&\StrutB 
$
	\alpha=  \dfrac{1}{3}\,i\, \left( 2 - \sqrt {2}\right) \tau_1,\enskip
       \beta= \dfrac{1}{3}\,i\, \left( \sqrt {2}+2 \right) \tau_1,\enskip
       \gamma=0,\enskip
       \epsilon=0,\enskip
       \kappa= \dfrac{3}{14}\,i \left(2\,\sqrt {2} -1 \right) \tau_1,\enskip
$
\\
&\StrutA
	
$	
	\lambda=0,\enskip
	\mu = 0,\enskip
	\nu= -{\dfrac {14}{27}}\, i  \left( 1+2\,\sqrt {2} \right) \tau_1 , \enskip
       \pi=i\,\tau_1,\enskip
       \rho=0,\enskip
       \sigma=0,\enskip
       \tau=i\tau_1
$
\\
&\StrutA 
$	
	\Lambda = -\dfrac{4}{3}\, \tau_1^2, \enskip
	F = i\,\tau_1\,( \Theta_K  + \dfrac{14}{9} \Theta_L) \wedge (\Theta_{\vphantom{\barMs}M}- \Theta_{\barMs}).
$ 
	\quad{\sc Petrov Type: I} \quad  {\sc Ref:}  \cite{Ozsvath:1965a} 

\\
\hline

\hline
\multicolumn{2}{|l|}{\StrutA\sc \normalsize Non-Inheriting, Non-Null Einstein-Maxwell Solutions}
\\
\multicolumn{2}{|l|}{\StrutA $G^{ab} = f^{ac}f^b{}_c - \frac{1}{4} f_{cd}f^{cd} h^{ab},\quad df = \omega \wedge {}^*f,$
$ d{}^*f =  -\omega \wedge f, \quad d\omega=0$}

\\
\hline
{\bf\SolnNo. heisLT }
\immediate\write5{"3.\theSolncounter" = "MaxHeisLT1",}

&
\Strut

$
	\alpha=0,\enskip
	\beta=0,\enskip
	\gamma=i\,\epsilon_1,\enskip
	\epsilon=i\,\epsilon_1,\enskip
	\kappa=0,\enskip
	\lambda=2\,\epsilon_1,\enskip
	\mu=2\,i\epsilon_1,\enskip
	\nu=0,\enskip
	\pi=0,\enskip	
$
\\
\Strut
&
$
	\rho=2\,i\,\epsilon_1,\enskip
	\sigma=2\,\epsilon_1,\enskip
	\tau=0, \enskip
	\omega = 4\,\epsilon_1(\Theta_K -\Theta_L), \enskip
$
\\
&
\Strut
$
	f =  -2a\,\Theta_K\wedge \Theta_L +2i\sqrt{\rule[-2pt]{0pt}{8pt}
	 8\epsilon_1^2 -a^2}\,\Theta_{\vphantom{\barMs}M} \wedge \Theta _{\barMs}.
$	
	
\\
\Strut &
{\sc Petrov Type: I} \quad  {\sc Ref:}  \cite{McLenaghan-Tariq:1975a};  \cite{Tupper:1976}; \cite{Stephani}, eq.~(12.21)

\\
\hline

\end{tabular}

%
%

\newcommand{\s}{\scriptsize}
\newcolumntype{L}{>{$\s} l <{$}}
\newcolumntype{K}{>{\scriptsize} l   } 
\newcolumntype{C}{>{$\s } c <{$}}

\newpage
\parindent = 12pt
\section{Further  Applications}\label{Apps}
\setlength{\baselineskip}{12pt}

	In this section we provide additional  applications of our classification of spacetime Lie groups.
	We begin by showing how a number of well-known spacetime groups are classified by our methods.
	We then illustrate how our classification results relate to one of the standard classifications of Lie algebras \cite{Snobol-Winternitz}.
	In Section \ref{PCSG} we enumerate all spacetime Lie groups which are algebraically special 
	for generic values of their spin coefficients. 
	For these spacetimes a complete classification by  Petrov type is given. 
	Finally, in Section \ref{CEMSG} we give examples of spacetime Lie groups, with Heisenberg or 3 dimensional abelian derived algebras, which admit 
	conformally Einstein metrics.
		
\par
\medskip
\noindent
\subsection{Classification of Spacetimes Groups in General Relativity}\label{CSGR}

To illustrate the ease with which our classification scheme may be invoked, we consider
	three examples of spacetime groups taken from the general relativity literature.

	We begin with the unique spacetime group (up to an overall scale) which solves the vacuum Einstein equations (see (3.1))  \cite{Petrov:1962}. The metric 
	is 
	defined in local coordinates $(x, y, z, t)$ by (\cite{Stephani}, see equation (12.14))
\begin{equation*}
h = \frac{1}{k^2}\bigl(dx \otimes dx + e^{-2x} dy \otimes dy + e^x\,\cos(\sqrt{3}x)\,(dz \otimes dz - dt \otimes dt) -2e^x \sin(\sqrt{3} x)\, dt \odot dz\bigr).
\end{equation*}
	A group-invariant orthonormal tetrad for $h$ 
	is (with $E_1$ time-like):
\begin{align*}
	E_1 &= ke^{-x/2}\bigl(\cos(\frac{\sqrt{3}}{2} x)\partial_t + \sin(\frac{\sqrt{3}}{2} x)\partial_z\bigr),
\quad
	E_2 = ke^{-x/2}\bigl(\sin(\frac{\sqrt{3}}{2} x)\partial_t - \cos(\frac{\sqrt{3}}{2} x)\partial_z\bigr),
\\[1\jot]
	E_3 &= ke^x\partial_y,\quad 
	E_4 = k \partial_x.
\end{align*}
	The non-zero structure equations are
\begin{equation*}
[E_1, E_4] = \frac{k}{2}\bigl(E_1 + \sqrt{3}E_2\bigr),\quad
[E_2, E_4] = \frac{k}{2}\bigl(-\sqrt{3}E_1 + E_2\bigr),\quad
[E_3, E_4]=-k E_3.
\end{equation*}
	It immediately follows that the derived algebra is the abelian 
	subalgebra spanned by $\{E_1, E_2, E_3\}$, and
	the metric restricted to this subspace is Lorentzian. 
	The adjoint transformation defined by $E_4$, restricted to this subspace and written as a covariant tensor using the metric, has components given by the symmetric matrix
\begin{equation*}
A_{ij} = 
\begin{bmatrix}
	\dfrac{1}{2} k   & -\dfrac{\sqrt{3}}{2} \,k & 0 \\[2\jot]
	 -\dfrac{\sqrt{3}}{2} \,k  & -\dfrac{{1}}{\sqrt{2}} \,k &0\\[2\jot]
	0 & 0&  k
\end{bmatrix}.
\end{equation*}
	Hence the vector $\zeta$ in \eqref{defzeta} vanishes,  
	and this spacetime group is of type {\bf abel3LZ}.  
	Moreover, the adjoint matrix  $A$ is exactly in the form II of \eqref{Segre}. Indeed,
	with respect to the null tetrad 
\begin{equation*}
	K = \frac{1}{\sqrt{2}}(E_1 - E_2),\enskip
	L =  \frac{1}{\sqrt{2}}(E_1 + E_2), \enskip
	M = \frac{1}{\sqrt{2}}(E_3 +  i\,E_4),\enskip
	\barM =  \frac{1}{\sqrt{2}}(E_3 - i\,E_4)
\end{equation*}
	the non-zero spin coefficients are 
\begin{equation*}
       \alpha=-i\frac{\sqrt{2}}{4}\, k,\enskip
       \beta=-i \frac{\sqrt {2}}{4} \, k,\enskip 
       \kappa=-i\frac{\sqrt {6}}{4}\,k,\enskip
       \nu = i\frac{\sqrt {6}}{4}\, k,\enskip
       \pi=-i\frac{\sqrt{2}}{4}\, k,\enskip
       \tau=-i\frac{\sqrt {2}}{4}\, k
\end{equation*}
	and the structure equations agree with (2.11.2).

	For our second example we consider
	the unique 1-parameter family of  spacetime groups which are  Einstein manifolds (see (3.2)) \cite{Kaigorodov}. It  
	is defined in local coordinates $(x, y, z, u)$ by the metric (\cite{Stephani}, see equation (12.35))
\begin{equation*}
	h = 
	\frac{3}{\Lambda} dz\otimes dz + e^{4z} dx \otimes dx + 4 e^zdx\odot dy + 2e^{-2z}(dy \otimes dy 
	+ du \odot dz),
\end{equation*}
	where the cosmological constant is $\Lambda <0$.
	A group-invariant orthonormal tetrad for $h$ is (with $E_1$ time-like):
\begin{align*}
	E_1 &= \frac{e^{4z}}{16} \partial_u - e^{-2z}\partial_x +\frac{5}{4} e^z \partial_y,\quad
	E_2 = -\frac{\sqrt{2}}{4}e^{4z}\partial_u -\frac{\sqrt{2}}{2}e^z\partial_y,
\\[1\jot]
	E_3 &= \frac{15}{16}e^{4z}\partial_u + e^{-2z}\partial_x - \frac{5}{4} e^z \partial_y,
\quad
	E_4 = \frac{\sqrt{-3\Lambda}}{3}\partial_z.
\end{align*}
	Setting $\chi = \sqrt{-3\Lambda}$, 
	the non-zero structure equations are
\begin{align*}
	[E_1, E_4] &= \chi\left(\frac{7}{6} E_1 +\frac{5\sqrt{2}}{4} E_2 +  \frac{1}{2} E_3\right),\quad
	[E_2, E_4] = \chi\left(\frac{\sqrt{2}}{4} E_1 - \frac{1}{3} E_2 + \frac{\sqrt{2}}{4}E_3\right),
\\[1\jot]
	[E_3, E_4] &= \chi\left(-\frac{5}{2} E_1 - \frac{5\sqrt{2}}{4}E_2 - \frac{11}{6}E_3\right).
\end{align*}
	We deduce that the derived algebra is the 3-dimensional abelian subalgebra spanned by 
	$\{E_1, E_2, E_3\}$, and  the metric restricted
	to this sub-algebra is Lorentzian.  
	A short calculation then shows  the direction $\zeta$ is  given by the space-like vector 
 \begin{equation*}
	\zeta = E_1 + \frac{2\sqrt{2}}{3} E_2 + E_3,
 \end{equation*}
	and therefore this spacetime group is of type {\bf abel3LS}. 
	To align the structure equations  with (2.9) we need a null tetrad for
	which  $\zeta$ is aligned with the vector $M + \barM$. Such a tetrad is given by
\begin{align*}
	K &= \frac{\sqrt{17}}{4}\,(E_1 + E_3), 
\quad 
	L = \frac{\sqrt{17}}{4}\,(E_1 + \frac{12\sqrt{2}}{17}E_2 + \frac{1}{17} E_3),
\\
	M &= \frac{3}{4}\,(E_1 +  E_3) + \frac{\sqrt{2}}{2} \,(E_2 +i E_4)
\quad
	\barM = \frac{3}{4}\,(E_1 +  E_3) +\frac{\sqrt{2}}{2}\, (E_2 -i E_4).
\end{align*}
	The non-zero spin coefficients for this tetrad are
\begin{equation*}
       \alpha = -i\frac{\sqrt {2}}{3}\,\chi,\enskip
       \beta =  i\frac{\sqrt {2}}{6}\,\chi,\enskip
        \gamma = i \frac{\sqrt {2}}{\sqrt {17}}\, \chi,\enskip   
	\lambda = i \frac{\sqrt {2}}{\sqrt {17}}\, \chi,\enskip
        \mu = i\frac{\sqrt {2}}{\sqrt {17}}\,\chi,\enskip
        \pi = i \frac{\sqrt {2}}{6}\, \chi,     \enskip
        \tau = i\frac{\sqrt {2}}{6}\, \chi.
\end{equation*}
	and the structure equations for this tetrad agree with (2.9).
\par
\medskip
	For our final example we consider a non-inheriting electrovacuum spacetime group (see (3.18)) \cite{McLenaghan-Tariq:1975a}, \cite{Tupper:1976}.
	The following metric, expressed in  local coordinates  $(t, x, y, z)$, 
	defines a 1-parameter family of spacetime groups which solve the Einstein-Maxwell equations 
	with non-inheriting electromagnetic field (see 
	 \cite{Stephani},  equation (12.21)):
\begin{equation*}
h = - dt \otimes dt + {a^2\over x^2}\, (dx \otimes dx + dy \otimes dy) + 2 y\, dt \odot dz   + (x^2 - 4y^2)\,dz\otimes dz.
\end{equation*}
	A group-invariant orthonormal tetrad $(E_1, E_2, E_3, E_4)$ (with $E_1$ timelike) 
	for this metric is given by
\begin{equation*}
	E_1 = \partial_t,\quad E_2 =  {x\over a}\,\partial_y,\quad 
	E_3 = {1\over x}(2\,y\,\partial_t + \partial_z),
	\quad E_4 = {x\over a}\,\partial_x.
\end{equation*}
	This tetrad has the following non-vanishing commutators:
\begin{equation*}
	[E_2, E_3] = {2\over a}E_1, \quad [E_2, E_4] = -{1\over a} E_2,\quad [E_3, E_4] = {1\over a} E_3,
\end{equation*}
	from which it follows that the derived sub-algebra,  spanned by $(E_1, E_2, E_3)$, 
	is the Heisenberg algebra. The center of this Heisenberg algebra is spanned by $E_1$, 
	which is timelike.  Thus the spacetime Lie algebra is {\bf heisLT}. With respect to the null tetrad
\begin{equation*}
	K = \frac{1}{\sqrt{2}}(E_1 + E_4),\enskip
	L =  \frac{1}{\sqrt{2}}(E_1-  E_4), \enskip
	M = \frac{1}{\sqrt{2}}(E_2 +  i\,E_3),\enskip
	\barM =  \frac{1}{\sqrt{2}}(E_2 - i\,E_3)
\end{equation*}
	the non-zero spin coefficients are 
\begin{equation*}
       \gamma = -i\frac{\sqrt {2}}{4\,a},\enskip
       \epsilon= -i \frac {\sqrt {2}}{ 4\,a},\enskip
       \lambda =  \frac {\sqrt {2}}{2 \, a},\enskip
       \mu = -i\frac {\sqrt {2}}{2\, a},\enskip
       \rho = -i \frac{\sqrt {2}}{2\, a},\enskip
       \sigma = \frac {\sqrt {2}}{2\, a},\enskip
\end{equation*}
	and the structure equations are of type (2.2).
\par
\bigskip
\noindent
\subsection{Lie Algebraic Classification of Spacetime Groups  } 

In this section we show how to relate our classification of spacetime Lie algebras to 
the purely Lie algebraic classification of \cites{Patera-Sharp-Winternitz-Zassenhaus:1976, Snobol-Winternitz} in a couple of illustrative examples.  We begin with the spacetimes of type
\heisLT.  We then look at spacetimes of  type \abeltwoLone.

The first step is to change from the null tetrad  to a basis adapted to the 
first and second derived algebras. For \heisLT, this basis is given by
\begin{equation*}
	E_1 = \frac12\,(K + L), \enskip
	E_2= \frac12\,(M + \barM), \enskip
	E_3 = \frac{i}2\,(M - \barM), \enskip
	E_4 = \frac12\,(K-L).
\end{equation*}
	The first and second derived algebras are now $\lieg'=\langle E_1, E_2, E_3\rangle$ and $\lieg'' = \langle E_1\rangle$.
	In terms of the multiplication table for the Lie algebra, the structure equations  become
\begin{center}
\begin{tabular}[t]{L| C  C  C  C }
	&\, E_1  &\, E_2   &\, E_3   &\, E_4 \\ 
\hline
	\, E_1 & . & 0 & 0 & -2\,\mu_0\, E_1\\
	\, E_2 &  & . & (2\,\epsilon_1+2\,\gamma_1)\, E_1 & 2\,\kappa_0\, E_1 -(\mu_0+\sigma_0)\, E_2+(\sigma_1-\epsilon_1+\gamma_1)\, E_3\\
\, E_3 &  &  & . & -2\,\kappa_1\, E_1+(\sigma_1+\epsilon_1-\gamma_1)\, E_2 -(\mu_0-\sigma_0)\, E_3\\
\, E_4 &  &  &  & .\\
\end{tabular}
\end{center}
	The spin-coefficient $\sigma_1$ may be set to zero by a rotation in the $E_3$-$E_4$ plane.
	The coefficient of $E_1$ in the bracket $[\,E_2, E_3\,]$ is transformed to 1 and 	
	the $E_1$ terms are eliminated from the brackets $[E_2,E_4]$, and $[E_3,E_4]$ by the 
	(non-orthonormal) change of basis
\begin{equation*}
	e_1 = 2(\epsilon_1 + \gamma_1)E_1, \enskip
	e_2 = E_2, \enskip
	e_3 = E_3, \enskip
	e_4 = E_4 - \frac{\kappa_1}{\epsilon_1 + \gamma_1} E_2 - \frac{\kappa_0}{\epsilon_1 + \gamma_1} E_3.
\end{equation*}
	
	The structure equations become 
\begin{center}
\begin{tabular}[t]{L| C  C  C  C }
&\, e_1  &\, e_2   &\, e_3   &\, e_4 \\ 
\hline
\, e_1 & . & 0 & 0 & -2\,\mu_0\, e_1\\
\, e_2 &  & . & \, e_1 & (-\mu_0-\sigma_0)\, e_2+(-\epsilon_1+\gamma_1)\, e_3\\
\, e_3 &  &  & . & (\epsilon_1-\gamma_1)\, e_2+(-\mu_0+\sigma_0)\, e_3\\
\, e_4 &  &  &  & .\\
\end{tabular}
\end{center}
	which are now aligned with the algebras [{\bf 4,\,7}] -- [{\bf 4,\,11}] in Appendix B.

	The group of unimodular transformations,  $e_2 \to a e_2 + ce_3$, $e_3 \to be_2 + de_3$, $ad- bc = 1$,
	and the scaling $e_4 \to ue_4$ preserve the form of these structure equations and can be used 
	to bring them to final form. The unimodular transformations act by similarity on 
	$A = 
\begin{bmatrix} 
	-\mu_0-\sigma_0 & -\epsilon_1+\gamma_1 
\\ 	\epsilon_1-\gamma_1 & -\mu_0+\sigma_0 
\end{bmatrix} 
$.
	If $\text{tr}(A) = -2\mu_0$ vanishes, then the spacetime algebra is isomorphic to [{\bf 4,\,7}] or [{\bf 4,\,8}].
	If $\text{tr}(A) \neq 0$ and the eigenvalues of $A$ are real and  distinct, then the spacetime algebra
	is [{\bf 4, 9}]. The other Jordan forms of $A$ lead to [{\bf 4,\,10}] or [{\bf 4,\,11}].
	As an example, the non-inheriting Einstein-Maxwell solution from Section \ref{CSGR} is isomorphic to [{\bf 4,\,7}].

We now turn our attention to the spacetime Lie algebras \abeltwoLone. 
	In terms of the adapted basis 
\begin{equation*}
	E_1 = K,\enskip
	E_2 = L, \enskip
	E_3 = \frac{i}{2}(M- \barM), \enskip
	E_4 = \frac{1}{2}(M + \barM)
\end{equation*}
	the structure equations for \abeltwoLone  become 
\begin{center}
\begin{tabular}[t]{L| C  C  C  C }
&\, E_1  &\, E_2   &\, E_3   &\, E_4 \\ 
\hline
\, E_1 & . & 0 & \tau_1\, E_1 & (-2\,\beta_0-\tau_0)\, E_1-\kappa_0\, E_2\\
\, E_2 &  & . & \tau_1\, E_2 & \nu_0\, E_1+(-\tau_0+2\,\beta_0)\, E_2\\
\, E_3 &  &  & . & -2\,\gamma_1\, E_1-2\,\epsilon_1\, E_2\\
\, E_4 &  &  &  & .\\
\end{tabular}
\end{center}
We now have two cases to consider.

\noindent {\bf Case 1:}	If $\tau_1 \neq 0$,  we make the change of basis 
\begin{equation*}
	e_1 = E_1, \enskip
	e_2 = E_2, \enskip
	e_3 = \frac{1}{\tau_1}  E_3, \enskip
	e_4 = E_4  - 2 \frac{\gamma_1}{\tau_1}\,E_2 -  2 \frac{\epsilon_1}{\tau_1}\, E_2 
	+ \frac{\tau_0}{\tau_1}\,E_3  
\end{equation*}
	and the structure equations become
\begin{center}
\begin{tabular}[t]{L| C  C  C  C }
&\, e_1  &\, e_2   &\, e_3   &\, e_4 \\ 
\hline
\, e_1 & . & . & \, e_1 & -2\,\beta_0\, e_1-\kappa_0\, e_2\\
\, e_2 &  & . & \, e_2 & \nu_0\, e_1+2\,\beta_0\, e_2\\
\, e_3 &  &  & . & 0\\
\, e_4 &  &  &  & .\\
\end{tabular}
\end{center}
	The form of these structure equations is preserved by  $e_1 \to a e_1 + ce_2$, $e_2 \to be_1 + de_2$,
	$ad - bc \neq 0$, and by a scaling of $e_4$. If $4\beta_0^2  - \nu_0\kappa_0 > 0$, 
	then the restriction of $\text{ad}(e_4)$ to $\langle e_1, e_2\rangle $ can be diagonalized, the algebra is decomposable, and is isomorphic to  {\bf [4, -2]}. If $4\beta_0^2 - \nu_0\kappa_0=0$, 		then the eigenvalues  of  $\text{ad}(e_4)$ are equal and the algebra {\bf abel2L1} becomes [{\bf 4,\,12}]. If  $4\beta_0^2  - \nu_0\kappa_0 <0$, the eigenvalues
	are pure imaginary and the algebra is isomorphic to {\bf [4,\,13]}.

\noindent{\bf Case 2:}
	If $\tau_1=0$, the nilradical becomes 3-dimensional; the center of the algebra is 
\begin{equation*}
	\langle(-4\,\gamma_1\,\beta_0+2\,\epsilon_1\,\nu_0+2\,\gamma_1\,\tau_0)\, e_1+(4\,\beta_0\epsilon_1+2\,\epsilon_1\,\tau_0-2\,\gamma_1\,\kappa_0)\, e_2+(4\,\beta_0^{2}-\nu_0\kappa_0-\tau_0^{2})\, e_3\rangle.
\end{equation*} 
	If $\Xi = 4\beta_0^2-\nu_0\kappa_0-\tau_0^2\neq0$, the algebra is decomposable and thus isomorphic to one of [{\bf 4, -4}], [{\bf 4, -5}], or [{\bf 4, -6}]. If $\Xi=0$, the center lies in the derived algebra and {\bf abel2L1} may be transformed to [{\bf 4, 4}].

	As an illustration, we remark that, generically, the scalar solutions (3.13) and (3.16) belong to Case 1, while (3.14) and (3.15) belong to Case 2.

\bigskip
\noindent
\subsection{Petrov Classification of Spacetime Groups}
\label{PCSG}
	The Petrov classification of the algebraic character of the Weyl tensor plays a central role in the study of 4-dimensional spacetimes (see, {\it e.g.,} \cite{Stephani}).
	The spacetime Lie groups which are algebraically  special for generic values of the spin coefficients have the algebras
	\heisLN{\bf2}, \abelthreeLN{\bf3}, \abelthreeLN{\bf4}, \abelthreeLZ{\bf3}, \abelthreeLZ{\bf4},
	\abelthreeNS, \LevitwoS{\bf 7}, \LevitwoN,   \abeltwoNtwo, 
	\abeltwoNthree, \abeltwoNfour, and \abeltwoNfive.  
	All the remaining spacetime Lie groups are of Petrov type I for generic values of the spin coefficients. 
	
	In this section we give a complete enumeration of the possible Petrov types and corresponding values of the spin coefficients for the generically algebraically special spacetimes listed above.  The spacetime Lie groups of Petrov type O shown here (see {\bf simpCN1} and {\bf abel2N3}) are in fact the only spacetime Lie groups of Petrov type O.
	As always, in this classification we  exclude those 
	spacetime groups for which the isometry 
	algebra is of dimension greater than 4.  
	
	Classification results for spacetime groups of type O can be found in \cite{Honda-Tsukada, Calvaruso-Zaeim}.  Our results are consistent with the results of these references, although 
	we have found that the isometry group of the spacetime given in Theorem 4.4, case (1) of ref. \cite{Calvaruso-Zaeim} has dimension 7, and is therefore a homogeneous Lorentz manifold rather  than a spacetime group.
	
	Aside from completely enumerating the algebras of all spacetime groups with Petrov type O, at this time we are 
	unable to give a full Petrov type analysis of the spacetime groups which are generically  algebraically general.	 	
\newpage
Table 4.1: Petrov Types
\begin{center}
\begin{tabular}[t]{|l|l|l|}
\hline
\multicolumn{3}{|l|}{\StrutA\sc \normalsize  Spacetime Lie Algebras of Generic Petrov Type II: Heisenberg Derived Algebra }
\\[2\jot]
\hline
\Strut
\heisLN{\bf 2} & {\bf II} & generic
\\
\cline{2-3}
\Strut
{\bf \PetrovNo.} 
\immediate\write5{"4.3.\thePetrovcounter" = "PTheisLN2IIIb",}& {\bf III}&
$
\alpha=3\,i\beta_1,\enskip
       \beta=i\beta_1,\enskip
       \gamma=i\gamma_1,\enskip
       \epsilon=0,\enskip
       \kappa=0,\enskip
       \lambda=2\,i\gamma_1,\enskip
	\mu=0,
$	
\\
\Strut
&& 
$
	\nu=\nu_0+i\nu_1,\enskip
       \pi=-4\,i\beta_1,\enskip
       \rho=0,\enskip
       \sigma=0,\enskip
	\tau=-4\,i\beta_1
$
\\
\cline{2-3}
\Strut
&{\bf D, N, O} & none
\\
\hline
\end{tabular}
\end{center}
\vspace{-20pt}
\begin{center}
\begin{tabular}[t]{|l|l|l|}
\hline
\multicolumn{3}{|l|}{\StrutA\sc \normalsize  Spacetime Lie Algebras of Generic Petrov Type II: 3D Abelian Derived Algebra}
\\[2\jot]
\hline
\Strut
\abelthreeLN{\bf3} & {\bf II} & generic
\\

\cline{2-3}
\Strut
{\bf \PetrovNo.} 
\immediate\write5{"4.3.\thePetrovcounter" = "PTabel3LN3D",}
& {\bf D} & 
$
    \alpha=i\alpha_1,\enskip
       \beta=i\alpha_1,\enskip
       \gamma=0,\enskip
       \epsilon=0,\enskip
       \kappa=0,\enskip
       \lambda=-i\mu_1,\enskip
       \mu=i\mu_1,    
$
\\
\Strut
&& 
$
\nu= -i{\dfrac {{\mu_1}^{2} \left( \tau_1+2\,\alpha_1 \right) }{2\alpha_1\, \left( \tau_1+\alpha_1 \right) }},\enskip
       \pi=i\tau_1,\enskip
       \rho=0,\enskip
       \sigma=0,\enskip
       \tau=i\tau_1.     
$\\
\cline{2-3}
\Strut
& {\bf III, N, O} & none
\\[1\jot]
\hline
\Strut
\abelthreeLN{\bf4} & {\bf III} & generic
\\
\cline{2-3}
\Strut & {\bf D, N, O}  & none 
\\
\hline
\Strut
\abelthreeLZ{\bf3} & {\bf II} & generic
\\
\cline{2-3}
\Strut
{\bf \PetrovNo.} 
\immediate\write5{"4.3.\thePetrovcounter" = "PTabel3LZ3D",}

& {\bf D} & 
$
       \alpha=i\alpha_1,\enskip
       \beta=i\alpha_1,\enskip
       \gamma=0,\enskip
       \epsilon=0,\enskip
       \kappa=0,\enskip
       \lambda=0,\enskip
       \mu=0,
$
\\
&&

\Strut
$      
 \nu=i\nu_1,\enskip
       \pi=-i\alpha_1,\enskip
       \rho=0,\enskip
       \sigma=0,\enskip
       \tau=-i\alpha_1.
$
\\
\cline{2-3}
\Strut & {\bf III, N, O}  & none 
\\
\hline
\Strut
\abelthreeLZ4 & {\bf III} & generic
\\
\cline{2-3}
\Strut & {\bf D, N, O}  & none
\\
\hline
\Strut
\abelthreeNS{\bf1} & {\bf II} & generic
\\
\cline{2-3}
\Strut
{\bf \PetrovNo.} 
\immediate\write5{"4.3.\thePetrovcounter" = "PTabel3NS1D",}& {\bf D}&
$
	\alpha=-i\beta_1,\enskip
	\beta=i\beta_1,\enskip
	\gamma= 0,\enskip
	\epsilon=0,\enskip
	\kappa=0,\enskip
	\lambda= \lambda_0 + i\lambda_1,\enskip
	\mu=\mu_0,$
\\
\Strut
&& 
$
\nu= \dfrac{1}{32\,\beta_1}\,[ -18\,\lambda_0\,\lambda_1-10\,\lambda_1\,\mu_0 +
i\,(9\,\lambda_0^2+10\,\lambda_0\,\mu_0-9\,\lambda_1^2+\mu_0^2)]
$
\\
\Strut
&&
$
	\pi=-2\,i\beta_1,\enskip
	\rho=0,\enskip
	\sigma=0,\enskip
	\tau=2\,i\beta_1
$
\\
\cline{2-3}
\Strut & {\bf III,  N, O}  & none 
\\
\hline
\end{tabular}
\vspace{-20 pt}
\end{center}
\newpage
Table 4.1 (cont.)
\begin{center}
\begin{tabular}[t]{|l|l|l|}
\hline
\multicolumn{3}{|l|}{\StrutA\sc \normalsize  Spacetime Lie Algebras of Generic Petrov Type II: Simple Derived Algebra}
\\[2\jot]

\hline
\Strut
\LevitwoS{\bf 7} & {\bf II} & generic
\\
\cline{2-3}
\Strut & {\bf III, D, N, O}  & none 
\\

\hline
\Strut
\LevitwoN{\bf1} & {\bf II} & generic
\\
\cline{2-3}
{\bf \PetrovNo.} 
\immediate\write5{"4.3.\thePetrovcounter" = "PTsimpCN1III",}
& \Strut {\bf III} & 
$	\alpha=0,\enskip
       \beta=0,\enskip
       \gamma=-3/2\,i\mu_1,\enskip
       \epsilon=i\epsilon_1,\enskip
       \kappa=0,\enskip
       \lambda=i\lambda_1,\enskip
       \mu=i\mu_1,
$
\\
&&
$	\nu=\nu_0+i\nu_1,\enskip
       \pi=0,\enskip
       \rho=2\,i\epsilon_1,\enskip
       \sigma=0,\enskip
       \tau=0
$
\\
\cline{2-3}
\StrutB
{\bf \PetrovNo.} 
\immediate\write5{"4.3.\thePetrovcounter" = "PTsimpCN1Da",}
& {\bf D} & 
$	  
	\alpha=0,\enskip
       \beta=0,\enskip
       \gamma=i\gamma_1,\enskip
       \epsilon={\dfrac {4\,i\lambda_1\,\gamma_1\, \left( 3\,\mu_1+2\,\gamma_1 \right) }{{9\nu_0}^{2}}},\enskip
       \kappa=0,\enskip
       \lambda=i\lambda_1,\enskip
       \mu=i\mu_1,
$
\\
\StrutA
&& 
$
	\nu=\nu_0,\enskip
       \pi=0,\enskip
       \rho=
	\dfrac {8\,i\lambda_1\,\gamma_1\,( 3\,\mu_1+2\,\gamma_1 ) }{9\nu_0^2},		\enskip
       \sigma=0,\enskip
       \tau=0
$
\\[2\jot]

\cline{3-3}
\StrutB
{\bf \PetrovNo.} 
\immediate\write5{"4.3.\thePetrovcounter" = "PTsimpCN1Db",}
&&
$
       \alpha=0,\enskip
       \beta=0,\enskip
       \gamma=0,\enskip
       \epsilon=i\epsilon_1,\enskip
       \kappa=0,\enskip
       \lambda=i\lambda_1,\enskip
       \mu=i\mu_1,
$
\\[.5\jot]
&&
$       \nu=0,\enskip
       \pi=0,\enskip
       \rho=2\,i\epsilon_1,\enskip
       \sigma=0,\enskip
       \tau=0
$

\\
\cline{2-3}
\StrutB
{\bf \PetrovNo.} 
\immediate\write5{"4.3.\thePetrovcounter" = "PTsimpCN1N",}
& {\bf N} & 
$
 \alpha=0,\enskip
       \beta=0,\enskip
       \gamma=-\dfrac{3}{2}\,i\mu_1,\enskip
       \epsilon=i\epsilon_1,\enskip
       \kappa=0,\enskip
       \lambda=i\lambda_1,\enskip
       \mu=i\mu_1,
$

\\
\Strut
&& 
$
 \nu=0,\enskip
       \pi=0,\enskip
       \rho=2\,i\epsilon_1,\enskip
       \sigma=0,\enskip
       \tau=0
$
\\
\cline{2-3}
{\bf \PetrovNo.} 
\immediate\write5{"4.3.\thePetrovcounter" = "PTsimpCN1O",}
\Strut & {\bf O}  & 
$
       \alpha=0,\enskip
       \beta=0,\enskip
       \gamma=0,\enskip
       \epsilon=i \epsilon_1,\enskip
       \kappa=0,\enskip
       \lambda=i\lambda_1,\enskip
       \mu=0,
$
\\
\Strut
&& 
$
       \nu=0,\enskip
       \pi=0,\enskip
       \rho=2\,i\epsilon_1,\enskip
       \sigma=0,\enskip
       \tau=0
$

\\
\hline
\Strut
\LevitwoN{\bf 2} & {\bf II} & generic
\\
\cline{2-3}
\StrutB
{\bf \PetrovNo.} 
\immediate\write5{"4.3.\thePetrovcounter" = "PTsimpCN2III",}& {\bf III}&
$
\alpha=0,\enskip
       \beta=0,\enskip
       \gamma=-\dfrac{3}{2}i\mu_1,\enskip
       \epsilon=i\epsilon_1,\enskip
       \kappa=0,\enskip
       \lambda=0,\enskip
	\mu=i\mu_1,
$	
\\
\Strut
&& 
$
	\nu=\nu_0+i\nu_1,\enskip
       \pi=0,\enskip
       \rho=2i\epsilon_1,\enskip
       \sigma=0,\enskip
	\tau=0
$
\\
\cline{2-3}
\Strut
&{\bf D, N, O} & none
\\
\hline
\Strut
\LevitwoN{\bf3} & {\bf II} & generic
\\
\cline{2-3}
{\bf \PetrovNo.} 
\immediate\write5{"4.3.\thePetrovcounter" = "PTsimpCN3D",}
& {\bf D} &
\Strut
$
       \alpha=-3\,\beta_0,\enskip
       \beta=\beta_0,\enskip
       \gamma=i\gamma_1,\enskip
       \epsilon=0,\enskip
       \kappa=0,\enskip
       \lambda=i \left( 2\,\gamma_1-\mu_1 \right) ,\enskip
       \mu=i\mu_1,
$
\\[1\jot]
\StrutA
&& 
$
       \nu={\dfrac {\gamma_1\, \left( 7\,\gamma_1-3\,\mu_1 \right) }{9\beta_0}},\enskip
       \pi=-2\,\beta_0,\enskip
       \rho=0,\enskip
       \sigma=0,\enskip
       \tau=2\,\beta_0
$
\\[1\jot]
\cline{2-3}
\Strut & {\bf III,  N, O}  & none 
\\
\hline
\end{tabular}
\end{center}
\newpage
Table 4.1 (cont.)
\begin{center}
\begin{tabular}[t]{|l|l|l|}
\hline
\multicolumn{3}{|l|}{\StrutA\sc \normalsize  Spacetime Lie Algebras of Generic Petrov Type II: 2D Abelian Derived Algebra}
\\[2\jot]
\hline
\Strut
\abeltwoNtwo & {\bf II} & generic
\\
\cline{2-3}
\Strut & {\bf III, D, N, O}  & none 
\\
\hline
\Strut
\abeltwoNthree & {\bf II} & generic
\\
\cline{2-3}
\Strut
{\bf \PetrovNo.} 
\immediate\write5{"4.3.\thePetrovcounter" = "PTabel2N3III",}& 
{\bf III} & 
$	\alpha=-i\,\beta_1, \enskip
	\beta= i\beta_1,\enskip\gamma=-\mu_0 +i\gamma_1, \enskip
	\epsilon=0,\enskip
	\kappa=0,\enskip
	\lambda=\mu_0+2\,i\gamma_1,\enskip
	\mu=\mu_0, \enskip
	
$
\\
&&
$
	\nu=i\nu_1, \enskip	
	\pi=2\,i\beta_1, \enskip
	\rho=0, \enskip
	\sigma=0, \enskip
	\tau=2\,i\beta_1
$
\\[1\jot]
\cline{2-3}
%
%
\StrutA
{\bf \PetrovNo.} 
\immediate\write5{"4.3.\thePetrovcounter" = "PTabel2N3Db",}
 & {\bf D} &
$
	\alpha=-i\tau_1+i\beta_1,\enskip
	\beta=i\beta_1,\enskip
	\gamma=i\gamma_1,\enskip
	\epsilon=0,\enskip
	\kappa=0,\enskip
	\lambda=2\,i\gamma_1,\enskip
	\mu=0,\enskip
$
\\
\StrutA
&&
$
	\nu= -8\,i{\dfrac {\tau_1\,\gamma_1^{2}}{ \left( \tau_1-2\,\beta_1 \right) ^{2}}},\enskip
	\pi=i\tau_1,\enskip
	\rho=0,\enskip
	\sigma=0,\enskip
	\tau=i\tau_1
$
\\[4\jot]
\cline{2-3}
\Strut
{\bf \PetrovNo.} 
\immediate\write5{"4.3.\thePetrovcounter" = "PTabel2N3N",}

 & {\bf N} & 
$
	\alpha=-i\tau_1, \enskip
	\beta=0,\enskip
	\gamma=-\mu_0+i\gamma_1,\enskip
	\epsilon=0,\enskip\kappa=0,\enskip
	\lambda=\mu_0+2\,i\gamma_1,\enskip
	\mu=\mu_0,\enskip
$
\\
&&
$	\nu=i\nu_1,\enskip
	\pi=i\tau_1,\enskip
	\rho=0,\enskip
	\sigma=0,\enskip
	\tau=i\tau_1
$
\\[1\jot]
\cline{2-3}
\Strut
{\bf \PetrovNo.} 
\immediate\write5{"4.3.\thePetrovcounter" = "PTabel2N3O",}

&  {\bf O} &

$
 \alpha=-i\tau_1,\enskip
       \beta=0,\enskip
       \gamma=i\gamma_1,\enskip
       \epsilon=0,\enskip
       \kappa=0,\enskip
       \lambda=2\,i\gamma_1,\enskip
       \mu=0,
$
\\
&&
$
       \nu= -8\,i{\dfrac{\gamma_1^{2}}{\tau_1}},\enskip
       \pi=i\tau_1,\enskip
       \rho=0,\enskip
       \sigma=0,\enskip
       \tau=i\tau_1
$

\\[2\jot]
\hline

\Strut
\abeltwoNfour & {\bf II} & generic
\\
\cline{2-3}
{\bf \PetrovNo.} 
\immediate\write5{"4.3.\thePetrovcounter" = "PTabel2N4D",}

\Strut
&  {\bf D} &
$
	\alpha=3\,i\beta_1,\enskip
	\beta=i\beta_1,\enskip
	\gamma=0,\enskip
	\epsilon=0,\enskip
	\kappa=0,\enskip
	\lambda=-i\mu_1,\enskip
	\mu=i\mu_1,
$
\\
\Strut
&&
$
	\nu= -i{ \dfrac {\mu_1^{2}}{4\beta_1}},\enskip
	\pi=-2\,i\beta_1,\enskip
	\rho=0,\enskip
	\sigma=0,\enskip
	\tau=-2\,i\beta_1
$
\\[3\jot]
\cline{2-3}
\Strut
&  {\bf III, N, O} & none

\\
\hline
\Strut
\abeltwoNfive & {\bf II} & generic
\\
\cline{2-3}
\Strut
& {\bf III, D, N, O} & none
\\
\hline
\end{tabular}
\end{center}

\par
\bigskip
\noindent
\subsection{Conformally Einstein Metrics on Spacetime Groups}\label{CEMSG}
	We conclude this  section  
	by providing some simple examples of spacetime groups $(G, h)$ for which 
	the metric $h$ is conformally equivalent to an Einstein metric.	
	It is shown in \cite{Grover-Nurowski:2006} that a pseudo-Riemannian metric  $h$  in four dimensions is conformally Einstein if, 
	granted certain genericity conditions,  $h$ is Bach-flat and if there is a closed 
	1-form $\chi= K_a dx^a$ satisfying 
\begin{equation}
	A_{abc} + K^dC_{dabc} = 0.
\label{GN}
\end{equation}
	Here $A_{abc}$ is the Cotton tensor and $C_{dabc} $ is the Weyl tensor for the metric $h$. The 
	 genericity conditions imply that if $h$ is an invariant metric on a Lie group then the 1-form $\chi$ 
	is  also  invariant. This observation allows one to search for
	conformally Einstein spacetime groups by taking $\chi$ to  be a closed 1-form on the 
	Lie algebra, solving \eqref{GN} for the  spin coefficients, 
	and then requiring the Bach tensor to vanish. Because the conformal factor will not be $G$-invariant, to explicitly find the conformal factor
	one must introduce local coordinates on the spacetime group. The conformal factor is then given by
	$e^{2\Upsilon}$, where $d\Upsilon =\chi$, so that $e^{2\Upsilon}h$ is an Einstein metric.  
	The following results use coordinates $(x, y, u, v)$ to define a basis $E_i$ of left invariant vector fields, with dual basis $\omega^i$, and the corresponding left invariant metric and conformal factor for the conformally Einstein spacetime groups of type {\bf heisLN2}, {\bf abel3RZ1} {\bf abel3LZ1}, and {\bf abel3LZ3}.
The spacetime groups of type {\bf abel2R2}, {\bf abel2R3}, {\bf abel2N2}, and {\bf abel2N3} also admit conformally Einstein metrics. We hope to provide a complete list of spacetime groups which are conformally Einstein and, more generally, Bach flat in the near future.
\newpage
\begin{center}
\begin{tabular}[t]{|l|l|}
\noalign{\vspace{-60pt}}
\hline
\multicolumn{2}{|l|}{\StrutA\sc \normalsize  Table 4.2: Conformally Einstein Metrics for spacetime Groups }
\\[2\jot]
\hline
\StrutB
\heisLN{\bf2} & 
	{\sc Spin Coefficients:}
$
	 \alpha=3\,i\beta_1,\enskip
       \beta=i\beta_1,\enskip
       \gamma=0,\enskip
       \epsilon=0,\enskip
       \kappa=0,\enskip
       \lambda=0,\enskip
       \mu=0,
$	
\\
{\bf \ConEinNo.} 
\immediate\write5{"4.4.\theConEincounter" = "CEheisLN1",} 
&
$
	\nu=\nu_0,\enskip
       \pi=-i\beta_1,\enskip
       \rho=0,\enskip
       \sigma=0,\enskip
       \tau=-i\beta_1.
$
\\ 
\StrutA
& {\sc Null Tetrad:} $K = E_1, \enskip L =E_2, \enskip M = E_3 - iE_4, \enskip \barM = E_3 + iE_4$
\\
\StrutA
&
	{\sc Inv. Vector Fields:} 	
$
	E_1 = \partial_x,\enskip
	E_2 = \partial_y,\enskip
	E_3 = \nu_0y\,\partial_x +\partial_u,\enskip
	E_4 = -\beta_1(3x\,\partial_x - y\,\partial_y+4u\,\partial_u)+\partial_v
	$
\\
&
\StrutA
	{\sc Metric:}  $h=-\omega^1 \otimes \omega^2  - \omega^2 \otimes \omega^1    +  
	\frac12\omega^3 \otimes \omega^3 +\frac12   \omega^4 \otimes \omega^4 $
\\
&
	 {\sc Conformal Factor:} $\Upsilon = 3\beta_1\, v$
\\[2\jot]
\hline
\StrutB
\abelthreeRZ\bf{1} & 
	{\sc Spin Coefficients:}
$
	\alpha=0,\enskip
	\beta=0,\enskip
	\gamma=-\epsilon_0,\enskip
	\epsilon=\epsilon_0,\enskip
	\kappa=0,\enskip
	\lambda=-\sigma_0, $
	
\\
{\bf \ConEinNo.} 
\immediate\write5{"4.4.\theConEincounter" = "CEabel3RZ1",} 
&
$
	\mu={\dfrac {\epsilon_0^{2}+\sigma_0^{2}}{2\epsilon_0}},\enskip
	\nu=0,\enskip
	\pi=0,\enskip
	\rho=-\,{\dfrac {\epsilon_0^{2}+\sigma_0^{2}}{2\epsilon_0}},\enskip
	\sigma=\sigma_0,\enskip
	\tau=0.
$
\\ 
\StrutA
& {\sc Null Tetrad:} $K = E_1 +E_4, \enskip L = -E_1 +E_4, \enskip M = E_2 - iE_3, \enskip \barM = E_2 + iE_3$
\\
\StrutA
&
	{\sc Inv. Vector Fields:} 	
$
	E_1 = \partial_x,\,\enskip
	E_2 = \partial_y,\, \enskip
	E_3 = \partial_u,\enskip
$
\\
&	
$
	E_4 = 2\,\epsilon_0x\, \partial_x 
	+ \dfrac{( \epsilon_0 -\sigma_0)^2}{2\epsilon_0} y\, \partial_y 
	+ \dfrac{( \epsilon_0 +\sigma_0)^2}{2\epsilon_0} u\, \partial_u+\, \partial_v
$

\\
&
\StrutA
	{\sc Metric:}  $h=\omega^1 \otimes \omega^1  + \omega^2 \otimes \omega^2    +  
	\omega^3 \otimes \omega^3 -   \omega^4 \otimes \omega^4 $
\\
&
	 {\sc Conformal Factor:} $\Upsilon = -\,{\dfrac {3\epsilon_0^{2}+\sigma_0^{2}}{2\epsilon_0}}\, v
$
\\[3\jot]
\hline
\StrutB
\abelthreeLZ{\bf1} & 
	{\sc Spin Coefficients:}
$
	 \alpha=i\alpha_1,\enskip
       \beta=i\alpha_1,\enskip
       \gamma=0,\enskip
       \epsilon=0,\enskip
       \kappa=i\kappa_1,\enskip
       \lambda=0,\enskip
       \mu=0,	
$
\\
\StrutB
{\bf \ConEinNo.} 
\immediate\write5{"4.4.\theConEincounter" = "CEabel3LZ11",}
&
$
	\nu=i\kappa_1,\enskip
       \pi={\dfrac {-i\left( \kappa_1^{2}+\alpha_1^{2} \right) }{2\alpha_1}},\enskip
       \rho=0,\enskip
       \sigma=0,\enskip
       \tau={\dfrac {-i\left( \kappa_1^{2}+\alpha_1^{2} \right) }{2\alpha_1}}
$	
\\
\StrutA
& {\sc Null Tetrad:} $K = E_1, \enskip L =E_2, \enskip M = E_3 - iE_4, \enskip \barM = E_3 + iE_4$
\\
&
	{\sc Inv. Vector Fields:}
$
	E_1 = \dfrac12\,( \partial_x+ \, \partial_y), \enskip
	E_2 = \dfrac12\,(-\partial_x+\, \partial_y),\enskip
	E_3 = \, \partial_u
$
\\
&
\StrutB
$
E_4 = -\dfrac {x (\alpha_1 + \kappa_1)^{2} }{2\alpha_1}\, \partial_x- \,\dfrac {y(\alpha_1 -\kappa_1)^2}{2\alpha_1}\, \partial_y-2\,\alpha_1\,u\, \partial_u+\, \partial_v
	
$
\\[3\jot]
&
{\sc Metric:}  $h=-\omega^1 \otimes \omega^2  - \omega^2 \otimes \omega^1    +  
	\dfrac12\omega^3 \otimes \omega^3 +\dfrac12   \omega^4 \otimes \omega^4 $
\\
\StrutB
&
{\sc Conformal Factor:} $\Upsilon = \dfrac {3\,\alpha_1^2+\kappa_1^2}{2\alpha_1}\,v $
\\[3\jot]
\hline
\StrutA
\abelthreeLZ{\bf3} & 
	{\sc Spin Coefficients:}
$
	\alpha=i\alpha_1,\enskip
	\beta=i\alpha_1,\enskip
	\gamma=0,\enskip
	\epsilon=0,\enskip
	\kappa=0,\enskip
	\lambda=0,\enskip
	\mu=0,\enskip	
$
\\
{\bf \ConEinNo.} 
\immediate\write5{"4.4.\theConEincounter" = "CEabel3LZ31",}
&
$
	\nu=i\nu_1,\enskip
	\pi=-\dfrac{i}{2}\alpha_1,\enskip
	\rho=0,\enskip
	\sigma=0,\enskip
	\tau=- \dfrac{i}{2}\alpha_1.
$
\\ 

\StrutA
& {\sc Null Tetrad:} $K = E_1, \enskip L = E_2, \enskip M = E_3 - iE_4, \enskip \barM = E_3 + iE_4$
\\
\StrutA
&
	{\sc Inv. Vector Fields:} 	
$
	E_1 = \partial_x, \enskip
	E_2 = \partial_y, \enskip 
	E_3 = \partial_u, \enskip
$
\\
&
$	E_4 = (-\dfrac12\,x\alpha_1+\nu_1\,y)\, \partial_x-\dfrac12\,\alpha_1\,y\, \partial_y-2\,\alpha_1\,u\, 
	\partial_u +	\, \partial_v.
$
\\	
 &
{\sc Metric: } $h=-\omega^1 \otimes \omega^2  - \omega^2 \otimes \omega^1    +  
	\dfrac12\omega^3 \otimes \omega^3 +\dfrac12   \omega^4 \otimes \omega^4 $
\\
\StrutB
&
{\sc Conformal Factor:} $\Upsilon = \dfrac32 \alpha_1\, v$.
\\[2\jot]
\hline
\end{tabular}
\end{center}

%
%
\newpage
\section{Proof of the Classification}\label{Classification}
\setlength{\parindent}{12pt}
	In this section we present the computations which establish the  enumeration of all spacetime 
	Lie algebras presented in Section \ref{Summary}. 
	We begin in \ref{Lorentz} with a short review of the properties of the Lorentz group and 
	the Newman-Penrose formalism that we shall need. In  Section \ref{NPPrel} we determine the 
	reductions of the  structure equations
	for spacetimes admitting a 3-dimensional derived algebra $\lieg'$, 
	based upon the signature of the induced inner product on $\lieg'$.  We use these results in sections   \ref{HeisDerived} and
	 \ref{AbelDerived} to classify the spacetime Lie algebras enumerated in Tables 1 and 2.  In section \ref{SimpleDerived} we classify the algebras appearing in Table 3. The final sections classify the spacetime  Lie algebras in Table 4 and dispose of algebras with one-dimensional derived algebras.
 
\setlength{\baselineskip}{12pt}
\subsection{Lorentz Transformations}\label{Lorentz}
	Let $V$ be a 4-dimensional vector space with Lorentz signature inner product $\eta$. 
	Fix a null tetrad $\{K,\, L,\, M,\, \barM\}$ so that, 
	with respect to the dual basis $\{\Theta_K,\, \Theta_L,\, \Theta_{\vphantom{\barMs}M},\, \Theta_{\barMs} \}$,
	the metric is
\begin{equation*}
	\eta = -\Theta_K \otimes \Theta_L - \Theta_L \otimes \Theta_K
	+ \Theta_M \otimes \Theta_{\bar{M}} + \Theta_{\bar{M}} \otimes \Theta_M.
\end{equation*}
	Note that the vector $K + L$ is time-like while $K-L$ is space-like.   The dual basis has the structure equations (see (\ref{masterNP})):
	\begin{align*}
	&d\Theta_K = (\gamma + \overline\gamma)\, \Theta_K\wedge \Theta_L
	+ (\overline\alpha +\beta-\overline\pi)\, \Theta_K \wedge \Theta_M + (\alpha +\overline\beta-\pi)\, \Theta_K \wedge \Theta_{\barMs }
	-\overline\nu\, \Theta_L\wedge \Theta_M -\nu\, \Theta_L\wedge \Theta_{\barMs}\\
	&\phantom{d\Theta_K =} - (\mu - \overline\mu)\,\Theta_M 	\wedge\Theta_{\barMs}\\
	&d\Theta_L = (\epsilon +\overline\epsilon)\, \Theta_K \wedge \Theta_L +\kappa\, \Theta_K \wedge \Theta_M +\overline\kappa\, \Theta_K \wedge 
	\Theta_{\barMs} + (\tau -\overline\alpha-\beta)\, \Theta_L \wedge \Theta_M + (\overline\tau -\alpha-\overline\beta)\, \Theta_L \wedge 
	\Theta_{\barMs}\\
	 &\phantom{d\Theta_K =}- (\rho - \overline\rho)\, \Theta_M\wedge\Theta_{\barMs}\\
	 &d\Theta_M = -(\pi + \overline\tau)\,\Theta_K\wedge\Theta_L + (\overline \epsilon -\epsilon -\overline\rho)\, \Theta_K\wedge\Theta_M
	 -\overline\sigma\, \Theta_K \wedge \Theta_{\barMs} +(\mu -\gamma+\overline\gamma)\,\Theta_L\wedge\Theta_M
	 +\lambda\,\Theta_L\wedge\Theta_{\barMs}\\
	  &\phantom{d\Theta_K =}+(\alpha - \overline\beta)\Theta_M\wedge\Theta_{\barMs}
	 \\
	\end{align*}
	
	The Lorentz group 
	$ \rm  O(\eta)$ is the group of real linear transformations on $V$ which fix  $\eta$. 
	We shall repeatedly use the 
	fact that the Lorentz group acts transitively on the sets of time-like, space-like, and null
	vectors.
	
	Various subgroups of the Lorentz group will play an important role in our analysis. 
	The subgroup of Euclidean rotations in the $M\barM$ plane is the real
	1-parameter subgroup $R_{M \barMs}$ given by
\begin{equation}
	K' = K, \quad L' = L, \enskip  M' = e^{i \theta}M, \enskip \barM' = e^{-i \theta}\barM .
\DGlabel{Rotation}
\end{equation}
	The real 1-parameter subgroup $B_{KL}$ of boosts in the $KL$ plane is 
\begin{equation}
	K' = \boostparm K, \enskip L' = \boostparm^{-1}L, \enskip  M' = M, \enskip \barM' =  \barM,
	\quad \text{where $\boostparm > 0.$} 
\DGlabel{Boost}
\end{equation}
	The sub-group  of null rotations $ N_{K}$ around the $K$ axis is the  
	real 2-parameter group of Lorentz transformations defined by
\begin{equation} 
	K' = K,\enskip 
	L' = \varphi \bar \varphi K + L +  \varphi M  + \bar \varphi\barM, \enskip
	M' = \bar \varphi K + M, \enskip
	\barM' = \varphi K +\barM,
\DGlabel{Null}
\end{equation}
	where  $\varphi=u+iv$ is a complex parameter. Note that  
\begin{equation*}
	M' + \barM'  = (\varphi + \bar \varphi)K + M + \barM, \quad\text{and}\quad
	i(M' - \barM') = i(\bar \varphi - \varphi)K  + i( M - \barM).
\end{equation*}
	The 2 planes $\langle\,K, M + \barM\, \rangle$ and $\langle\,K, i(M - \barM)\, \rangle$
	are preserved by all null rotations $N_{K}$.  In addition,
	with $\varphi$ real, the  vector  $i(M  - \barM)$ is fixed and the group is denoted by $N_{K,u}$;
	with $\varphi$ imaginary,  the  vector  $M  + \barM$ is fixed and the group is denoted by $N_{K,iv}$. 
	

The rotations, boosts, and null rotations all belong to the
    connected  component of the identity in ${\rm O}(\eta)$.
    To describe the various residual groups we shall need the the following discrete transformations, some of which are in the disconnected components of ${\rm O}(\eta)$:
\begin{alignat*}{2}
    \mathcal R : K'&= K, \enskip L' = L, \enskip M' = - M, \enskip 
\barM' = - \barM ; \qquad
&
    \mathcal T : K'&= -K, \enskip L' = -L, \enskip M' = M, \enskip 
\barM' = \barM ;
\\
\mathcal Y : K'&= K, \enskip L' = L, \enskip M' = \barM, 
\enskip\hphantom{-} \barM' = M;
&
\mathcal Z : K'&= L, \enskip \hphantom{-}  L'  = K, \enskip  \hphantom{-}
M' = M, \enskip \barM' = \barM .
\end{alignat*}

    We shall also need the discrete group which permutes the three 
spacelike vectors
     $\{K-L,  M + \barM , i(M - \barM) \}$ up to sign. This group is generated by $\mathcal Y$, $\mathcal T$, and
\begin{align*}
    \mathcal U : K'&= K, \enskip L' = L, \enskip M' = i\, M, \enskip 
\barM' = -i\,\barM ;
\\
    \mathcal V : K' &= \dfrac12( K+ L+ M + \barM),
    \enskip\quad L' = \dfrac12( K+ L- M - \barM),
\\      M' &= \dfrac12( -K+ L+ M - \barM),
    \enskip
    \barM' = \dfrac12(-K+ L- M + \barM) .
\end{align*}

	The group consisting of boosts  $B_{KL}$ and spatio-temporal  reflections $\mathcal T$ is denoted by $B_{KL}^*$.  This group can be defined by 			extending	\eqref{Boost} to all values $\boostparm \neq 0$.  We will also need the group of boosts in the $K+L$, $M+\barM$ plane and an analogous extension to negative values of the boost parameter, which will be 	denoted by $B_{K+L,M+\barMs}$ and $B^*_{K+L,M+\barMs}$, respectively.  See equation (\ref{bigBoost}) for the explicit definition of these groups.

	We will need the formulas for the action of the foregoing Lorentz transformations on the spin coefficients 
	\cite{Newman-Penrose:1962a, Stephani, Stewart:1991}.  
	Under the $M\barM$ rotations and $KL$ boosts the spin coefficients transform as
\begin{equation}
\begin{aligned}
       \alpha' &= e^{-i\theta}\alpha, \enskip
       \beta' = e^{i\theta}\beta, \enskip
       \gamma' =  \boostparm^{-1}\gamma, \enskip
       \epsilon' = \boostparm\,\epsilon, \enskip
       \kappa' =  \boostparm^{2}e^{i\theta}\kappa, \enskip
       \lambda' = \boostparm^{-1}e^{-2\,i\theta}\lambda, \enskip
       \mu' =  \boostparm^{-1}\,\mu,
\\[.5\jot]
        \nu' & = \boostparm^{-2}\, e^{-i\theta} \nu,\enskip
	\pi' = e^{-i\theta}\,\pi,\enskip
	\rho' = \boostparm\,\rho,\enskip  
	\sigma' = \boostparm\,e^{2\,i\theta}\sigma, \enskip
       	\tau'= e^{i\theta}\tau.
\end{aligned}
\DGlabel{NPspintranformRB}
\end{equation}
	Under the null rotations \eqref{Null} they transform as
\begin{equation}
\begin{aligned}
       \kappa'& = \kappa,\qquad
       \epsilon' = \epsilon +\varphi\,\kappa,\qquad
       \sigma'=  \sigma + \bar\varphi\,\kappa,\qquad
       \rho' = \rho + \varphi\,\kappa,\qquad
       \tau' = \tau 
	+ \bar\varphi\,\varphi\,\kappa 
	+ \bar\varphi\,\rho 
	+ \varphi\, \sigma,
\\[1\jot]
       \alpha' & = \alpha 
	+ \varphi^2\,\kappa 
	+ \varphi\,\epsilon 
	+ \varphi\,\rho,
	\qquad
	\beta' = \beta 
	+ \varphi\,\bar\varphi\,\kappa 
	+ \bar\varphi\,\epsilon 
	+ \varphi\,\sigma, 
	\qquad
        \pi'  = \pi + \varphi^2\,\kappa + 2 \,\varphi\,\epsilon ,
\\[1\jot]
       \gamma' &= \gamma 
	+ \varphi^2\bar\varphi\,\kappa\ 
	+ \varphi\,\bar\varphi\,\epsilon
	+ \varphi^2\sigma 
	+ \varphi\,\bar\varphi\,\rho
	+ \bar\varphi\,\alpha
	+ \varphi\,\beta
	+\varphi\,\tau,
\\[1\jot]
       \lambda' &= \lambda 
	+ \varphi^3\,\kappa
	+ 2\,\varphi^2\epsilon
	+ \varphi^2\,\rho
	+ 2\,\varphi\,\alpha
	+ \varphi\,\pi,
	\qquad
        \mu' = \mu 
	+ \varphi^2\bar\varphi\,\kappa
	+ 2\varphi\,\bar\varphi\,\epsilon
	+ \varphi^2\sigma
	+ 2\,\varphi\,\beta
	+ \bar\varphi\,\pi,
\\[1\jot]
       \nu'&= \nu 
	+  \varphi^3\bar\varphi \kappa
	+2\varphi^2\bar\varphi\,\epsilon
	+ \varphi^3\sigma
	+ \varphi^2\bar\varphi\,\rho
	+ 2\,\varphi\,\bar \varphi\,\alpha
	+ 2\,\varphi^2\beta\,
	+ \varphi^2\tau
	+ \varphi\,\bar\varphi\,\pi
	+ 2\varphi\,\gamma
	+ \bar\varphi\,\lambda 
	+ \varphi\,\mu.
\end{aligned}
\DGlabel{NPspintranformN}
\end{equation}
	The derivative terms that usually appear in these formulas are absent since the spin coefficients
	are constant for any $G$-invariant null tetrad on a spacetime group.

	For the discrete transformations we have 
\begin{align*}
\mathcal R : \quad 
       \alpha'&=-\alpha,\enskip
       \beta'=-\beta,\enskip
       \gamma'=\gamma,\enskip
       \epsilon'=\epsilon,\enskip
       \kappa'=-\kappa,\enskip
       \lambda'=\lambda,\enskip
       \mu'=\mu,\enskip
       \nu'=-\nu,\enskip
       \pi'=-\pi,\enskip
\\
       \rho'&=\rho,\enskip
       \sigma'=\sigma,\enskip
       \tau'=-\tau;
\\[2\jot]
\mathcal Y  : \quad
       \alpha'&=\bar\alpha,\enskip
       \beta'=\bar\beta,\enskip
       \gamma'=\bar\gamma,\enskip
       \epsilon'=\bar\epsilon,\enskip
       \kappa'= \bar\kappa,\enskip
       \lambda'= \bar\lambda,\enskip
       \mu'=\bar\mu,\enskip
       \nu'=\bar\nu,\enskip
       \pi'=\bar\pi,\enskip
       \rho'=\bar\rho,\enskip
       \sigma'=\bar\sigma,\enskip
       \tau'= \bar\tau;
\\[2\jot]
\mathcal Z  : \quad\ 
	\alpha'&=-\bar\beta,\enskip
       \beta'=-\bar\alpha,\enskip
       \gamma'=-\bar\epsilon,\enskip
       \epsilon'=-\bar\gamma,\enskip
       \kappa'=-\bar\nu,\enskip
       \lambda'=-\bar\sigma,\enskip
       \mu'=-\bar\rho,\enskip
       \nu'=-\bar\kappa,\enskip
       \pi'=-\bar\tau,\enskip
\\
       \rho&=-\bar\mu,\enskip
       \sigma=-\bar\lambda,\enskip
       \tau=-\bar\pi;
\\[2\jot]
\mathcal T  : \quad
       \alpha'&=\alpha,\enskip
       \beta'=\beta,\enskip
       \gamma'=-\gamma,\enskip
       \epsilon'=-\epsilon,\enskip
       \kappa'=\kappa,\enskip
       \lambda'=-\lambda,\enskip
       \mu'=-\mu,\enskip
       \nu'=\nu,\enskip
       \pi'=\pi,\enskip
\\
	\rho' &= - \rho,\enskip
	\sigma' = -\sigma, \enskip
	\tau' =  \tau; 
\\[2\jot]
     \mathcal U : \quad
	\alpha'&=-i\alpha,\enskip
       \beta'=i\beta,\enskip
       \gamma'=\gamma,\enskip
       \epsilon'=\epsilon,\enskip
       \kappa'=i\kappa,\enskip
       \lambda'=-\lambda,\enskip
       \mu'=\mu, \enskip
       \nu'=-i\nu,\enskip
       \pi'=-i\pi,\enskip
\\
       \rho'&=\rho,\enskip
       \sigma'=-\sigma,\enskip
       \tau'=i\tau.
\\[2\jot]
\mathcal V  : \quad
      \alpha'&= (-\pi-\tau+\sigma+\kappa-\rho+\nu+\lambda-\mu)/4,\enskip
       \beta'=
      (-\pi-\tau-\sigma+\kappa+\rho+\nu-\lambda+\mu)/4,\enskip
\\
      \gamma'&= (\pi-\tau+\sigma-\kappa+\rho+\nu-\lambda-\mu)/4,\enskip
       \epsilon' =(\pi-\tau-\sigma-\kappa-\rho+\nu+\lambda+\mu)/4,\enskip
\\       
	\kappa' & =(2\gamma+2\alpha+2\beta+\pi+\tau+\sigma+\kappa+2\epsilon+\rho+\nu+\lambda+\mu)/4,\enskip
\\ 
	\lambda' & =(-2\gamma-2\alpha+2\beta-\pi+\tau-\sigma-\kappa+2\epsilon+\rho+\nu+\lambda-\mu)/4,\enskip
\\	
        \mu'&=(-2\gamma+2\alpha-2\beta-\pi+\tau+\sigma-\kappa+\epsilon-\rho+\nu-\lambda+\mu)/4,
\\
        \nu'&=(-2\gamma+2\alpha+2\beta+\pi+\tau-\sigma+\kappa-2\epsilon-\rho+\nu-\lambda-\mu)/4,\enskip
\\
      \pi' &=(-2\gamma-2\alpha-2\beta+\pi+\tau+\sigma+\kappa-2\epsilon+\rho+\nu+\lambda+\mu)/4,\enskip
\\	
       \rho'&=(2\gamma+2\alpha-2\beta-\pi+\tau-\sigma-\kappa-\epsilon+\rho+\nu+\lambda-\mu)/4,\enskip
\\
       \sigma'&=(2\gamma-2\alpha+2\beta-\pi+\tau+\sigma-\kappa-2\epsilon-\rho+\nu-\lambda+\mu)/4,\enskip
\\
       \tau' & =(2\gamma-2\alpha-2\beta+\pi+\tau-\sigma+\kappa+2\epsilon-\rho+\nu-\lambda-\mu)/4.
\end{align*}

\medskip

	Finally, the subgroups of $\rm O(\eta)$ which stabilize  various linear subspaces of the vector space $V$ are the residual groups for the corresponding spacetime Lie algebra class. Generators for these subgroups are given in the following table.
\begin{center}
\begin{table}[h]
\caption{\sc  Residual Groups }\label{LeviTable}
\medskip
\begin{tabular}{|l|@{\enskip} l||l|@{\enskip}l|}
\hline
\Strut
	Subspace Pair & Residual Group & Subspace & Residual Group
\\
\hline
\Strut	
	1. $\langle K - L \rangle, \langle \, K - L,\, M,\, \barM \, \rangle$   
&	$R_{M \barMs},\ \mathcal T,\ \mathcal Y,\ \mathcal Z$
&	6.  $\langle  M,\, \barM \, \rangle$
& 	$R_{M \barMs},\enskip B_{KL}^*, \ \mathcal Y,\ \mathcal Z$
\\
\hline
\Strut
	2. $\langle K + L \rangle$, $\langle \, K + L,\, M,\, \barM \, \rangle$   
&	$R_{M \barMs},\ \mathcal T,\ \mathcal Y,\ \mathcal Z$
&	7. $\langle K, L \rangle$
&	$R_{M \barMs}, \enskip B_{KL}^*, \ \mathcal Y,\ \mathcal Z$
\\
\hline
\Strut
	3. $\langle M + \barM \rangle$, $\langle \, K, L,\, M + \barM \, \rangle$   
& 	$B_{KL}^*,\  \mathcal R,\ \mathcal Y,\ \mathcal Z\   $
&	8. $\langle K,  M + \barM \, \rangle$
&	$B_{KL}^*\, ,\enskip N_{K},\  \mathcal R,\  \mathcal Y\ $
\\
\hline
\Strut
	4. $\langle K \rangle$, $\langle \, K, L,\, M + \barM \, \rangle$   
	& $B_{KL}^*,\  N_{K,u},\ \mathcal R,\ \mathcal Y$
&	9. $\langle K+L \rangle$
&	${\rm O}(3) \times \mathcal T$
\\
\hline
\Strut
	5. $\langle M + \barM \rangle$, $\langle \, K, M,  \barM \, \rangle$   
	&  $B_{KL}^*,\ N_{K,iv},\ \mathcal R, \ \mathcal Y$
&	10. $\langle K-L \rangle$
&	${\rm O}(2,1) \times \mathcal Z$

\\
\hline
\Strut
	  
	& 
&	11. $\langle K \rangle$
&	$R_{M\barMs}, B_{KL}, N_K, \mathcal T, \mathcal Y$

\\
\hline
\end{tabular}
\end{table}
\end{center}

\subsection{Preliminary Simplification of the Structure Equations}\label{NPPrel}
	In this section we derive the conditions on the spin coefficients which are implied by
	the assumption that the derived algebra $\lieg'$ is 3-dimensional. We consider 3 cases 
	according to the signature of the induced inner product $\eta'$ on $\lieg'$, 
	where $\eta'(x, y) = \eta(x,y)$ for $x, y \in \lieg'$. 

	By way of preliminary remarks, we note   the derived algebra is an
	ideal, so that $[x, y] \in \lieg'$ for all $x \in \lieg'$ and $y \in \lieg$. If $\omega$ is 
	the annihilating 1-form for $\lieg'$, that is, $\omega(x) = 0$ for all $x\in \lieg'$, 
	then the fact that $\lieg'$ is an ideal is equivalent to $d\,\omega = 0$. 
	The Jacobi tensor is the $(3,1)$ tensor  defined on $\lieg$  by 
\begin{equation*}
	J(x, y, z) = [[x,y],z] + [z,x], y] + [[y,z], x],
\end{equation*}
	so that the structure equations define a Lie algebra if and only if
	$J= 0$. We label the components of $J$ with respect to the null tetrad $\{\,K,\, L,\, M,\,\barM\,\}$ 
	by $J^{d}_{abc}$.  Finally, if  $x, y \in \lieg$, then we write 
	$x\wedge y = \dfrac12(x \otimes y - y \otimes x)$.

\par
\noindent
{\bf Case I. 3-Dimensional Riemannian Derived Algebra.\ } 
	If  the inner product $\eta'$ on $\lieg'$ is Riemannian, then the normal subspace is time-like and
	so can be rotated by a Lorentz transformation to $\langle\, K+L\, \rangle$. Accordingly, 
	the derived algebra is given by
\begin{equation}
	\lieg' = \langle K- L,\ M,\ \barM \rangle.
\DGlabel{derivedR}
\end{equation}
	The conditions $d(\Theta_K + \Theta_L) =0$, which encode the fact that $\lieg'$ is an ideal,
	are solved for the variables
	$\left\{\, \alpha,\, \gamma,\, \mu,\, \pi\, \right\}$ 
	to deduce
\begin{equation*}
	\alpha= -\bar\beta -\nu+\bar\tau, \enskip
	\gamma=i\gamma_1-\dfrac12\epsilon- \dfrac12\bar\epsilon, \enskip
        \mu=\mu_0+ \dfrac12\,\bar\rho- \dfrac12\,\rho, \enskip
	\pi= \bar\kappa -\nu+\bar\tau
\end{equation*}
	and the structure equations \eqref{masterNP} become
\begin{equation}
\begin{alignedat}{1}
	[\, K,\, L\,] & = (\epsilon+\bar\epsilon)(K- L)
	+(\bar\kappa  -\nu + 2\,\bar\tau)\, M+(\kappa\,-\bar\nu + 2\,\tau)\, \barM
\\[1\jot]
	[\, K,\, M\,] & = \kappa\,(K- L) + (\epsilon-\bar\epsilon  +\bar\rho)\, M
	+\sigma\, \barM
\\
	[\, L,\, M\,] & = \bar\nu\,(K- L)+(2\,i\gamma_1-\mu_0- \frac12\,\bar\rho + \frac12\,\rho)\, M
	-\bar\lambda\, \barM
\\
	[\, M,\, \barM\,] & = (\bar\rho-\rho)\, (K- L)+(2\,\bar\beta +\nu-\bar\tau)\, M
	-(2\,\beta +\bar\nu-\tau)\, \barM.
\end{alignedat}
\DGlabel{prelimNPR}
\end{equation}
	These equations will be the starting point for the analysis of \heisRS, \abelthreeRS,  \abelthreeRZ. 
\par
\medskip
\noindent
{\bf Case II. 3-Dimensional Lorentzian Derived Algebra.\ } If the induced metric is 
	Lorentzian, then the normal subspace to $\lieg'$ is space-like and 
	can be rotated by a Lorentz transformation to either $\langle\, K-L\, \rangle$ 
	or $\langle\, i\,(M -\barM)\, \rangle$. Thus, the derived algebra is given by 
\begin{equation}
	\lieg' = \langle\, K + L,\, M, \, \barM\, \rangle \quad\text{or}\quad
	\lieg' = \langle\, K,\,  L,\,  M  + \barM\, \rangle .
\DGlabel{derivedL}
\end{equation}
	The second case will be more useful in those situations
	where the privileged vector in  $\lieg'$ is a null vector.

	In the first instance, the conditions $d(\Theta_K - \Theta_L) =0$ are solved for the variables
	$\left\{\, \alpha,\, \gamma,\,\mu,\ \pi \right\}$ 
	to yield
\begin{equation*} 
	\alpha= -\bar\beta + \nu+\bar\tau, \enskip
	\gamma=i\gamma_1+ \dfrac12\, \epsilon + \dfrac12\,\bar\epsilon, \enskip
	\mu=\mu_0+ \dfrac12\,\rho -\dfrac12\,\bar\rho, \enskip
	\pi= -\bar\kappa + \nu+\bar\tau,
\end{equation*}
	and the structure equations \eqref{masterNP} become
\begin{equation}
\begin{alignedat}{1}
	[\, K,\, L\,] & = -(\epsilon +\bar\epsilon)\, (K + L)-(\bar\kappa -\nu -2\,\bar\tau)\, M
	-(\kappa -\bar\nu  -2\,\tau)\, \barM
\\[1\jot]
	[\, K,\, M\,] & = -\kappa\, (K + L)+(\epsilon-\bar\epsilon + \bar\rho)\, M+\sigma\, \barM
\\[1\jot]
	[\, L,\, M\,] & = \bar\nu\,(K + L) +(2\,i\gamma_1-\mu_0-\dfrac12\rho + \dfrac12\,\bar\rho)\, M-\bar\lambda\, \barM
\\[1\jot]
	[\, M,\, \barM\,] & = (\rho-\bar\rho)( K + L)+(2\,\bar\beta-\nu-\bar\tau)\, M
	-(2\,\beta -\bar\nu-\tau)\, \barM.
\end{alignedat}
\DGlabel{prelimNPL1}
\end{equation}
	The analysis of these structure equations is continued in   \heisLT\ and  \abelthreeLT.

	In the latter instance we solve $d(\Theta_{\vphantom{\barMs}M} -\Theta_{\barMs}) = 0$ for the variables
	$\left\{\, \alpha,\, \lambda,\, \pi,\, \rho \right\}$ 
	to arrive at
\begin{equation*}
	\alpha=i\alpha_1+ \dfrac12\, \beta + \dfrac12\,\bar\beta,\enskip
	\lambda=\bar\mu + \gamma-\bar\gamma,\enskip
	\pi=\pi_0 + \dfrac12\,\tau - \dfrac12\,\bar\tau ,\enskip
	\rho=\bar\sigma +\epsilon -\bar\epsilon.
\end{equation*}
	This results in the structure equations
\begin{equation}
\begin{alignedat}{1}
	[\, K,\, L\,] & = -(\gamma + \bar\gamma)\, K -(\epsilon +\bar\epsilon)\, L+(\pi_0 + \dfrac12\,\bar\tau + \dfrac12\,\tau)\, (M+\barM)
\\[.5\jot]
	[\, K,\, M\,] & = (i\alpha_1-\dfrac32\,\beta- \dfrac12\bar\beta +\pi_0+ \dfrac12\bar\tau- \dfrac12\tau)\, K-\kappa\, L+\sigma\,( M+ \barM)
	\\[.5\jot]
	[\, L,\, M\,] & = \bar\nu\, K+(-i\alpha_1+ \dfrac32\,\beta+ \dfrac12\bar\beta -\tau)\, L+(\gamma-\bar\gamma -\mu)\, ( M+ \barM)
\\[.5\jot]
	[\, M,\, \barM\,] & = (\mu-\bar\mu)\, K+(2\,\epsilon-2\,\bar\epsilon-\sigma+\bar\sigma)\, L - (i\alpha_1+ \dfrac12\,\beta - \dfrac12\,\bar\beta)\, ( M+ \barM).
\end{alignedat}
\DGlabel{prelimNPL2}
\end{equation}

\noindent
	These structure equations are used in \heisLS, \heisLN, \abelthreeLS, \abelthreeLN,  \abelthreeLZ.
\par                          
\noindent
\medskip
{\bf Case III. 3-Dimensional Degenerate Derived Algebra.}	
	If the metric $\eta'$  on $\mathfrak g'$ is degenerate, then $\mathfrak g'$ 
	contains a null vector which, by  a Lorentz transformation,  we may suppose to be $K$ and hence 
\begin{equation}
	\lieg' = \langle\, K,\, M, \, \barM \,\rangle.
\DGlabel{derivedN}
\end{equation}
	The conditions $d\,\Theta_L =0$ are solved for the variables 
	$\left\{\,\alpha,\, \epsilon,\, \kappa,\, \rho\,\right\}$
	to conclude that
\begin{equation*} 
	\alpha=\bar\tau-\bar\beta,\enskip
	\epsilon=i \epsilon_1,\enskip
	\kappa=0,\enskip
	\rho=\rho_0.
\end{equation*}
	In this case the  structure equations \eqref{masterNP} reduce to 
\begin{equation}
\begin{alignedat}{1}
	[\, K,\, L\,] & = -(\gamma +\bar\gamma)\, K+(\pi + \bar\tau)\, M+(\bar\pi + \tau)\, \barM
\\[.5\jot]
	[\, K,\, M\,] & = (\bar\pi-\tau)\, K+(\rho_0+2 i\epsilon_1)\, M+\sigma\, \barM
\\[.5\jot]
	[\, L,\, M\,] & = \bar\nu\, K+(\gamma-\bar\gamma -\mu)\, M-\bar\lambda\, \barM
\\[.5\jot]
	[\, M,\, \barM\, ] & = (\mu-\bar\mu)\, K+(2\bar\beta-\bar\tau\,)\, M- (2\,\beta  -\tau)\, \barM.
\end{alignedat}
\DGlabel{prelimNPN}
\end{equation}
	The analysis of the spacetime Lie algebras \heisNS\ and \abelthreeNS\  
	begins with these structure equations.

	The cases where the derived algebra is 2-dimensional are analyzed in Section \ref{TwoDDerived}.
	The one-dimensional case is considered in Section \ref{OneDDerived}.

\newcommand\summaryref[1]{2.\ref{#1}}

\subsection{Spacetime Groups with  Heisenberg Derived Algebra}\label{HeisDerived}

	In this section  we study the 4-dimensional spacetime Lie algebras $\lieg$ whose derived algebra 
	$\lieg'$ is the 3-dimensional Heisenberg algebra.  The Heisenberg algebra is nilpotent.
	We first remark that, for this case,
	the structure equations for  $\lieg'$ are fixed once the  basis for $\lieg''$ is fixed. Indeed,
	if  $\lieg''$ is spanned by a vector $e_1$,  then for any complementary basis vectors 
	$e_2$, $e_3$ in  $\lieg'$,  the structure equations for $\lieg'$ are
\begin{equation}
	[e_1, e_2] = ae_1  \quad 
	[e_1, e_3] = be_1  \quad  
	[e_2, e_3] = ce_1.
\DGlabel{ThreeDheisenberg}
\end{equation}
	But this algebra is nilpotent if and only if $a = b = 0$. 
	The center of $\lieg'$ is then $\langle e_1 \rangle$.
	
	The cases to be considered are given in Table \ref{hiesTable}.
	
\par
\noindent
{\bf \ref{HeisRS}.  \heisRS.\ } 
	In accordance with \eqref{derivedR}, the derived algebra is written as 
	$\lieg' = \langle \, K - L,\, M,\, \barM\, \rangle$.
	We perform a 3-dimensional Euclidean rotation in $\lieg'$ so that  
	$\lieg''$ is spanned by $K - L$. 
	As noted in \eqref{ThreeDheisenberg}, this implies that
\begin{equation*}
	[\,K - L,\, M + \barM\,] = 0, \quad [\,K -L,\, M - \barM\,] = 0, \quad [\,M,\, \barM\,]\wedge (K -L) = 0,
\end{equation*}
	where the Lie brackets are computed using  \eqref{prelimNPR}. 
	We solve these  equations for 
	$\{ \lambda,\, \nu,\, \rho,\, \tau\}$ to obtain
\begin{equation*}
	\lambda = -\bar\sigma, \quad
	\nu = \bar\kappa, \quad
	\rho =-i\gamma_1+ \dfrac12 \epsilon - \dfrac12\bar\epsilon  -\mu_0, \quad
	\tau = 2\,\beta + \kappa
\end{equation*}
	and substitute these results back into \eqref{prelimNPR}. These structure equations become
\begin{equation}
\begin{alignedat}{1}
	[\, K,\, L\,] & = (\epsilon+\bar\epsilon)\, (K - L) + 2\,(2\,\bar\beta + \bar\kappa)\, M + 
	2 \, (2\,\beta + \kappa)\, \barM
\\[.5\jot]
[\, K,\, M\,] & = \kappa\,( K- L) +(i\gamma_1+ \dfrac12 \epsilon- \dfrac12 \bar\epsilon -\mu_0)\, M+\sigma\, \barM
\\[.5\jot]
[\, L,\, M\,] & = \kappa\, (K- L) + (i\gamma_1+ \dfrac12 \epsilon - \dfrac12 \bar\epsilon -\mu_0)\, M+\sigma\, \barM
\\[.5\jot]
[\, M,\, \barM] & = (2\,i\gamma_1-\epsilon+\bar\epsilon)\,( K - L).
\end{alignedat}
\DGlabel{NPheisRS1}
\end{equation}				
	The  condition  $\epsilon_1 -\gamma_1 \neq 0$ ensures that $\lieg'$ is non-abelian. 
	The   Jacobi tensor components  $J_{123}^1$ and $J_{134}^1$ are
\begin{equation*}
	J_{123}^1 = -4i(\kappa + 2\beta)(\epsilon_1 -\gamma_1) 
	\quad \text{and}\quad  
	J_{134}^1 =  4i (\epsilon_0 - \mu_0)(\epsilon_1 -\gamma_1),
\end{equation*}
	so that we must have $\beta = - \dfrac12\kappa$ and $\epsilon_0 = \mu_0$. 
	The Jacobi identities are then all satisfied and the final structure equations 
	are given in  \summaryref{HeisRS}. The free spin coefficients  for \heisRS\ are
	$\{\,\epsilon_1, \,\gamma_1,\, \mu_0, \, \kappa,\,\sigma\, \}$ with
\begin{align*}                         
	\alpha&=-\frac12\bar\kappa, \enskip
	\beta=- \frac12\kappa,\enskip
	\gamma=-\mu_0+i\gamma_1,\enskip
	\epsilon =\mu_0+i\epsilon_1,\enskip
	\lambda=-\bar\sigma,\enskip	
\\[.5\jot]
	\mu&=i\gamma_1+\mu_0-i\epsilon_1,\enskip
	\nu=\bar\kappa,\enskip
	\pi=0,\enskip
	\rho=-\mu_0-i\gamma_1+i\epsilon_1,\enskip
	\tau=0.
\end{align*}           
	
\par
\smallskip
\noindent
{\bf  \ref{HeisLT}. \heisLT.\ }
	By using the first option in \eqref{derivedL}, we may suppose that 
	$\lieg' = \langle K + L,\, M,\, \barM \rangle $.  
	If  $\lieg''$ is time-like (with respect to the induced inner product on  $\lieg'$),
	then we  perform a 3-dimensional Lorentz transformation  in $\lieg'$ 
	so that $\lieg''$ is spanned by $K + L$. 
	The structure equations \eqref{ThreeDheisenberg} become
\begin{equation*}
	[\, K + L,\, M + \barM\,] = 0, \quad [\,K + L,\, M - \barM\,] = 0, 
	\quad [\,M,\, \barM\,]\wedge (K + L)= 0.
\end{equation*}
	These brackets are computed from  \eqref{prelimNPL1}. 
	We solve the resulting equations for
	$\left\{\lambda,\, \nu,\,\rho,\, \tau \right\}$
	to conclude that
\begin{equation*}
	\lambda  =\bar\sigma, \enskip
	\nu = \bar\kappa, \enskip
	\rho =i\gamma_1+ \frac12\epsilon - \frac12 \bar\epsilon +\mu_0, \enskip
	\tau =2\,\beta  -\kappa
\end{equation*}
	and then substitute these values back into \eqref{prelimNPL1}.  The result is 
\begin{equation}
\begin{alignedat}{1}
	[\, K,\, L\,] & = -(\epsilon +\bar\epsilon)\,(K+ L)
	+2\,(2\bar\beta - \bar\kappa)\, M+ 2\,(2\,\beta-\kappa)\, \barM
\\[.5\jot]
	[\, K,\, M\,] & = -\kappa\, (K + L) +(-i\gamma_1+ \dfrac12\epsilon - \dfrac12\bar\epsilon+\mu_0)\, M+\sigma\, \barM
\\[.5\jot]
	[\, L,\, M\,] & = \kappa\,( K+ L)+(i\gamma_1-\dfrac12 \epsilon+ \dfrac12 \bar\epsilon-\mu_0)\, M-\sigma\, \barM
\\[.5\jot]
	[\, M,\, \barM\,] & = (2\,i\gamma_1+\epsilon-\bar\epsilon)\, (K+ L).
\end{alignedat}
\DGlabel{NPheisLT}	
\end{equation}
	For the second  derived algebra $\lieg''$ to be 1-dimensional, $[\, M,\,\barM\,]$ must be non-zero
	and hence $2i{\gamma_1}+\epsilon-\bar\epsilon  = 2i(\gamma_1 + \epsilon_1)\neq 0$.
	The remaining non-zero components of the Jacobi tensor are 
\begin{equation*}
	J^1_{123} = 2\, \left(2\,\beta -\kappa \right)  \left( 2\,i{\gamma_1}+\epsilon-
	\bar\epsilon \right) 
	\quad\text{and}\quad
	J^1_{234} =\left( 2\,i{\gamma_1}+\epsilon-\bar\epsilon \right)  \left( \epsilon+ \bar\epsilon +
	2\,\mu_0 \right)
\end{equation*}
	so that we must have
	$\beta = \dfrac12 \kappa$ and $\epsilon = - \mu_0 + i\epsilon_1$.
	All Jacobi identities are now satisfied. 
	The structure equations for \heisLT\   are \summaryref{HeisLT}. The free spin coefficients are
	$\{\gamma_1,\, \epsilon_1,\, \mu_0,\, \kappa,\, \,\sigma\}$ with 
\begin{align*}
      \alpha&= \frac12\bar\kappa,\enskip
      \beta= \frac12\kappa,\enskip
      \gamma=i\,\gamma_1 -\mu_0,\enskip
      \epsilon=i\,\epsilon_1-\mu_0,\enskip
      \lambda=\bar\sigma,\enskip
\\[.5\jot]
       \mu&= i\,\gamma_1 + i\,\epsilon_1+\mu_0,\enskip	
      \nu=\bar\kappa,\enskip
      \pi=0,\enskip
      \rho= i\,\gamma_1 + i\,\epsilon_1+\mu_0,\enskip
      \tau=0.
\end{align*}

\par
\smallskip
\noindent
{\bf \ref{HeisLS}. \heisLS.\ } 
	Our starting point is the second case in \eqref{derivedL}, that is,
	$\lieg' = \langle K,\, L,\, M + \barM \rangle$.
	If the second derived algebra $\lieg''$ (or the  center  of $\lieg'$) is space-like 
	with respect to the inner product on $\lieg'$, then we can
	perform a 3-dimensional Lorentz transformation of the basis for $\lieg'$  so that $\lieg''$ 
	is spanned by $M + \barM $. This implies the structure equations
\begin{equation*}
	[\,K , M + \barM\,] = 0, \enskip	
	[\,L, M + \barM\,]  =0, \enskip	 
	[\,K, L\,] \wedge (M + \barM) = 0.
\end{equation*}
	We evaluate these brackets using \eqref{prelimNPL2}.
	These equations then show that the real parts of 
	$\epsilon,\gamma,\kappa, \mu, \nu,  \sigma$ vanish,
	$\beta_0 = \pi_0/2$, and $\tau_0 = \pi_0$, that is,
\begin{equation*}
	\beta= \frac12\pi_0 + i\beta_1,\enskip
	\epsilon=i\epsilon_1,\enskip
	\gamma=i\gamma_1,\enskip
	\kappa=i\kappa_1, \enskip
	\mu=i\mu_1,\enskip
	\nu=i\nu_1,\enskip
	\sigma=i\sigma_1,\enskip
	\tau=\pi_0+i\tau_1.
\end{equation*}
	Consequently, the structure equations  \eqref{prelimNPL2} reduce to 
\begin{equation}
\begin{alignedat}{1}
	[\, K,\, L\,] & = 2\,\pi_0\,( M+ \barM)
\\[.5\jot]
	[\, K,\, M] & = i\,(\alpha_1-\beta_1-\tau_1)\, K-i\kappa_1\, L+i\sigma_1\,(M+ \barM)
\\[.5\jot]
	[\, L,\, M\,] & = -i\nu_1\, K - i\,(\alpha_1 -\beta_1 + \tau_1)\, L+ i\,(2\,\gamma_1 -\mu_1)\,(M +  \barM)
\\[.5\jot]
	[\, M,\, \barM\,] & = 2\,i\mu_1\, K+ 2\, i\,(2\epsilon_1-\sigma_1)\, L
	-i(\alpha_1 +\beta_1)\, (M+ \barM).
\end{alignedat}
\DGlabel{NPheisLS}
\end{equation}
	For the second derived algebra $\lieg''$ to be 1-dimensional, we must have $\pi_0 \neq 0$.
	The independent components of the Jacobi tensor are
\begin{equation*}
	J^1_{123} =  4\,i\,\pi_0\, \mu_1, \enskip
	J^2_{123} =  -4\,i\,\pi_0\,( - 2 \epsilon_1 + \sigma_1), \enskip 
	J^3_{123} = -2\,i \,\pi_0\,(\alpha_1+\beta_1 + 2\,\tau_1) \enskip
\end{equation*}
	and hence
	$\mu_1 =0$, $\sigma_1 =2 \,\epsilon_1$ and $\alpha_1 = - 2\,\tau_1 - \beta_1$.
	The Jacobi identities are thereby satisfied and the final structure equations in this case are
	\summaryref{HeisLS}. The free spin coefficients are 
	$\{\beta_1,\, \gamma_1,\, \epsilon_1,\, \kappa_1,\, \nu_1,\, \pi_0,\, \tau_1\}$ with
\begin{align*}
      \alpha&=-i\,\beta_1 + \pi_0/2-2\,i\,\tau_1,\enskip
      \beta= i\,\beta_1+ \pi_0/2,\enskip
      \gamma=i\,\gamma_1,\enskip
      \epsilon=i\,\epsilon_1,\enskip
      \kappa=i\,\kappa_1,\enskip
      \lambda=2\,i\,\gamma_1,\enskip
\\[.5\jot]
      \mu & = 0, \enskip	
      \nu=i\nu_1,\enskip
      \pi=\pi_0+i\tau_1,\enskip
      \rho=0,\enskip
      \sigma=2\,i\,\epsilon_1,\enskip
      \tau=\pi_0+i\tau_1.
\end{align*}

\par
\smallskip
\noindent
{\bf \ref{HeisLN}. \heisLN.\ }
	For this case we use \eqref{derivedL} to take
	$\lieg' = \langle K,\,  L,\, M + \barM \rangle$ as our starting point.    
	We assume here that the second derived algebra $\lieg''$ is null.
	Using a  Lorentz transformation in $\lieg'$, we can suppose that the 
	second derived algebra $\lieg''= \langle K \rangle$. 
	The structure equations for $\lieg'$ are therefore 
\begin{equation*}
	[\,K, L\,] = 0,\enskip  [\,K, M + \barM\,] = 0, \enskip [\,L, M + \barM \,]\wedge K = 0. 
\end{equation*}
	These equations show, on account of \eqref{prelimNPL2}, that the real parts of 
	$\beta, \epsilon, \gamma,  \kappa, \mu, \pi,  \sigma,\tau$ vanish, 
	that is,
\begin{equation*}
	\beta=i\,\beta_1, \enskip 
	\epsilon=i\,\epsilon_1,\enskip 
	\gamma=i\,\gamma_1, \enskip 
	\kappa=i\,\kappa_1, \enskip 
	\mu=i\,\mu_1,\enskip 
	\pi_0=0,\enskip 
	\sigma=\,i\,\sigma_1,\enskip 
	\tau=i\,\tau_1 .
\end{equation*}
	Consequently, the structure equations  \eqref{prelimNPL2} simplify to  
\begin{equation}
\begin{alignedat}{1}
	[\, K,\, L\,] & = 0
\\[.5\jot]
	[\, K,\, M\,] & = i\,(\alpha_1-\beta_1-\tau_1)\, K-i\,\kappa_1\, L+i\,\sigma_1\, (M + \barM)
\\[.5\jot]
	[\, L,\, M\,] & = \bar\nu\, K -i ( \alpha_1 - \beta_1 + \tau_1)\, L+ 
	i\,(2\,\gamma_1 -\mu_1)\,( M + \barM)
\\[.5\jot]
	[\, M,\, \barM\, ] & = 2\,i\,\mu_1\, K+2\,i\,( 2\,\epsilon_1-\sigma_1)\, L 
	-i\,(\alpha_1 +\beta_1)\,( M + \barM).
\end{alignedat}
\DGlabel{NPheisLN}
\end{equation}
	The algebra $\lieg''$ is non-zero provided $\nu_0 \neq 0$.
	The non-trivial components of the Jacobi tensor are
\begin{equation*}
	J^1_{123} = -2\,i\,\sigma_1\,\nu_0, \quad 
	J^2_{234} = -2\,i\,\kappa_1\,\nu_0, \quad
	J^1_{234} = 2\,i\,(\alpha_1 - 3\,\beta_1)\,\nu_0
\end{equation*}
	so that $\sigma_1 = 0$,  $\kappa_1 = 0$,  $\alpha_1 = 3\,\beta_1$. 
	All the Jacobi identities hold and consequently the  final structure equations for \heisLN\ are 
	\summaryref{HeisLN}. The free spin coefficients are 
	$\{\, \beta_1,\, \gamma_1,\,\epsilon_1,\,\mu_1,\, \nu,\, \tau_1\, \}$, with
\begin{align*}
	\alpha& = 3\,i\,\beta_1,\enskip
	\beta = i\,\beta_1,\enskip
	\gamma = i\,\gamma_1,\enskip
	\epsilon = i\,\epsilon_1,\enskip
	\kappa = 0,\enskip
\\[.5\jot]
	\lambda& = 2\,i\gamma_1 -i\,\mu_1,\enskip
	\mu = i\mu_1,\enskip
	\pi = i\tau_1,\enskip
	\rho = 2\,i\epsilon_1,\enskip
	\sigma = 0,\enskip
	\tau=i\tau_1.
\end{align*}
	
\par
\smallskip
\noindent
{\bf \ref{HeisNS}. \heisNS.\ }
	In accordance with \eqref{derivedN}, we start with $\lieg' = \langle\, K,\, M ,\, \barM \,\rangle$ 
	and the structure equations \eqref{prelimNPN}.
	The subgroup of the Lorentz group preserving $\lieg'$ includes the $M\barM$ rotations, 
	the $KL$ boosts, and the complex null rotations about the $K$ axis.
	If the second derived algebra $\lieg''$ is space-like, then, by a spatial rotation \eqref{Rotation}  
	and a null rotation \eqref{Null} (with $\varphi$ real),  we may take 
	$\lieg'' =  \langle\,  M + \barM \,\rangle$. 
	The structure equations for $\lieg'$ are therefore
\begin{equation*}
	[\,M + \barM, K\,] = 0,\enskip 
	[\,M + \barM,  i\,(M - \barM)\,] = 0, \enskip
	[\,K, i\,(M - \barM)\,] \wedge (M + \barM)  = 0.
\end{equation*}	
	We solve these equations for 
	$\{ \beta,\, \mu,\, \pi,\, \, \rho_0,\ \sigma \}$ 
	to find 
	$\beta= \dfrac12 \tau$,  
	$\mu=\mu_0$,
	$\pi =\bar\tau$,  
	$\rho_0 = 0$,
	$\sigma=2\, i\,\epsilon_1$.
	The structure equations \eqref{prelimNPN} become
\begin{equation}
\begin{alignedat}{1}
	[\, K,\, L\,] & = -(\gamma +\bar\gamma)\, K+2\,\bar\tau\,M +  2\,\tau\barM
\\[.5\jot]
	[\, K,\, M\,] & = 2\,i\epsilon_1\,( M+  \barM)
\\[.5\jot]
	[\, L,\, M\,] & = \bar\nu\, K + (\gamma-\bar\gamma - \mu_0)\, M - \bar\lambda\, \barM
\\[.5\jot]
	[\, M,\, \barM\,] & = 0.
\end{alignedat}
\DGlabel{NPheisNS}
\end{equation}
	The derived algebra $\lieg'$ is the  Heisenberg algebra provided $\epsilon_1 \neq 0$.
	The non-zero components of the Jacobi tensor are
\begin{equation*}
	J^1_{123} = -2\,i\,(\nu + \bar \nu)\,\epsilon_1, \quad
	J^3_{123} = 2\,i\,(\gamma + \bar\gamma + \lambda + \bar\lambda)\epsilon_1,\quad
	J^4_{123} =  2\,i\,(3\, \gamma - \bar\gamma + 2\,\bar \lambda)\,\epsilon_1,\quad
	J^3_{124} =  2\,i\,(\gamma - 3\,\bar\gamma -  2\,\lambda)\, \epsilon_1.
\end{equation*} 
	We solve the Jacobi identities for $\{\lambda, \nu \}$ to deduce that 
	 $\lambda = \dfrac12\, \gamma- \dfrac32\,\bar\gamma$ and $\nu=i\nu_1$.
	The structure equations for a spacetime algebra of type  \heisNS\  are then \summaryref{HeisNS}.
	The free spin coefficients are 
	$\{\,\gamma, \,\epsilon_1,\, \mu_0,\, \nu_1,\, \tau\,\}$
	with
\begin{align*}
      \alpha&= \frac12\bar\tau,\enskip
      \beta= \frac12\tau ,\enskip
      \epsilon=i\epsilon_1,\enskip
      \kappa=0,\enskip
      \lambda=-\gamma_0+2\,i\,\gamma_1,\enskip
\\[.5\jot]
      \mu & =\mu_0,\enskip
      \nu =i\,\nu_1,\enskip
      \pi=\bar\tau,\enskip
      \rho=0,\enskip
      \sigma=2\,i\,\epsilon_1.\enskip
\end{align*}
\par
\smallskip
	The remaining case is \heisNN, which we omitted from the summary in Section 2
	since the isometry algebra of such a spacetime is 6-dimensional.
	Here are the details. As in the case \heisNS, we begin with 
	$\lieg' = \langle K, M , \barM \rangle$ and structure equations \eqref{prelimNPN}.
	Since the derived algebra is  assumed to be null,  we  take  
	$\lieg'' =  \langle K \rangle$.
	The structure equations  for $\lieg'$  become 
\begin{equation*}
	[\, K, M + \barM \,] = 0,\quad  
	[\, K,  i\, (M - \barM)\,] = 0, \quad
	[\, M + \barM, i\, (M - \barM) \,] \wedge K  = 0.
\end{equation*}	
	We solve these for
	$ \{\, \beta,\, \epsilon_1,\, \pi,  \rho_0,\, \sigma\}$ 
	to arrive at
	$\beta= \dfrac12\, \tau$, $\epsilon_1=0$, $\pi=\bar\tau$, $\rho_0=0$ and  $\sigma=0$.
	The structure equations  \eqref{prelimNPN} become
\begin{equation}
\begin{alignedat}{2}
	[\, K,\, L] & = -(\gamma+\bar\gamma)\, K+2\,\bar\tau\, M+2\,\tau\, \barM,
&\qquad
	[\, K,\, M] & = 0,
\\[.5\jot]
	[\, L,\, M] & = \bar\nu\, K+(\gamma-\bar\gamma -\mu)\, M-\bar\lambda\, \barM,
&\qquad
	[\, M,\, \barM] & = (\mu-\bar\mu)\, K.
\end{alignedat}
\DGlabel{NPheisNN}
\end{equation}
	The requirement that $\lieg'$ is not abelian becomes $\mu_1 \neq 0$. 
	From the components of the  Jacobi tensor
\begin{equation*}
	J^1_{123} = 4\,i\,\tau\,\mu_1 \quad\text{and}\quad
	J^1_{234}= 2\,i\,(\gamma + \bar\gamma +\mu +\bar\mu)\,\mu_1
\end{equation*}
	we arrive at $\tau = 0$ and $\mu = -\dfrac12\,(\gamma + \bar\gamma) + i\,\mu_1$. 
	The Jacobi identities are now satisfied. A spacetime Lie group with this Lie algebra admits a 6 dimensional 
	isometry algebra. 
%
%
%
\subsection{Spacetime Groups with  3 Dimensional Abelian Derived Algebra}\label{AbelDerived}
	In this section  we study the 4-dimensional spacetime Lie algebras $\lieg$ whose derived algebra 
	$\lieg'$ is the 3-dimensional abelian algebra. Again our analysis is organized by the 
	signature of the inner product $\eta'$ on $\lieg'$ and then by the spacetime character of the 
	line defined by $\zeta \in \lieg'$ (see  \eqref{defzeta}) with respect to the inner product $\eta'$.
	The cases to be considered are given in Table \ref{abel3Table}.
\par
\smallskip
\noindent
{\bf \ref{IIIdAbelianRS}. \abelthreeRS.\ } 
	We begin by aligning the null tetrad to the Riemannian derived algebra $\lieg'$, that is, we take 
	$\lieg' = \langle\, K-L,\,  M,\, \barM\, \rangle$.
	The initial structure equations are \eqref{prelimNPR}. We solve the equations 
\begin{equation*}
	[\,K-L, M + \barM\,] = 0, \quad [\, K - L, i\,(M - \barM)\,]  =0, \quad [\,M + \barM,  i\,(M - \barM)\,] = 0
\end{equation*}
	for the variables 
	$\{\gamma_1,\,  \lambda,\ \mu_0,\, \nu,\,\rho, \, \tau\, \}$
	to find that
	$\gamma_1 = \epsilon_1$, 
	$\lambda = -\bar\sigma$,
	$\mu_0 = -\rho_0$, 
	$\nu = \bar\kappa$,
	$\rho = \rho_0$, and 
	$\tau = \kappa + 2\beta$.
	The structure equations at this point are  
\begin{equation}
\begin{alignedat}{1}
	[\, K,\, L] & = (\epsilon+\bar\epsilon)\,(K -L) 
	+2\,( 2\,\bar\beta + \bar\kappa)\,M+ 2\,(2\,\beta + \kappa)\, \barM
\\[.5\jot]
	[\, K,\, M\, ] & = \kappa\,(K- L) +(\epsilon-\bar\epsilon + \rho_0)\, M+\sigma\, \barM
\\[.5\jot]
	[\, L,\, M] & = \kappa\, (K  - L) + (\epsilon-\bar\epsilon + \rho_0)\, M + \sigma\, \barM
\\[.5\jot]
	[\, M,\, \barM] & = 0.
\end{alignedat}
\DGlabel{NPabel3RS}
\end{equation}
	The Jacobi identities are satisfied. The vector $\zeta$ defined in  \eqref{defzeta} (which may be calculated using the orthogonal complement  $\langle K +L\rangle)$ 
	 is 
\begin{equation*} 
	\zeta  =  i\,(\epsilon-\bar\epsilon)\,( K - L)   
	+ 2\,i\,(\bar\beta +\bar\kappa)\,M - 2\,i\,(\beta+\kappa)\, \barM.
\end{equation*}
	Since we assume that $\zeta$ is space-like, the null tetrad can be rotated so that $\zeta$ is a multiple of $K-L$. 
	In the rotated null tetrad, one has $\beta = - \kappa$.
	This gives \summaryref{IIIdAbelianRS}. The free spin coefficients are
	$\{\, \epsilon,\, \kappa,\, \rho_0,\,  \sigma\,\}$
	with 
\begin{align*}
      \alpha&=-\bar\kappa,\enskip
      \beta=-\kappa,\enskip
      \gamma=-\epsilon_0+i\epsilon_1,\enskip
      \lambda=-\bar\sigma,\enskip
      \mu=-\rho_0,\enskip
     \nu =\bar\kappa,
\\[.5\jot]
      \pi&=-\bar\kappa,\enskip
      \rho=\rho_0,\enskip
      \tau=-\kappa.
\end{align*}

\smallskip
\noindent
{\bf \ref{IIIdAbelianRZ}. \abelthreeRZ.\ }	
	In this case the structure equations \summaryref{IIIdAbelianRZ}  are derived 
	from \eqref{NPabel3RS}  with  $\epsilon = \epsilon_0$ and $\beta = - \kappa$. 
\par
\medskip
\noindent
{\bf \ref{IIIdAbelianLT}. \abelthreeLT.\ }
	In accordance with the first possibility in \eqref{derivedL}
	we take  $\lieg' = \langle \, K + L,\, M,\, \barM \,\rangle$ 
	and begin with the structure equations \eqref{prelimNPL1}. 
	We require that  $\lieg'$ be abelian
	and solve the resulting equations for
	$\{\, \epsilon,\, \mu_0,\,   \lambda,\,\nu,\, \rho,\, \tau\,\}$ 
	to conclude that 
	$ \epsilon = \epsilon_0 - i\gamma_1$, 
	$\mu_0 = \rho_0$,
	$\lambda=\bar\sigma$,
	$\nu = \bar\kappa$, 
	$\rho=\rho_0$, 
	$\tau = -\kappa + 2\,\beta$.
	The structure equations  \eqref{prelimNPL1}  become
\begin{equation}
\begin{alignedat}{1}
	[\, K,\, L\,] & = -2\,\epsilon_0\,( K +  L)+ 2(2\,\bar\beta-\,\bar\kappa)\, M 
	+ 2(2\,\beta -\,\kappa)\, \barM
\\[.5\jot]
	[\, K,\, M\, ] & = -\kappa\,( K + L)+(\rho_0-2\,i\gamma_1)\, M+\sigma\, \barM
\\[.5\jot]
	[\, L,\, M\,] & = \kappa\, (K+  L) -(\rho_0-2\,i\gamma_1)\, M-\sigma\, \barM
\\[.5\jot]
	[\, M,\, \barM] & = 0.
\end{alignedat}
\DGlabel{NPabel3LT1}
\end{equation}
	At this point the Jacobi identities all hold. The direction $\zeta$ defined in \eqref{defzeta}  (which may be calculated using the orthogonal complement $\langle K - L\rangle$) is 
\begin{equation*}
	\zeta  = \gamma_1\, (K +L) - 
	i\,(\bar\beta- \bar\kappa)\,M + i\,(\beta -\kappa)\, \barM.
\end{equation*}
	Since we assume $\zeta$ is time-like, by a Lorentz transformation we may align this vector with  $K + L$ 
	so that $\beta = \kappa$ and $\gamma_1 \neq 0$. The structure equations \eqref{NPabel3LT1}
	become \summaryref{IIIdAbelianLT}. The free spin coefficients for  \abelthreeLT\ are
	$\{\gamma_1,\, \epsilon_0,\, \kappa,\, \rho_0,\, \sigma\,\}$
	with
\begin{align*}
      \alpha& =\bar\kappa,\enskip
      \beta=\kappa,\enskip
      \gamma=\epsilon_0+i\gamma_1,\enskip
      \epsilon=\epsilon_0-i\gamma_1,\enskip
      \lambda=\bar\sigma,\enskip
      \mu=\rho_0,\enskip
       \nu= \bar\kappa,\enskip		
\\[.5\jot]      
      \pi &=\bar\kappa,\enskip
      \rho=\rho_0,\enskip
      \tau=\kappa.
\end{align*}
\smallskip
\noindent
{\bf \ref{IIIdAbelianLS}. \abelthreeLS.\ }
	In this case we start with the second possibility in \eqref{derivedL}, namely, 
	$\lieg' = \langle K, L, M + \barM \rangle$ and the structure 
	equations \eqref{prelimNPL2}. The equations 
	$[K, L] = [K, M +\barM] = [L, M + \barM] = 0$
	are solved for
	$\{ \beta, \epsilon, \gamma, \kappa,  \mu, \nu,  \pi_0, \sigma, \tau \,\}$
	to  yield
\begin{equation*}
	\beta = i\,\beta_1,\enskip
	\epsilon = i\,\epsilon_1, \enskip
	\gamma = i\,\gamma_1, \enskip
	\kappa = i\,\kappa_1,\enskip
	\mu = i\,\mu_1, \enskip
	\nu = i\,\nu_1, \enskip
	\pi_0=0,\enskip
	\sigma=i\,\sigma_1,\enskip
	\tau=i\,\tau_1.
\end{equation*}
	The structure equations \eqref{prelimNPL2} therefore become
\begin{equation}
\begin{alignedat}{1}
	[\, K,\, L\,] & = 0
\\[.5\jot]
	[\, K,\, M\,] & = i\,(\alpha_1-\beta_1-\tau_1)\, K-i\,\kappa_1\, L+i\,\sigma_1\, (M+ \barM)
\\[.5\jot]
	[\, L,\, M\,] & = -i\nu_1\, K - i\,(\alpha_1 -\beta_1 + \tau_1)\, L+ i\,(2\, \gamma_1 -\mu_1)\, (M +\barM) 
\\[.5\jot]
	[\, M,\, \barM\,] & = 2\,i\mu_1\, K+ 2\,i\,(2\,\epsilon_1- \sigma_1)\, L -i\,(\alpha_1 + \beta_1)\,( M 
	+\barM).
\end{alignedat}
\DGlabel{NPabel3LS2}
\end{equation}
	The direction $\zeta$, which may be calculated using the orthogonal complement $\langle i(M-\barM)\rangle$, is
\begin{equation}
	\zeta = 2\,(\gamma_1 -\mu_1)\, K  -2\,(\sigma_1-\epsilon_1)\, L - (\,\alpha_1-\beta_1)\,( M+ \barM).
\DGlabel{zetaLS}
\end{equation}
	In this case  we are assuming that $\zeta$ is space-like and therefore   
	we can rotate the null tetrad so that $\zeta$ becomes a multiple of $M + \barM$.  Hence 
	$\gamma_1=\mu_1$, $\sigma_1=\epsilon_1$  and  $\alpha_1-\beta_1 \neq 0$.
	The final structure equations for \abelthreeLS\ are \summaryref{IIIdAbelianLS}.  
	The free spin coefficients are 
	$\{\, \alpha_1,\, \beta_1,\, \epsilon_1,\, \kappa_1,\, \mu_1,\, \nu_1,\, \tau_1\, \} $
	with
\begin{align*}
      \alpha&=i\,\alpha_1,\enskip
      \beta=i\,\beta_1,\enskip
      \gamma=i\,\mu_1,\enskip
      \epsilon=i\,\epsilon_1,\enskip
      \kappa=i\,\kappa_1,\enskip
      \lambda=i\,\mu_1,\enskip
      \mu=i\,\mu_1,
\\[.5\jot]
      \nu&=i\,\nu_1,\enskip
      \pi=i\,\tau_1,\enskip
      \rho=i\,\epsilon_1,\enskip
      \sigma=i\,\epsilon_1,\enskip
      \tau=i\,\tau_1.
\end{align*}
                               
\par
\smallskip
\noindent
{\bf \ref{IIIdAbelianLN}. \abelthreeLN.\ }
	As in the case \abelthreeLS,  we start with $\lieg' = \langle\, K, L, M + \barM \,\rangle$, 
	the structure 
	equations are \eqref{NPabel3LS2}, and the vector \eqref{zetaLS} is assumed to be null.
	We rotate $\zeta$ to lie along $K$ so that, for this null tetrad,
	$\beta_1=\alpha_1$, $\sigma_1=\epsilon_1$  and  $\gamma_1 - \mu_1 \neq 0$.
	The structure equations \eqref{NPabel3LS2} become  \summaryref{IIIdAbelianLN}.
	The free spin coefficients are 
	$\{\alpha_1,\, \gamma_1,\, \epsilon_1,\,  \kappa_1,\, \mu_1,\, \nu_1,\, \tau_1\,\}$  
	 with	
\begin{align*}
      \alpha&=i\,\alpha_1,\enskip
      \beta=i\,\alpha_1,\enskip
      \gamma=i\,\gamma_1,\enskip
      \epsilon=i\,\epsilon_1,\enskip
      \kappa=i\,\kappa_1,\enskip
      \lambda=-i\,\mu_1+2\,i\gamma_1,\enskip
      \mu=i\,\mu_1,
\\[.5\jot]
      \nu&=i\,\nu_1,\enskip
      \pi=i\,\tau_1,\enskip
      \rho=i\,\epsilon_1,\enskip
      \sigma=i\,\epsilon_1,\enskip
      \tau= i\,\tau_1.
\end{align*}

\par
\smallskip
\noindent
{\bf \ref{IIIdAbelianLZ}. \abelthreeLZ.\ }
	Again we take  $\lieg' = \langle\, K, L, M + \barM \,\rangle$ with  structure 
	equations \eqref{NPabel3LS2}. Now we require that the vector \eqref{zetaLS}
	vanishes and therefore $\beta_1=\alpha_1$, $\sigma_1=\epsilon_1$  and  $\mu_1 = \gamma_1$.
	This leads to the structure  equations \summaryref{IIIdAbelianLZ}. The free 
	spin coefficients are
	$\{\alpha_1,\, \gamma_1,\, \epsilon_1,\, \kappa_1,\, \nu_1,\, \tau_1\, \}$ and
\begin{align*}
       \alpha&=i\,\alpha_1,\enskip
       \beta=i\,\alpha_1,\enskip
       \gamma=i\,\gamma_1,\enskip
       \epsilon=i\,\epsilon_1,\enskip
       \kappa=i\,\kappa_1,\enskip
       \lambda=i\,\gamma_1,\enskip
       \mu=i\gamma_1,
\\[.5\jot]
       \nu&=i\,\nu_1,\enskip
       \pi=i\,\tau_1,\enskip
       \rho=i\,\epsilon_1,\enskip
       \sigma=i\,\epsilon_1,\enskip
       \tau=i\,\tau_1.
\end{align*}

\par
\smallskip
\noindent
{\bf \ref{IIIdAbelianNS}.  \abelthreeNS.\ }
	In view of \eqref{derivedN}, we may suppose that the null tetrad is chosen so that
	$\lieg'  = \langle K, M,  \barM \rangle$ with 
	structure equations \eqref{prelimNPN}. The conditions  for $\lieg'$ to be abelian
	are solved for $ \{\epsilon_1,\, \mu,\, \pi,\, \rho_0,\, \sigma,\, \tau\, \}$;
	the result being
	$\epsilon_1 = 0$,
	$\mu=\mu_0,$ 
	$\pi=2\,\bar\beta$, 
	$\rho_0 = 0$,
	$\sigma=0$, 
	$\tau = 2\,\beta$. 
	By virtue of these equations, the structure equations  \eqref{prelimNPN} simplify to
\begin{equation}
\begin{alignedat}{2}
	[\, K,\, L\,] & = -(\gamma+\bar\gamma)\, K+4\,\bar\beta\, M+4\,\beta\, \barM,
&\qquad	
	[\, K,\, M\,] & = 0,
\\[.5\jot]
	[\, L,\, M\,] & = \bar\nu\, K+(\gamma-\bar\gamma -\mu_0)\, M-\bar\lambda\, \barM,
&\qquad
	[\, M,\, \barM\,] & = 0.
\end{alignedat}
\DGlabel{NPabel3NS}
\end{equation}
	The Jacobi identities hold. The vector $\zeta$ in \eqref{defzeta} can be computed using  $v =L$ and $N = K$; this yields
\begin{equation*}
	\zeta =  i\,(\gamma  - \bar \gamma)\,K + 2\,i\,\beta\,M - 2\,i\,\bar \beta\,\barM.
\end{equation*}
	We suppose here that $\zeta$ is space-like. Therefore 
	the null tetrad may be transformed, by a null rotation and a Euclidean rotation, so that 
	$\zeta$ is a multiple of $M + \barM$.  For the transformed tetrad
	$\gamma = \gamma_0$  and $\beta = i\,\beta_1$.
	This gives the final result \summaryref{IIIdAbelianNS}. The  free spin coefficients for 
	\abelthreeNS\ are $\{\, \beta_1,\, \gamma_1,\, \lambda,\, \mu_0,\, \nu\, \}$, with 
\begin{align*}
       \alpha&=-i\,\beta_1,\enskip
       \beta=i\,\beta_1,\enskip
       \gamma=\gamma_0,\enskip
       \epsilon=0,\enskip
       \kappa=0,\enskip
       \mu=\mu_0,\enskip
	\pi=-2\,i\,\beta_1,\enskip
\\[.5\jot]
       \rho & = 0,\enskip
       \sigma=0,\enskip
       \tau=2\,i\,\beta_1.
\end{align*}

\par
\smallskip
	The remaining cases are \abelthreeNN\ and \abelthreeNZ, whose structure equations can be obtained from  \eqref{NPabel3NS} by assuming $\zeta$ is null or zero.  
	In the former case, by a Lorentz transformation it is possible to align $K$ with $\zeta$.  Using the transformed tetrad it follows that $\beta = 0$.  
	However this implies the isometry group of the spacetime 	 is 6-dimensional.  Similarly, if $\zeta =0$ then $\beta=0$ and $\gamma_1=0$; again the 		isometry group of the metric is 6.
	Consequently, we have omitted these cases from the summary in Section 2.

\subsection{Spacetime Algebras with Simple Derived Algebra}\label{SimpleDerived}
	In this section we classify spacetime Lie algebras whose derived algebra is semi-simple. All such algebras have a 
	1-dimensional center $\liez$  -- our analysis is based on the spacetime character of $\liez$.
\par 
\noindent
{\bf \ref{simpT}. \LevitwoT.} \quad 
	If the center $\liez$ of our spacetime Lie algebra $\lieg$  is time-like, 
	then we may rotate the null tetrad so that $\liez = \langle\, K +L \,\rangle$.
	The conditions that $\liez$  has vanishing brackets with $L$, $M$, $\barM$ lead to a set of 
	linear equations for the spin coefficients, which we solve for 
	$\{\alpha,\, \gamma,\, \epsilon,\, \kappa,\,  \pi,\, \rho,\,   \sigma\}$.
	We find 
\begin{equation*}
	\alpha = -\bar\beta +\nu -\bar\tau,\quad
	\gamma = i\,\gamma_1,\enskip
	\epsilon = i\,\epsilon_1,\enskip
	\kappa = \bar\nu-2\,\tau, \enskip
	\pi = -\bar\tau,\enskip
	\rho = 2\,i\,\epsilon_1+2\,i\,\gamma_1 + \bar \mu,\enskip
	\sigma = \bar\lambda.
\end{equation*}
	The structure equations  \eqref{masterNP} become 
\begin{equation}
\begin{alignedat}{1}
	[\, K,\, L\,] & = 0,
\quad 
	[\, K,\, M\,] =  - [\, L,\, M\,] 
	= -\bar\nu\, K -(\bar\nu-2\,\tau )\, L -(2\,i\,\gamma_1 -\mu)\, M+\bar\lambda\, \barM,
\\[1\jot]
	[\, M,\, \barM\,] & = (\mu-\bar\mu)\, K+(4\,i\,\epsilon_1+4\,i\,\gamma_1 -\mu+\bar\mu)\, L +
	(2\,\bar\beta-\nu  +\bar\tau)\, M -(2\,\beta  -\bar\nu +\tau )\, \barM.
\end{alignedat}
\DGlabel{NPsimpCT1}
\end{equation}

	Next we require that the quotient algebra $\tilde \lieg = \lieg/\liez$ be a semi-simple 
	(in fact, simple) Lie algebra. This implies that $\tilde \lieg  = (\tilde \lieg)' $
	which, in turn, implies  that the adjoint matrices $\text{ad}(x)$ are trace-free for all $x\in \tilde \lieg$. 
	These conditions yield $\mu = i\,\mu_1$  and  $\tau = 2\,\beta - \bar \nu$. 
	The Jacobi identities are now all satisfied and the structure equations are 
	\summaryref{simpT}.  In this case the free spin coefficients are 
	$\{\,\beta,\, \gamma_1,\, \epsilon_1,\, \lambda,\, \mu_1,\, \nu\, \}$ with
\begin{align*}
       \alpha&=2\,\nu-3\,\bar\beta,\enskip
       \gamma=i\,\gamma_1,\enskip
       \epsilon=i\,\epsilon_1,\enskip
       \kappa=-4\,\beta+3\,\bar\nu,\enskip
       \mu=i\,\mu_1,\enskip
	\pi=-2\,\bar\beta+\nu,
\\[.5\jot]
       \rho&=2\,i\,\epsilon_1-i\,\mu_1+2\,i\,\gamma_1,\enskip
       \sigma=\bar\lambda,\enskip
       \tau=2\,\beta-\bar\nu.
\end{align*}	
{\bf \ref{simpS}. \LevitwoS.}\quad If the center $\liez$ of our spacetime Lie algebra $\lieg$  is space-like,
	then we may rotate the null tetrad to make $\liez = \langle\, K -L \,\rangle$.
	From the conditions that $\liez$ has vanishing brackets with $L$, $M$, $\barM$ we deduce that
\begin{equation*}
	\alpha = -\bar\beta-\bar\tau-\nu,\enskip
	\gamma = i\,\gamma_1, \enskip
	\epsilon = i\,\epsilon_1,\enskip
	\kappa = \bar\nu + 2\,\tau, \enskip
	\pi = -\bar\tau, \enskip
	\rho = -2\,i\,\gamma_1 +2\,i\,\epsilon_1  -\bar\mu, \enskip
	\sigma = -\bar\lambda.
\end{equation*}
	The structure equations  \eqref{masterNP} become
\begin{equation}
\begin{alignedat}{1}
	[\, K,\, L\,] & = 0,
\quad
	[\, K,\, M\,] = -[\, L,\, M\,]   = \bar\nu\, K -(\bar\nu +2\,\tau  )\, L
	+(2\,i\,\gamma_1 -\mu)\, M-\bar\lambda\, \barM,
\\[1\jot]
	[\, M,\, \barM] & = 
	(\mu-\bar\mu)\, K -  (4\,i\,\gamma_1 -4\,i\,\epsilon_1 -\mu +\bar\mu)\, L+(2\,\bar\beta+\nu +\bar\tau)\, M 
	-(2\,\beta +\bar\nu +\tau)\, \barM.
\end{alignedat}
\DGlabel{NPsimpCS1}
\end{equation}

	Next we require that the adjoint matrices for the quotient algebra $\tilde \lieg = \lieg/\liez$ 
	be trace-free. These conditions require $\mu = i\,\mu_1$ and $\tau =  \bar \nu+ 2\,\beta$. 
	The Jacobi identities are satisfied and the structure equations  \eqref{NPsimpCS1}
	become \summaryref{simpS}. The free spin coefficients for \LevitwoS\ are
	$\{\,\beta,\, \gamma_1,\, \epsilon_1,\, \lambda,\, \mu_1,\, \nu\, \}$, with
\begin{align*}
       \alpha&=-2\,\nu-3\,\bar\beta,\enskip
       \gamma=i\,\gamma_1,\enskip
       \epsilon=i\,\epsilon_1,\enskip
       \kappa=4\,\beta+3\,\bar\nu,\enskip
       \mu=i\,\mu_1,\enskip
	 \pi=-2\,\bar\beta-\nu,
\\[.5\jot]
       \rho&=2\,i\,\epsilon_1+i\,\mu_1-2\,i\,\gamma_1,\enskip
       \sigma=-\bar\lambda,\enskip
       \tau=2\,\beta+\bar\nu.
\end{align*}
\par
\smallskip
\noindent
{\bf \ref{simpN}. \LevitwoN.} \quad If the center $\liez$ of our spacetime Lie algebra $\lieg$  is null,
	then we may rotate the null tetrad to give $\liez = \langle\, K \,\rangle$.
	From the conditions that $\liez$ has vanishing brackets with $L$, $M$, $\barM$ we deduce that
\begin{equation*}	
	\alpha = -\bar\beta-\bar\tau, \enskip
	\gamma=i\gamma_1, \enskip
	\epsilon=i\epsilon_1,\enskip
	\kappa=0, \enskip
	\pi=-\bar\tau, \enskip
	\rho = 2\,i\epsilon_1,  \enskip
	\sigma=0,
\end{equation*}	
	in which case the structure equations \eqref{masterNP} reduce to
\begin{equation}
\begin{alignedat}{1}
	[\, K,\, L\,] & = 0,
\quad 
	[\, K,\, M\,]  = 0,
\quad 
	[\, L,\, M\,]  = \bar\nu\, K-2\,\tau\, L +(2\,i\gamma_1-\mu)\, M-\bar\lambda\, \barM,
\\[1\jot]
	[\, M,\, \barM] & = (\mu-\bar\mu)\, K+4\,i\epsilon_1\, L+(2\,\bar\beta+\bar\tau)\, M 
	- (2\,\beta +\tau)\, \barM.
\DGlabel{NPsimpCN1}
\end{alignedat}
\end{equation}

	The requirement that the adjoint matrices for the quotient algebra $\tilde \lieg = \lieg/\liez$ 
	be trace-free leads to $\mu = i\mu_1$ and  $\tau = 2\, \beta$. 
	The Jacobi identities hold and the structure equations \eqref{NPsimpCN1} 
	become \summaryref{simpN}. The free spin coefficients for \LevitwoN\ are
	$\{\,\beta,\, \gamma_1,\, \epsilon_1,\, \lambda,\, \nu\,\}$,
	with
\begin{align*}
       \alpha&=-3\,\bar\beta,\enskip
       \gamma=i\,\gamma_1,\enskip
       \epsilon=i\,\epsilon_1,\enskip
       \kappa=0,\enskip
       \mu=i\mu_1,\enskip
	\pi=-2\,\bar\beta,
\\[.5\jot]
       \rho & = 2\,i\,\epsilon_1,\enskip
       \sigma=0,\enskip
       \tau=2\,\beta.
\end{align*}
%
%
\subsection{Spacetime Groups with  2-Dimensional Derived Algebra}\label{TwoDDerived} 
	If $\lieg$ is a 4-dimensional Lie algebra with a 2-dimensional derived algebra, 
	then $\lieg$ is necessarily solvable.  But the derived algebra of a solvable algebra is 
	always nilpotent. Hence $\lieg'$ is a 2-dimensional nilpotent algebra
	and therefore $\lieg'$ is abelian. Our analysis is based on the spacetime character of the 2-dimensional abelian algebra $\lieg'$.
	The cases are listed in Table \ref{abel2Table}.
\par
\noindent
{\bf \ref{IIdAbelR1}--\ref{IIdAbelR3}. \abeltwoRone, \abeltwoRtwo, \abeltwoRthree.\ }
	If the derived algebra is a 2-dimensional Riemannian subspace, 
	then we may take $\lieg' = \langle\, M, \barM \,\rangle$. 
	We solve the equations
\begin{equation*}
	[\,M, \barM\,] =0 , \quad [\,x,y\,] \wedge M \wedge \barM = 0 \quad 
	\text{for all $x, y \in \mathfrak g$}
\end{equation*}
	for the spin coefficients 
	$\{\alpha,\, \epsilon_0,\, \gamma_0,\, \kappa,\, \mu,\,\nu,\,\pi,\,\rho,\,\tau\}$ to find
\begin{equation*}
	\alpha=\bar\beta,\quad
	\epsilon=i\,\epsilon_1,\quad
	\gamma =i\,\gamma_1, \quad
	\kappa=0, \quad
	\mu=\mu_0, \quad
	\nu=0,\quad
	\pi=2\,\bar\beta,\quad
	\rho=\rho_0,\quad
	\tau=2\,\beta. 
\end{equation*}
	The structure equations are then
\begin{equation}
\begin{alignedat}{2}
	[\, K,\, L\, ] & = 4\,\bar\beta\, M+4\,\beta\, \barM,
&\quad
	[\, K,\, M\,] & = (2\,i\,\epsilon_1   + \rho_0)\, M+\sigma\, \barM,
\\[.5\jot]
	[\, L,\, M\, ] & = (2\,i\,\gamma_1- \mu_0)\, M-\bar\lambda\, \barM,
&\quad
	[\, M,\, \barM\, ] & = 0,
\end{alignedat}
\DGlabel{NPabel2R1}
\end{equation}
	and the  independent components of the Jacobi tensor are
\begin{equation*}
	J^3_{123} = \lambda\,\sigma-\bar\lambda\,\bar\sigma,
\quad
	J^4_{123} = 4\,i \,(\epsilon_1 \bar\lambda +\gamma_1 \sigma).
\end{equation*}
	The sub-group of the Lorentz group which stabilizes $\lieg'$ 
	is the 2-dimensional group generated by $KL$ boosts, $M\barM$ rotations, and  $\{\mathcal Y,\ \mathcal Z,\ \mathcal T \}$.

	To solve the Jacobi identities we consider 4 cases. If $\sigma \neq 0$ and $\lambda \neq0$,  we use 
	the $M\barM$ rotation to make $\lambda$ real. The Jacobi identity
	$J^3_{123} =0 $ then implies that $\sigma$ is real. Next, we use a boost to make $\lambda = q\sigma$, 
	where $q^2 =1$. Then $J^4_{123} =0 $ gives $\gamma_1 = -q\epsilon_1$. This case is \abeltwoRone; the structure equations are 
	\summaryref{IIdAbelR1}. 
	If $\sigma \neq 0$ and $\lambda = 0$, then  we have $\gamma_1 = 0$. This case is \abeltwoRtwo; the
	 structure equations become \summaryref{IIdAbelR2}.
	The case where $\sigma = 0$ and $\lambda \neq 0$ matches the previous case with a $KL$ swap, 
	that is, after making the residual transformation $\mathcal Z$.  It is therefore not a new case.
	The Jacobi identities are immediately satisfied when $\sigma = \lambda = 0$, which leads to the case \abeltwoRthree\ with structure equations
	 \summaryref{IIdAbelR3}.

	For \abeltwoRone\ the residual group is generated by 
	$\mathcal  T$, $\mathcal U$, $\mathcal Y$, $\mathcal Z$.
	The independent spin coefficients are
	$\{\beta,\, \epsilon_1,\, \mu_0,\, \rho_0,\, \sigma_0\,\}$,
	with
\begin{align*}
       \alpha&=\bar\beta,\enskip
       \gamma=-i\epsilon_1,\enskip
       \epsilon=i\epsilon_1,\enskip
       \kappa=0,\enskip
       \lambda=\sigma_0,\enskip
       \mu=\mu_0,
\\[.5\jot]
       \nu&=0,\enskip
       \pi=2\,\bar\beta,\enskip
       \rho=\rho_0,\enskip
       \sigma=\sigma_0,\enskip
       \tau=2\,\beta.
\end{align*}
	For \abeltwoRtwo\ the residual group is the 2-dimensional group generated by 
	rotations and boosts, $\mathcal Y$ and  $\mathcal T$.
	The independent spin  coefficients are
	$\{\beta,\ \epsilon_0,\ \mu_0,\ \rho_0,\ \sigma\,\}$,
	with
\begin{align*}
       \alpha&=\bar\beta,\enskip
       \gamma=0,\enskip
       \epsilon=i\epsilon_1,\enskip
       \kappa=0,\enskip
       \lambda=0,\enskip
       \mu=\mu_0,\enskip
	\nu =0,
\\[.5\jot]
       \pi&=2\,\bar\beta,\enskip
       \rho=\rho_0,\enskip
       \tau=2\,\beta.
\end{align*}
	For \abeltwoRthree\ the residual group is  the 2-dimensional group generated by rotations, boosts,
	$\mathcal Y$, $\mathcal Z$ and  $\mathcal T$.
	The independent spin coefficients are
	$\{\beta,\, \gamma_1,\, \epsilon_1,\,  \mu_0,\, \rho_0\}$,
	with
\begin{align*}
       \alpha&=\bar\beta,\enskip
       \gamma=i\gamma_1,\enskip
       \epsilon=i\epsilon_1,\enskip
       \kappa=0,\enskip
       \lambda=0,\enskip
       \mu=\mu_0,
\\[.5\jot]
       \nu&=0,\enskip
       \pi=2\,\bar\beta,\enskip
       \rho=\rho_0,\enskip
       \sigma=0,\enskip
       \tau=2\,\beta.
\end{align*}
\medskip
\noindent
{\bf  \ref{IIdAbelL1}--\ref{IIdAbelL2}. \abeltwoLone, \abeltwoLtwo.\ } 
	If the derived algebra is a 2-dimensional Lorentzian subspace, 
	then we may take $\lieg' = \langle\, K, L \,\rangle $. We solve the equations
\begin{equation*}
	[\,K, L\,] =0 , \quad [x,y] \wedge K \wedge L = 0 \text{ for all $x, y \in \mathfrak g$}
\end{equation*}
	for the spin coefficients 
	$\{\alpha,\, \epsilon_0,\, \gamma_0,\,  \lambda,\, \mu,\, \pi,\, \rho,\, \sigma\}$ 
	to find
\begin{equation*}
	\alpha = \bar\beta, \enskip
	\epsilon = i\, \epsilon_1,\enskip
	\gamma=i\,\gamma_1,\enskip
	\lambda=0, \enskip
	\mu=2\,i\,\gamma_1,\enskip
	\pi=-\bar\tau,	\enskip
	\rho=2\,i\,\epsilon_1,\enskip 
	\sigma=0.
\end{equation*}
	The structure equations are then 
\begin{equation}
\begin{alignedat}{2}
	[\, K,\, L\,] & = 0,
&\quad
	[\, K,\, M\,] & = -(2\,\beta +\tau)\, K-\kappa\, L,
\\[.5\jot]
	[\, L,\, M\,] & = \bar\nu\, K+(2\,\beta -\tau)\, L,
&\quad
	[\, M,\, \barM\,] & = 4\,i\,\gamma_1\, K+4\,i\,\epsilon_1\, L.
\end{alignedat}
\DGlabel{NPabel2L1}
\end{equation}
	The non-zero components of the Jacobi tensor are 
\begin{equation*}
	J^1_{134} = \kappa\, \nu - \bar \kappa\, \bar\nu, \quad     
	J^1_{234} = -4\,(\beta\,\nu - \bar \beta\, \bar\nu), \quad
	J^2_{134} = -4\,(\beta\,\bar \kappa - \bar\beta\, \kappa).
\end{equation*}
	The sub-group of the Lorentz group which stabilizes $\lieg'$ 
	is the 2-dimensional group generated by $KL$ boosts, $M\barM$ rotations, and  $\{\mathcal Y,\ \mathcal Z,\ \mathcal T \}$.

	The Jacobi identities imply 
	(after a  rotation in the $M\barM$ plane)  that 
	{\it all} the complex spin coefficients $ \beta, \kappa, \nu$  may be taken to be real, 
	that is, 
	$\beta = \beta_0$, 
	$\kappa = \kappa_0$,
	$\nu = \nu_0$.
	Here are the details. 
	If all these complex numbers are zero, then there is nothing to prove so 
	we assume that one of them, say $\kappa$, is non-zero. 
	Rotate the null tetrad to make $\kappa$ real. Then $J^1_{134} =  0$ implies that 
	$\nu$ is real while $J^2_{134} = 0 $  implies $\beta$ is real.
	If we take $\nu$ non-zero and real, then $J^1_{134} =0 $ and $J^1_{234} =0$ 
	imply $\kappa$ and $\beta$ are real.
	If we take $\beta$ non-zero and real, then $J^1_{234}=0$ and $J^2_{134} =0$ 
	force $\nu$  and $\kappa$ to be real.

	In summary, if $\lieg'$ is Lorentzian, then we have just 2 cases. If
	$\kappa \bar \kappa + \nu \bar \nu + \beta \bar \beta\neq 0 $,    
	we may take these coefficients to be real and arrive at  \summaryref{IIdAbelL1}.
	The generators of the residual group are  the $KL$ boosts, $\mathcal R$,
	$\mathcal T$,  $\mathcal Y$,   $\mathcal Z$.   The free spin coefficients are
	$\{\beta_0,\, \gamma_1,\, \epsilon_1,\, \kappa_0,\, \nu_0,\, \tau\, \}$, with
\begin{align*}
       \alpha&=\beta_0,\enskip
       \beta=\beta_0,\enskip
       \gamma=i\,\gamma_1,\enskip
       \epsilon=i\epsilon_1,\enskip
       \kappa=\kappa_0,\enskip
       \lambda=0,\enskip
       \mu=2\,i\,\gamma_1,
\\[.5\jot]
       \nu&=\nu_0,\enskip
       \pi=-\bar\tau,\enskip
       \rho=2\,i\epsilon_1,\enskip
       \sigma=0,\enskip
       \tau=\tau.
\end{align*}
	If $\kappa_0 = \nu_0 = \beta_0 =0$, the structure equations are  \summaryref{IIdAbelL2} and the residual group is generated by
	boosts, rotations,  $\mathcal T$, $\mathcal Y$, and $\mathcal Z$.
	
\par
\medskip
\noindent
{\bf  \ref{IIAbelN1}--\ref{IIAbelN5}. \abeltwoNone, \abeltwoNtwo, \dots, \abeltwoNfive.\ }
	If the derived algebra  $\lieg'$  is a 2-dimensional degenerate subspace, 
	then we may take $\mathfrak g' = \langle\, K, M + \barM\, \rangle $. 
	We solve the equations 
\begin{equation*}
	[\,K, M + \barM\,] =0 , \quad [\,x,y\,] \wedge K \wedge (M + \barM) = 0 \text{\quad for all $x, y \in \mathfrak g$}
\end{equation*}
	(with Lie brackets given by \eqref{masterNP}) for $\{\alpha,\, \beta_0,\, \epsilon_0,\, \kappa,\, \lambda,\, \pi,\,  \rho,\, \sigma\}$ to deduce that
\begin{equation}
	\alpha= i\, \beta_1 + \dfrac12\,\tau_0 -i\,\tau_1,\enskip
	\beta= i\,\beta_1 + \dfrac12\,\tau_0,\enskip
	\epsilon =i\,\epsilon_1,\enskip
	\kappa=0,\enskip
	\lambda= \gamma -\bar\gamma + \bar\mu,\enskip
	\pi=\tau,\enskip
	\rho=0,\enskip
	\sigma=2\,i\,\epsilon_1.
\DGlabel{abel2Na}
\end{equation}
	For our subsequent analysis, we note that $\alpha + \bar \beta = \bar\tau$.
	The structure equations are then
\begin{equation}
\begin{alignedat}{1}
	[\, K,\, L\,] & = -(\gamma +\bar\gamma)\, K+2\,\tau_0( M+  \barM), 
\quad
	[\, K,\, M\,]  = -2\,i\,\tau_1\, K+2\,i\,\epsilon_1 (M+  \barM),
\\[.5\jot]
	[\, L,\, M\,]  &= \bar\nu\, K+(\gamma-\bar\gamma -\mu)\, (M+  \barM),
\quad
	[\, M,\, \barM\,]  = (\mu - \bar\mu)\, K 
	-i\,(2\,\beta_1-\tau_1)\, (M + \barM),
\end{alignedat}
\DGlabel{NPabel2N1}
\end{equation}
	and the independent components of the Jacobi tensor  are 
\begin{align*}
	J_{123}^4 &= -2\,i \left( \tau_1\,\tau_0-2\,\gamma_0\epsilon_1+2\,\tau_0\,\beta_1-2\,\epsilon_1\,\mu_0 \right), \quad
	J_{123}^1 = -4\,i \left( \epsilon_1\,\nu_0-\tau_0\mu_1 \right),\quad 
\\[.5\jot] 
\quad 
	J_{234}^1 & = -2\,i\bigl((\tau_1 + 2\beta_1)\nu_0  - 2(\gamma_0 + \mu_0) \mu_1\bigr).
\end{align*}
\indent To fully analyze the Jacobi identities we  first simplify the structure equations 
	using the complex null rotation \eqref{Null}.
	The spin coefficients  $\epsilon = i\,\epsilon_1$, $\kappa =0$,  $\rho =0$, and $\sigma = 2\,i \,\epsilon_1$,
	are unchanged by this transformation, while $\tau' = \tau + 2\,i \varphi\,\epsilon_1$.
	This transformation leads us to consider the two possibilities, $\epsilon_1 \neq 0$ and 
	$\epsilon _1 = 0$. 
	
	If $\epsilon_1 \neq 0$, we can use the null rotation to set $\tau = 0$ in which case 
	the Jacobi identity $J^1_{123}=0$ implies $\nu_0 = 0$ while $J^4_{123}=0$
	gives $\gamma_0 = - \mu_0$. 
	The structure equations \eqref{NPabel2N1} become \summaryref{IIAbelN1} (\abeltwoNone ).
	The residual group is generated by the $KL$ boost, $\mathcal R$, $\mathcal Y$, $\mathcal T$.
	The free spin coefficients for \abeltwoNone\ are
	$\{\, \beta_1,\,  \gamma_1,\, \epsilon_1,\, \mu,\, \nu_1\,  \}$, with
\begin{align*}
       \alpha&=i\,\beta_1,\enskip
       \beta=i\,\beta_1,\enskip
       \gamma=-\mu_0+i\gamma_1,\enskip
       \epsilon=i\,\epsilon_1,\enskip
       \kappa=0,\enskip
       \lambda=-i\,\mu_1+\mu_0+2\,i\,\gamma_1,\enskip
       \nu=i\,\nu_1
\\[.5\jot]
       \pi&=0,\enskip
       \rho=0,\enskip
       \sigma=2\,i\,\epsilon_1,\enskip
       \tau=0.
\end{align*}
	If $\epsilon_1 = 0$ then, on account of \eqref{abel2Na}, 
	we have $ \epsilon = \kappa  = \rho = \sigma  = 0$ and the structure equations \eqref{NPabel2N1}
	simplify to
\begin{equation}
\begin{alignedat}{1}
	[\, K,\, L\,] & 
	=  -(\gamma +\bar\gamma)\, K+2\,\tau_0\, M+2\,\tau_0\, \barM, \quad [\, K,\, M\,]  = -2\,i\,\tau_1 K,
\\[1\jot]
	[\, L,\, M\,] & = \bar\nu\, K+(\gamma-\bar\gamma -\mu)\, (M+  \barM),
\quad
	[\, M,\, \barM\,]  = (\mu-\bar\mu)\, K -i\,(2\,\beta_1 -\tau_1)\,( M + \barM).
\end{alignedat}
\DGlabel{NPabel2N2}
\end{equation}
	The Jacobi tensors simplify to 
\begin{equation*}
	J^4_{123} =  -2\,i\tau_0(\tau_1 + 2\beta_1) \quad \text{and} \quad J^1_{123} = -4\,i\tau_0\mu_1.
\end{equation*}
	
	The spin coefficients $\alpha,\, \beta,\,  \pi,\, \tau$  are all invariant under \eqref{Null} 
	while, again by \eqref{abel2Na},
\begin{equation}
	\gamma' = \gamma  + \bar\varphi\,\alpha + \varphi\,(\tau + \beta)  
	=  \gamma  + \bar\varphi\,(\bar\tau - \bar \beta) + \varphi\,(\tau + \beta)
\quad\text{and}\quad
	\mu' = \mu + 2\varphi\,\beta + \bar\varphi \,\pi   = \mu + 2\varphi\,\beta + \bar\varphi \,\tau .
\DGlabel{spintrans1}  
\end{equation}
	We now consider   two sub-cases 
	-- {\bf I:} $\epsilon_1 = 0$, $\tau_0 \neq 0$, and  {\bf II:} $\epsilon_1 = 0$, $\tau_0 = 0 $.

\noindent
{\bf Sub-case I:}
	If  $\epsilon_1 = 0$ and  $\tau_0 \neq 0$, then the Jacobi identities $J{}_{123}^1 = 0$ and 
	$J^4_{123} = 0$ imply $\mu_1 =0$ and $\tau_1  =-2\beta_1$. Then
	$\tau  = 2\,\bar \beta$ and the structure equations  \eqref{NPabel2N2} reduce to
\begin{equation}
\begin{alignedat}{1}
	[\, K,\, L\,] & = -(\gamma + \bar\gamma) K+2\,\tau_0\, M+2\,\tau_0 \barM,
\quad
	[\, K,\, M\,]  = 4\,i\beta_1 K,
\\[1\jot]
[\, L,\, M\,] & = \bar\nu\, K +(\gamma-\bar\gamma- \mu_0)\, (M + \barM),
\quad
	[\, M,\, \barM\,]  = -4\,i\,\beta_1 (M+ \barM).
\end{alignedat}
\DGlabel{NPabel2N2a}
\end{equation}
	The spin coefficient transformations  \eqref{spintrans1} become 
\begin{equation*}
	\gamma' = 
	 \gamma  + \bar\varphi\,(\bar \tau - \frac12\tau) + \varphi\,(\tau +\frac12\bar\tau)
	=  \gamma  + (\bar\varphi\,\bar \tau + \varphi\,\tau) - \frac12(\bar\varphi \tau - \varphi\bar\tau)
\quad\text{and}\quad
	\mu' = \mu + \varphi\,\bar \tau + \bar\varphi \,\tau
\end{equation*}
	or, with $\varphi = u +  i\,v$,
\begin{equation}
	\gamma_0' = \gamma_0  +  2\, \tau_0\,u - 2\, \tau_1\, v , \quad
	\gamma_1' = \gamma_1  -   \tau_1\, u + \tau_0\, v   \quad
	\mu_0' =  \mu_0 + 2\,\tau_0\, u + 2\, \tau_1\,v.
\DGlabel{spintrans2}  
\end{equation}
	In this sub-case we are assuming that $\tau_0 \neq 0$ and, therefore, 
	one can choose $\varphi$ so that $\gamma_1' = \mu_0' = 0$. 
	The structure equations \eqref{NPabel2N2a}  then simplify to \summaryref{IIAbelN2} (\abeltwoNtwo).
	The residual group is generated by $KL$ boosts, $\mathcal R$, $\mathcal Y$, $\mathcal T$.
	The free spin coefficients for \abeltwoNtwo\ are
	$\{\,\beta_1,\, \gamma_0,\, \nu,\, \tau_0\, \}$, with
\begin{align*}
       \alpha&= \frac12\tau_0 + 3\,i\,\beta_1,\enskip
       \beta= \frac12\tau_0+i\,\beta_1,\enskip
       \gamma=\gamma_0,\enskip
       \epsilon=0,\enskip
       \kappa=0,\enskip
       \lambda=0,\enskip
       \mu=0,
\\[.5\jot]
       \pi &=\tau_0-2\,i\,\beta_1,\enskip
       \rho=0,\enskip
       \sigma=0,\enskip
       \tau=\tau_0-2\,i\,\beta_1.
\end{align*}

\noindent
{\bf Sub-case II:}
	We now turn to the  sub-case where $\epsilon_1 = 0$ and $\tau_0 = 0$. From (\ref{abel2Na}),  this implies 
	$\beta_0 = 0$. The structure equations  
	\eqref{NPabel2N2} simplify to 
\begin{equation}
\begin{alignedat}{1}
	[\, K,\, L\,] & = -(\gamma +\bar\gamma)\, K, 
\quad
	[\, K,\, M\,]  = -2\,i\,\tau_1 K,
\\[1\jot]
	[\, L,\, M\,] & = \bar\nu\, K+(\gamma-\bar\gamma - \mu)\, (M+  \barM),
\quad
	[\, M,\, \barM\,]  = (\mu-\bar\mu)\, K - i\,(2\,\beta_1 -\tau_1)\, (M  + \barM),
\end{alignedat}
\DGlabel{NPabel2N2b}
\end{equation}
	and the spin coefficient
	transformations \eqref{spintrans2} become
\begin{equation}
	\gamma_0'  = \gamma_0 -2\,\tau_1\,v,\quad
	\gamma_1' = \gamma_1  +2\,\beta_1\,u, \quad
	\mu_0' = \mu_0  + (\tau_1-2\beta_1)\,v,\quad
	\mu_1' = \mu_1 + (\tau_1 +2\beta_1)\, u.
\DGlabel{spintrans3}
\end{equation}
	Note that $\gamma_0' + \mu_0' = \gamma_0 + \mu_0  -(\tau_1 + 2\beta_1) \,v$. 
	Consequently, we case split once more according  to 
	$\tau_1 + 2\beta_1 \neq 0$ or  $\tau_1 + 2\beta_1 = 0$.
\par
\noindent
{\bf Sub-case II${}'$:} Our starting point is \eqref{NPabel2N2b}. If $\tau_1 + 2\,\beta_1 \neq 0$, 
	we can use the null rotation  \eqref{spintrans3} to transform $\mu_1 = 0$ 
	and $\gamma_0= - \mu_0$. Both $J^1_{123}$ and $J^4_{123}$ vanish and the remaining Jacobi identity 
	$J^1_{234} =0$ gives $\nu_0 = 0$.  The structure 
	equations \eqref{NPabel2N2b} reduce to \summaryref{IIAbelN3} (\abeltwoNthree). The residual group for \abeltwoNthree\
	is generated by $KL$ boosts, $\mathcal R$, $\mathcal Y$, $\mathcal T$.  The  free spin coefficients are
	$\{\beta_1, \gamma_1, \mu_0, \nu_1, \tau_1\}$, with
\begin{align*}
       \alpha&=-i\tau_1+i\,\beta_1,\enskip
       \beta=i\,\beta_1,\enskip
	\gamma = -\mu_0 + i\,\gamma_1, \enskip
       \epsilon=0,\enskip
       \kappa=0,\enskip
       \lambda=\mu_0+2\,i\,\gamma_1,\enskip       
\\[.5\jot]
	\mu&=\mu_0,\enskip
       \nu =i\,\nu_1,\enskip
       \pi=i\,\tau_1,\enskip
       \rho=0,\enskip
       \sigma=0,\enskip
       \tau=i\,\tau_1.
\end{align*}

\noindent
{\bf Sub-case II${}''$:}
	Finally, if  $\tau_1 = -2\,\beta_1$, then  \eqref{NPabel2N2b} becomes
\begin{equation}
\begin{alignedat}{1}
	[\, K,\, L\,] & = -(\gamma +\bar\gamma)\, K,
\quad
	[\, K,\, M\,]  = 4\,i\,\beta_1 K,
\\[1\jot]
	[\, L,\, M\,] & = \bar\nu\, K+ (\gamma-\bar\gamma -\mu)\, (M +  \barM),
\quad
	[\, M,\, \barM \,]  = (\mu-\bar\mu)\, K-4\,i\,\beta_1 (M +\barM).
\end{alignedat}
\DGlabel{NPabel2N2c}
\end{equation}

	The Jacobi identity $J_{234}^1 = 0$ implies either $\gamma_0 =-\mu_0$
	(\abeltwoNfour\ and \summaryref{IIAbelN4}) or  $\mu_1 =0$ (\abeltwoNfive\ and \summaryref{IIAbelN5}). 
	In either situation, if $\beta_1 = 0$ then the isometry algebra becomes 6-dimensional. 
	With $\beta_1 \neq 0$, the null rotation \eqref{spintrans3} can be used to set 
	$\gamma_1 =0$ and $\mu_0 =0$. 
	For both of these cases the residual group is generated by the $KL$ boosts, $\mathcal R$, $\mathcal Y$, $\mathcal T$. The free spin coefficients for  	\abeltwoNfour\ are 
	$\{\,\beta_1, \mu_1, \nu \,\}$, with
\begin{align*}
       \alpha&=3\,i\,\beta_1,\enskip
       \beta=i\,\beta_1,\enskip
       \gamma=0,\enskip
       \epsilon=0,\enskip
       \kappa=0,\enskip
       \lambda=-i\,\mu_1,\enskip
       \mu=i\,\mu_1,
\\[.5\jot]
       \pi&=-2\,i\,\beta_1,\enskip
       \rho=0,\enskip
       \sigma=0,\enskip
       \tau=-2\,i\,\beta_1.
\end{align*}
	The free spin coefficients for \abeltwoNfive\ are
	$\{\beta_1, \gamma_0\}$, with
\begin{align*}
       \alpha& =3\,i\,\beta_1,\enskip
       \beta=i\,\beta_1,\enskip
       \gamma=\gamma_0,\enskip
       \epsilon=0,\enskip
       \kappa=0,\enskip
       \lambda=0,\enskip
       \mu=0,
\\[.5\jot]
       \pi& =-2\,i\,\beta_1,\enskip
       \rho=0,\enskip
       \sigma=0,\enskip
       \tau=-2\,i\,\beta_1.
\end{align*}
	
	Finally, we remark that the spacetime Lie algebras \abeltwoNtwo, \abeltwoNthree, \abeltwoNfour, \abeltwoNfive\  
	are all of Petrov type II.

%
%

\subsection{Spacetime Groups with  1-Dimensional Derived Algebra}\label{OneDDerived}
	If the derived algebra $\lieg'$ is 1-dimensional, then it is either a 
	time-like, space-like, or null subspace of $\lieg$. In each case we find that 
	once the Jacobi identities are imposed, the isometry algebra of spacetime 
	has dimension greater that  4.  Thus  there are no {\it spacetime groups with 1-dimensional 
	derived algebras}.  The details follow.

\par\noindent
{\bf abel1T.} 
	If the derived algebra is 1-dimensional and time-like,
	we may rotate the null tetrad so that 
	$\lieg' = \langle K + L \rangle$. The annihilating forms for $\lieg'$ are 
	$\langle \Theta_K -\Theta_L, \Theta_{\vphantom{\barMs}M}, \Theta_{\barMs} \rangle$.
	We solve the equations $d(\Theta_K -\Theta_L) = d\,\Theta_{\vphantom{\barMs}M} = d\,\Theta_{\barMs} =0$
	(using the structure equations \eqref{masterNP}) for 
	$\{\alpha,\, \beta,\, \epsilon,\, \kappa,\, \lambda,\, \mu,\, \pi,\, \rho,\, \sigma\}$
	to yield
\begin{equation*}
	\alpha= \frac12\nu + \frac12 \bar\tau,\enskip
	\beta= \frac12\bar\nu + \frac12 \tau,\enskip
	\epsilon=\gamma,\enskip
	\kappa=2\,\tau+\bar\nu,\enskip
	\lambda=0,\enskip
	\mu=\gamma-\bar\gamma,\enskip
	\pi=-\bar\tau,\enskip
	\rho=\gamma-\bar\gamma,\enskip
	\sigma=0.
\end{equation*}
The structure equations become
\begin{equation}
\begin{alignedat}{2}
	[\, K,\, L\,]  &= -(\gamma +\bar\gamma)\,(K+ L),
&\quad
	[\, K,\, M\,]  &= -(2\,\tau +\bar\nu)\, (K+L),
\\[.5\jot]
	[\, L,\, M\,]  &= \bar\nu\, (K+ L),
&\quad
	[\, M,\, \barM\,] & = 2\,(\gamma-\,\bar\gamma)\, (K + L),
\end{alignedat}
\DGlabel{NPabel1T1}
\end{equation}
	and the Jacobi identities reduce to $\tau_0\nu_1 + \,\tau_1\,\nu_0 - 2\,\gamma_0\gamma_1 = 0$.

\par
\noindent
{\bf abel1S.} 
	If the derived algebra is 1-dimensional and space-like,
	we may rotate the null tetrad so that 
	$\lieg' = \langle K - L \rangle$, in which case
	 $d(\Theta_K + \Theta_L) = d\,\Theta_{\vphantom{\barMs}M} = d\,\Theta_{\barMs} =0$.   
	Using the structure equations \eqref{masterNP}, these equations are solved for 
	$\{\alpha,\, \beta,\, \epsilon,\, \kappa,\, \lambda,\, \mu,\, \pi,\, \rho,\, \sigma\}$
	to yield
\begin{equation*}
	\alpha=-\frac12\nu + \frac12\bar\tau,\enskip
	\beta= - \frac12 \bar\nu + \frac12\tau,\enskip
	\epsilon=-\gamma,\enskip
	\kappa=-2\,\tau+\bar\nu,\enskip
	\lambda=0,\enskip
	\mu=\gamma-\bar\gamma,\enskip
	\pi=-\bar\tau,\enskip
	\rho=-\gamma+\bar\gamma,\enskip
	\sigma=0.
\end{equation*}
	The structure equations are then
\begin{equation}
\begin{alignedat}{2}
[\, K,\, L\,]  &= -(\gamma+\bar\gamma)\,( K - L),
&\quad
[\, K,\, M\,]  &= (-2\,\tau+\bar\nu)\, (K- L),
\\[.5\jot]
[\, L,\, M\,] & = \bar\nu\,( K- L),
&\quad
[\, M,\, \barM]  &= 2(\gamma-\bar\gamma)\, (K- L),
\end{alignedat}
\DGlabel{NPabel1S1}
\end{equation}
	and the Jacobi identities reduce again to $\nu_1\,\tau_0+\tau_1\,\nu_0-2\,\gamma_0\gamma_1  = 0$.
\par
\noindent
{\bf abel1N.}
	If the derived algebra is 1-dimensional and null,
	we may rotate the null tetrad so that 
	$\lieg' = \langle\, K\,\rangle$ 
	and  therefore $d\,\Theta_L = d\,\Theta_{\vphantom{\barMs}M} = d\,\Theta_{\barMs} =0$.   
	Using the structure equations \eqref{masterNP}, 
	these equations are solved for 
	$\{\alpha,\, \beta,\, \epsilon,\, \kappa,\, 
	\lambda,\, \mu,\, \pi,\, \rho,\, \sigma\}$
	to yield
\begin{equation*}
	\alpha= \frac12\,\bar\tau,\enskip
	\beta= \frac12\tau,\enskip
	\epsilon=0,\enskip
	\kappa=0,\enskip
	\lambda=0,\enskip
	\mu=\gamma-\bar\gamma,	\enskip
	\pi=-\bar\tau,\enskip
	\rho=0,\enskip
	\sigma=0.
\end{equation*}
The structure equations are 
\begin{equation}
[\, K,\, L\,]  = -(\gamma +\bar\gamma)\, K,
\quad
[\, K,\, M\,]  = -2\,\tau\, K,
\quad
[\, L,\, M\,]  = \bar\nu\, K,
\quad
[\, M,\, \barM\,]  = 2(\gamma- \bar\gamma)\, K
\DGlabel{NPabel1N1}
\end{equation}
	and the Jacobi identities reduce again to $\nu_1\,\tau_0+\tau_1\,\nu_0-2\,\gamma_0\gamma_1 = 0$.

	Using the results of Appendix A, it  follows that the isometry algebras of {\bf abel1T} and  {\bf abel1S} have 
	dimension 5 and the isometry algebra of {\bf abel1N} has dimension  $\geq 7$.

\section {The Equivalence Problem For Spacetime Lie Algebras}\label{Equivalence}
	The equivalence problem for spacetime Lie algebras was formulated in Section  \ref{EquivalenceMethod}. 
	The details are provided here.  In particular, we show  how the residual group can be reduced to a finite discrete group in each case by suitable normalization (``gauge 		fixing'') conditions.

	The cases \abelthreeRZ,  \abelthreeLZ, \LevitwoT ,  \LevitwoS , and {\bf simpCN} require special attention.
	For each of these  cases we identify a  
	symmetric tensor $Q$, defined on a 3-dimensional vector space $W$,
	and reduce the residual group by  transforming $Q$ to normal form.
	For the cases \abelthreeRZ\ and \abelthreeLZ,  $W$ is the derived algebra $\lieg'$  
	of the spacetime Lie algebra, $Q$ is obtained from the linear transformation 
	$\text{ad}(v)$ for any vector $v$ complementary to $W$ (see the remark after eq. (\ref{defzeta})). 
	For the cases \LevitwoT,  \LevitwoS, and {\bf simpCN} the symmetric tensor $Q$ is the Killing form 
	for the quotient algebra $W= \lieg/\liez$, where $\liez$ is the center of $\lieg$.

	For the cases $\abelthreeRZ$ and   $\LevitwoT$, the residual group includes $\text{O}(\tilde\eta)=\text{O}(3)$, where $\tilde \eta$ is the induced inner product on $W$.  We use this group to align our basis for $W$  with the principal axes of the symmetric tensor $Q$.
	For the cases $\abelthreeLZ$ and   $\LevitwoS$,  the residual group includes $\text{O}(\tilde\eta)=\text{O}(2, 1)$. 
	The normal forms for a symmetric tensor $Q$ on a 3 dimensional vector space $W$, 
	with  Lorentz inner product $\tilde \eta$, with respect to the group $\text{O}(\tilde\eta)$ are found 
	in \cite{Hall-Morgan-Perjes:1987a} (page 1141). To describe these normal  forms, let 
	$\{\bfk, \bfell, \bfm\}$ be a real triad for $W$ such that the vectors $\bfk$ and  $\bfell$ 
	are null vectors; $\bfm$ is orthogonal to $\bfk$, $\bfell$; $\tilde\eta(\bfk, \bfell) = -1$; 
	and $\tilde\eta(\bfm, \bfm) = \dfrac12$. (The factor of $\dfrac12$  will 
	help simplify some of our subsequent formulas).  In terms of the dual basis 
	$\{\bfalpha, \bfbeta, \bfgamma \}$, the normal forms are as follows: 
\begin{alignat*}{2}
&\text{\bf I} &  \quad Q &= -2\,s_1\,\bfalpha \odot \bfbeta  +s_2\, (\bfalpha \odot \bfalpha + \bfbeta \odot \bfbeta) 
	+ \frac12\,s_3\, \bfgamma \odot \bfgamma 
\\
&\text{\bf II} &  \quad  Q &= -2\,s_1\,\bfalpha \odot \bfbeta  +s_2\, (\bfalpha \odot \bfalpha - \bfbeta \odot \bfbeta) 
	+\frac12\,s_3\, \bfgamma \odot \bfgamma 
\\
&\text{\bf III} & \quad Q &= -2\,s_1\,\bfalpha \odot \bfbeta  +s_2\,\bfbeta \odot \bfbeta  + \frac12\,s_3\, \bfgamma \odot \bfgamma 
\\
&\text{\bf IV} &\quad Q &= -2\,s_1\bfalpha\, \odot \bfbeta  + 2s_2\,\bfbeta \odot \bfgamma  + 
	\frac12\,s_1\, \bfgamma \odot \bfgamma .
\end{alignat*}
	Here $\bfalpha \odot \bfbeta= \dfrac12 (\bfalpha \otimes \bfbeta + \bfbeta\otimes \bfalpha)$. 
	As matrices, these normal forms are
\begin{equation}
I = \begin{bmatrix}
	s_2 & -s_1  & 0
\\
	-s_1  & s_2 & 0
\\	0 & 0  &\dfrac12  s_3
\end{bmatrix}
\quad
II = \begin{bmatrix}
	s_2 & -s_1  & 0
\\
	-s_1  & -s_2 & 0
\\	0 & 0  & \dfrac12s_3
\end{bmatrix}
\quad
III = \begin{bmatrix}
	0 & -s_1  & 0
\\
	-s_1  & s_2 & 0
\\	0 & 0  & \dfrac12s_3
\end{bmatrix}
\quad
IV = \begin{bmatrix}
	0 & -s_1  & 0
\\
	-s_1  & 0 & s_2
\\	 0 &  s_2 & \dfrac12 s_1
\end{bmatrix}.
\DGlabel{Segre}
\end{equation}
	In Cases II, III, IV, one assumes $s_2 \neq 0$; otherwise these matrices take the form of Case I. 

	Having used $\text{O}(\tilde \eta)$ to transform a given symmetric tensor into one of the four normal forms above, we must characterize the group which preserves the 		normal form.  	This is the remaining residual group. In each of the four cases, write $Q = \sum s_i Q_i$ and  let $\mathcal Q = \text{span}\{Q_i \}$. 
	Then the residual group consists of those 
	Lorentz transformations $\tilde \varphi \in \text{O}(\tilde\eta)$ for which $\tilde \varphi^*\mathcal Q = \mathcal Q$. 
	Each such $\tilde\varphi$ can be identified with a 4-dimensional 
	Lorentz transformation acting on the spacetime Lie algebra $\lieg$.
	
	The normal forms can be characterized by solutions 
	to the eigenvector problem $Q(X, \cdot \,) = \xi\, \tilde \eta(X, \cdot \,)$ where  $X \in W$ and 
\begin{equation*} 
	\tilde\eta = -2\, \alpha \odot \beta + \frac{1}{2}\, \gamma \otimes\gamma. 
\end{equation*}
	In each of the  cases I-IV, the (real) eigenspaces are independent of the parameters $s_i$ and 
	so these eigenspaces are   preserved by the residual group.
	In Cases III and IV there is a repeated eigenvalue with a single  eigenvector which is null. 
	Let $\mathcal K$ be the generalized eigenspace for the repeated eigenvalue;	
	$\mathcal K$ is a  2-dimensional subspace which is  preserved by the 
	residual group.
	The details in each case  will be important to us and are as follows. 
	
\smallskip
\noindent
{\bf Case I.} In this case $Q$ is diagonalizable with  eigenvalues/eigenvectors
\begin{equation*}
	\xi_1 = s_1 + s_2, \enskip X_1 = \bfk-\bfell; 
\qquad  	 \xi_2 = s_1 - s_2, \enskip X_2 = \bfk + \bfell;
 \qquad    \xi_3 = s_3, \enskip X_3 = \bfm.
\end{equation*}
	If the eigenvalues are distinct, the residual group is discrete.
		If $\xi_1 = \xi_2$, the residual group  includes the group of boosts in the $\bfk\bfell$ plane.
	If $\xi_2 = \xi_3$, the residual group includes the group of boosts in the $\bfk+\bfell , \bfm$ plane.
	If $\xi_1 = \xi_3$, the residual group includes the group of boosts in the $\bfk-\bfell , \bfm$ plane.

\noindent
{\bf Case II.} In this case there  is a complex conjugate eigenvalue pair. The eigenvalues/eigenvectors are 
\begin{equation*}
	\xi_1 = s_1 + i\,s_2, \enskip X_1 = \bfk + i\, \bfell; 
\quad  	 \xi_2 = s_1 - i\,s_2, \enskip X_2 = \bfk -i\,\bfell;
 \quad    \xi_3 = s_3, \enskip X_3 = \bfm.
\end{equation*}
The real null vectors $\bfk$ and $\bfell$ associated to this normal form can be recovered from the complex eigenvectors once the real and imaginary parts of these eigenvectors are chosen (by the appropriate complex scaling) to have the same inner products as $\bfk$ and $\pm\bfell$. Since $s_2 \neq 0$, the residual group must preserve the matrix $Q_2$ and is therefore a discrete group.

\noindent
{\bf Case III.}
	The eigenvalues/eigenvectors are 
\begin{equation*}
	\xi_1 = \xi_2=s_1,  \enskip X_1 = \bfk, \enskip \mathcal K =  \langle \bfk, \bfell \rangle; 
\quad	\xi_3 = s_3, \enskip  X_3 = \bfm. 
\end{equation*}
	If the eigenvalues are distinct, the residual group includes $\bfk\bfell$ boosts. If the eigenvalues
	coincide, $\mathcal K = W$ and the residual group also includes null rotations about the $\bfk$ axis.

\noindent
{\bf Case IV.} The eigenvalues/eigenvectors are 
\begin{equation*}
	\xi_1 = \xi_2=  \xi_3 = s_1, \enskip X_1 = \bfk, \enskip \mathcal K =\langle \bfk,\bfm \rangle.
\end{equation*}
	The basis adapted to this normal form can be obtained from the eigenspaces as follows. Set $\bfk = X_1$. Fix   any  vector $\bfm$ complementary to $\bfk$ in $\bfm \in\mathcal K$ and any vector $\bfell$ in the complement of $\mathcal K$. Require these three vectors  (i) have the inner products of $\bfk, \bfell, \bfm$, and (ii) satisfy $Q(\bfell, \bfell)=0$.	Note that for this normal form $Q =  s_1\eta  + 2s_2\bfbeta \odot \bfgamma$.  The residual group must therefore preserve the tensor 
	$\bfbeta \odot \bfgamma$ 
	and therefore is, aside from discrete transformations, just the group of $\bfk\bfell$
	boosts.

The remaining special case is {\bf simpCN}.  Here we need a new normal form classification, which we provide in Section 6.3.15.
\subsection{Spacetime Groups with  Heisenberg Derived Algebra}

{\bf \ref{HeisRS}.  \heisRS.\ } The subgroup of the Lorentz group preserving the flag  
	$\langle K - L \rangle \subset \langle \, K - L,\, M,\, \barM\, \rangle$ includes 
	the group \eqref{Rotation} of rotations in the $M\barM$ plane (see Table \ref{LeviTable}). The independent spin coefficients
	transform as 
\begin{equation*}
	\kappa' = \exp(i\theta)\,\kappa, \enskip
	\sigma' = \exp(2\,i\,\theta)\,\sigma, \enskip
	\epsilon'_1 = \epsilon_1,\enskip
	\gamma'_1 = \gamma_1,\enskip
	\mu_0' = \mu_0.
\end{equation*}
	As long  as $\kappa$ or $\sigma$ is nonzero these rotations are gauge-fixed 
	by the  condition $\kappa = \kappa_0 >0$ or $\sigma = \sigma_0 >0$. The residual group is
	then a discrete group. If $\kappa = \sigma$ = 0
	then the isometry group is 5 dimensionsal. 

\par
\smallskip
\noindent
{\bf \ref{HeisLT}.  \heisLT.\ }
	The subgroup of the Lorentz group preserving the flag 
	$\langle K + L \rangle \subset \langle \, K + L,\, M,\, \barM\, \rangle$ includes 
	the group \eqref{Rotation} of rotations in the $M\barM$ plane (see Table \ref{LeviTable}).
	The independent  spin coefficients are  
	$\left\{ \kappa, \sigma, \epsilon_1,\gamma_1, \mu_0\right\}$ 
	and the rotations can be gauge-fixed in the same way as \heisRS, leaving a discrete residual group. 
\par
\medskip
\noindent
{\bf \ref{HeisLS}.  \heisLS.\ }
	The subgroup of the Lorentz group preserving the flag 
	$\langle M + \barM  \rangle \subset \langle \, K,\, L,\, M + \barM\, \rangle$ includes 
	the group  \eqref{Boost} of $KL$ boosts. The  independent spin coefficients transform as
\begin{equation*} 
	\epsilon_1' = \boostparm\, \epsilon_1,\enskip
	\gamma_1 =  \boostparm^{-1}\gamma_1,\enskip
	\kappa_1'= \boostparm^2 \kappa_1,\enskip
	\nu_1'=  \boostparm^{-2}\nu_1,\enskip
	\tau_1'= \tau_1,\enskip
	\beta_1' = \beta_1, \enskip
	\pi_0'=\pi_0 .
\end{equation*}
	The boosts can be gauge-fixed, leaving a discrete residual group, so long as one of 
	$\epsilon_1$, $\gamma_1$, $\kappa_1$ or $\nu_1$ is non-zero. If all these 
	spin coefficients are zero, the isometry algebra becomes  5-dimensional.
\par
\smallskip
\noindent
{\bf \ref{HeisLN}.  \heisLN.\ }
	According to Table \ref{LeviTable}, the subgroup of the Lorentz group preserving the flag 
	$\langle K \rangle  \subset \langle\, K, \,L,\,  M + \barM \,\rangle$
	includes the boosts \eqref{Boost} and null rotations
	\eqref{Null}, with $\varphi =u$ (real).
	The independent spin coefficients are  
	$\{\beta_1,\, \epsilon_1,\, \gamma_1,\, \mu_1,\, \nu,\, \tau_1 \}$. 
	Since $\kappa = \sigma =0$, $\pi = i\,\pi_1$, and $\rho = 2\,i\,\epsilon_1$,
	the spin coefficient $\tau_1$ transforms under the 
	null rotation \eqref{Null} as $\tau_1' =  \tau_1 + 2u \,\epsilon_1$.  
	Thus, if $\epsilon_1 \neq 0$, the 
	null rotation can be gauge-fixed by setting $\tau_1=0$ (see 2.\ref{HeisLN}.1).
	If  $\epsilon_1 =0$, then $\mu_1$ transforms as $\mu_1' = \mu_1 + u( \tau_1 + 2\beta_1 )$
	and the null transformation is normalized by the gauge fixing condition $\mu_1 = 0$ (see 2.\ref{HeisLN}.2).  
	The spin coefficients  $\epsilon_1$ and $\tau_1 + 2\beta_1$ cannot vanish  simultaneously
	since otherwise  the 
	derived algebra is 2-dimensional. In this way the null rotation subgroup is gauge-fixed.
	The independent spin coefficients are transformed under  boosts by
\begin{equation*}
	\epsilon_1' =  \boostparm\,\epsilon_1, \enskip
	\gamma_1' = \boostparm^{-1}\,\gamma_1, \enskip
	\mu_1'= \boostparm^{-1\,} \mu_1,\enskip
	\nu'= \boostparm^{-2} \nu\,\enskip
	\tau_1' = \tau_1,\enskip
	\beta_1'= \beta_1.
\end{equation*}
	The boosts can always be gauge-fixed since the isometry algebra jumps to 5 dimensions
	if $\epsilon_1 = \gamma_1 = \mu_1  = \nu  =0$.  Thus the residual group is reduced to a discrete 
	group.
\par
\smallskip
\noindent
{\bf \ref{HeisNS}.  \heisNS.\ } 
	The subgroup of the Lorentz group preserving the flag 
	$\langle M + \barM \rangle  \subset \langle\, K,\, M ,\, \barM \,\rangle$	
	includes the boosts \eqref{Boost} and null rotations 
	\eqref{Null} with $\varphi = iv$ (imaginary). The independent spin coefficients are 
	$\{\gamma,\, \epsilon_1,\, \mu_0,\, \nu_1,\, \tau\}$. 
	Since $\kappa = \rho =0$ and $\sigma = 2\,i\,\epsilon_1$,
	the spin coefficient $\tau_0$ transforms under the 
	null rotation as $\tau_0' =  \tau_0 - 2v\,\epsilon_1$ ($\tau_1$ is invariant).  
	Since $\epsilon_1\neq 0$ (the Heisenberg condition), the null rotation can be 
	fixed by the gauge $\tau_0 =0$ (see 2.\ref{HeisNS}.1).
	The spin coefficients are transformed under the boosts by
\begin{equation*}
	\gamma' = \boostparm^{-1}\gamma, \enskip
	\epsilon_1' = \boostparm\, \epsilon_1, \enskip
	\mu_0 ' = \boostparm^{-1}\mu_0, \enskip
	\nu_1' = \boostparm^{-2}\nu_1, \enskip
	\tau' = \tau.
\end{equation*}
	The boosts can always be gauge-fixed since $\epsilon_1 \neq 0$.  Thus the residual group is reduced to a discrete 
	group.

\subsection{Spacetime Groups with  3-Dimensional Abelian Derived Algebra}
{\bf \ref{IIIdAbelianRS}. \abelthreeRS.\ } 
	The flag here is the same as for \heisRS\ 
	so the residual group includes \eqref{Rotation} and the independent spin coefficients 
	transform as
\begin{equation*}
	\kappa' = e ^{i\,\theta} \kappa, \enskip
	\sigma' = e ^{2i\,\theta} \sigma, \enskip
	\epsilon_1' = \epsilon_1, \enskip
	\gamma_1' = \gamma_1, \enskip
	\rho_0' = \rho_0.
\end{equation*}
	The residual group can always be reduced to a discrete group by normalizing $\kappa$ or $\sigma$
	since the isometry algebra jumps to dimension 5 if $\kappa = \sigma = 0$. 
\par
\smallskip
\noindent
{\bf \ref{IIIdAbelianRZ}. \abelthreeRZ.\ } 
	With respect to the adapted basis
	$\{\,\dfrac12\,( K - L),\, \dfrac12 \,(M + \barM),\, \dfrac{i}{2}\,(M -\barM)\,\}$
	  for the derived subalgebra, the contravariant components of the symmetric tensor  $A^{ij}$ 
	for $\text{ad}(\dfrac12(K +L))$ (see \ref{defzeta}) are 
\begin{equation*}
	A = \begin{bmatrix}
	-4\,\epsilon_0&4\,\kappa_0&-4\,\kappa_1
\\  	4\,\kappa_0&2\,\rho_0+2\,\sigma_0&-2\,\sigma_1
\\	-4\,\kappa_1&-2\,\sigma_1&2\,\rho_0-2\,\sigma_0
\end{bmatrix}.
\end{equation*} 
	The residual group is ${\rm O}(3) \times \mathcal T $, with  ${\rm O}(3)$ acting on the derived subalgebra, and hence we may rotate the adapted basis 
	to transform the matrix $A$ to diagonal form.  
	This gives the gauge fixing conditions $\kappa=0$ and $\sigma_1 = 0$ (see 2.\ref{IIIdAbelianRZ}.1).

\par
        In this gauge, the eigenvalues of $A$ are $\xi_1 = -4\,\epsilon_0$, 
	$\xi_2 = 2\,\rho_0 + 2 \sigma_0$  and  $\xi_3 =  2\,\rho_0 - 2\, \sigma_0$. 
	If two of the eigenvalues coincide, then either 
	$\rho_0 = \pm\sigma_0 -2\epsilon_0$ or $\sigma_0= 0$. In either case
	the isometry algebra is 5-dimensional. 
	Accordingly, we may assume that $\sigma_0\neq 0$ and $\rho_0 \neq \pm \sigma_0 - 2 \epsilon_0$
	so that the eigenvectors correspond to distinct eigenvalues. The residual group is generated by $\mathcal T$ and the group of scalings and 
	permutations of the eigenvectors, 	generated by $\mathcal R$, $\mathcal U$,  $\mathcal V$, $\mathcal Y$, $\mathcal Z$.
	
\par
\smallskip
\noindent
{\bf \ref{IIIdAbelianLT}. \abelthreeLT.\ } 
	The flag here is the same as for \heisLT\ 
	so the residual group includes the rotations \eqref{Rotation}. The independent spin coefficients
	transform as
\begin{equation*}
	\kappa' = e ^{i\,\theta}\kappa,\enskip
	\sigma' = e ^{2i\,\theta} \sigma,\enskip
	\epsilon_0 = \epsilon_0, \enskip 
	\gamma_1= \gamma_1,\enskip
	\rho_0 = \rho_0.
\end{equation*}
	The rotation group  can always be gauge-fixed, leaving a discrete residual group, since the isometry algebra is  5-dimensional if 
	$\kappa = \sigma = 0$. 
\par
\smallskip
\noindent
{\bf \ref{IIIdAbelianLS}. \abelthreeLS.\ } 
	The flag here is the same as for \heisLS\  
	so the residual group includes \eqref{Boost}, under which the spin coefficients 
	transform as
\begin{equation*}
	\epsilon_1' =  \boostparm\, \epsilon_1, \enskip
	\kappa_1' = \boostparm^2 \kappa_1, \enskip
	\mu_1' = \boostparm^{-1}\mu_1, \enskip
	\nu_1' = \boostparm^{-2}\nu_1, \enskip
	\tau_1' = \tau_1, \enskip
	\alpha_1' = \alpha_1, \enskip
	\beta_1' = \beta_1.
\end{equation*}
	The gauge can always be fixed, leaving a discrete residual group, since the isometry algebra is  5-dimensional if
	$\epsilon_1 = \kappa_1 = \mu_1 = \nu_1 = 0$.
\par
\smallskip
\noindent
{\bf \ref{IIIdAbelianLN}. \abelthreeLN.\ }
	The subgroup of the Lorentz group fixing the flag  
	$\langle K \rangle  \subset \langle\, K,\, L,\, M + \barM \, \rangle$ 
	includes the $KL$ boosts  and the null rotations \eqref{Null} with $\varphi = u$ (real) (see Table \ref{LeviTable}).
	The independent spin coefficients are
	$ \left\{\alpha_1,\, \gamma_1,\, \epsilon_1,\,\kappa_1,\, \mu_1,\,  \nu_1,\,\tau_1 \right\}$.
	while
\begin{equation*}
	\beta =i\,\alpha_1, \enskip
	\epsilon_0 = 0, \enskip
	\lambda = -i\,(\mu_1 -2\,\gamma_1),\enskip
	\pi = i\,\tau_1,\enskip
	\sigma = i\epsilon_1, \enskip
	\rho = i\,\epsilon_1.
\end{equation*}	

	The null rotations can be gauge-fixed as follows. The spin-coefficient $\epsilon_1$ transforms as
	$\epsilon_1' = \epsilon_1 + u\kappa_1$ so that if $\kappa_1\neq 0$, we may gauge fix by setting
	$\epsilon_1 = 0$ (see 2.\ref{IIIdAbelianLN}.1).
	
	If $\kappa_1 =0$ and $\epsilon_1 \neq 0$, we have 
	$\tau_1' + 2\,\alpha_1' =  \tau_1 + 2\,\alpha_1 + 6\, u\epsilon_1$ and we gauge fix by setting 
	$\tau_1 + 2\,\alpha_1 =0$ (see 2.\ref{IIIdAbelianLN}.2). 

	If $\kappa_1 =\epsilon_1 = 0$ and $\tau_1 + 2\,\alpha_1  \neq 0$, then 
	$\gamma_1' = \gamma_1 + u(\tau_1 + 2\alpha_1)$ and the gauge  $\gamma_1 = 0$ fixes the null rotations 
	(see (2.\ref{IIIdAbelianLN}.3)).

	Finally, if $\kappa_1 =\epsilon_1 = 0$,  $\tau_1 = -2\alpha_1$, and $\gamma_1 \neq  0$, 
	then the null rotation induces $\nu_1' = \nu_1 + 4u \gamma_1$.  Hence   $\nu_1 = 0$ gauge fixes the null rotations provided $\gamma_1\neq0$
	(see 2.\ref{IIIdAbelianLN}.4).
	If  $\gamma_1=0$, then the isometry algebra becomes  5-dimensional.

	With respect to the boosts, the independent spin coefficients transform as
\begin{equation*}
	\epsilon_1' = \boostparm \epsilon_1, \enskip
	\gamma_1' = \boostparm^{-1} \gamma_1, \enskip
	\kappa_1' =  \boostparm^{2} \kappa_1, \enskip
	\mu_1' = \boostparm^{-1} \mu_1, \enskip
	\nu_1' = \boostparm^{-2}\tau_1,\enskip
	\alpha_1' = \alpha_1, \enskip
	\tau_1' = \tau_1.
\end{equation*} 
	Since the isometry algebra becomes 5-dimensional 
	if the spin coefficients $\{\epsilon_1, \gamma_1, \kappa_1,\mu_1, \nu_1\}$ all vanish, the boosts 
	can always be gauge-fixed, {\it e.g.,} by normalizing one of these spin coefficients to 1.  Having gauge-fixed the null rotations 
	and the boosts the residual group is discrete.

\par
\smallskip
\noindent
{\bf \ref{IIIdAbelianLZ}. \abelthreeLZ.\ } With respect to the adapted basis  
	$\{K,\, L,\, \dfrac12 (M + \barM) \,\}$ for $\lieg'$,
	the covariant components $A_{ij}$ of $\text{ad}(\dfrac{i}{2}(M -\barM))$  
	are 
\begin{equation*}
	A_{ij}  = \begin{bmatrix}
\kappa_1&\tau_1&\epsilon_1
\\ \noalign{\medskip}\tau_1&\nu_1&\gamma_1\\ \noalign{\medskip}
\epsilon_1&\gamma_1&\alpha_1
\end{bmatrix}.
\end{equation*}

	The residual group is ${\rm O}(2,1) \times \mathcal Y$, with ${\rm O}(2,1)$ acting on $\lieg'$.
	The basis for $\lieg'$ may be rotated by a Lorentz transformation to bring 
	the matrix  $A$ to  one of the normal forms  in \eqref{Segre}. 
\par
\noindent
	{\bf Case I.}  This implies that in the rotated basis $ \epsilon_1 = \gamma_1 =0$, 
	$\nu_1 = \kappa_1$, $s_1 = -\tau_1$, $s_2 = \kappa_1$, and $s_3 = 2\alpha_1$. 
	The eigenvalues and eigenvectors of $A$ (relative to $\tilde \eta$) are then
\begin{equation*} 
	\xi_1 = -\tau_1 +\kappa_1, \enskip    X_1 = K-L; \quad
	\xi_2 = -\tau_1 - \kappa_1, \enskip X_2 = K +L; \quad
	\xi_3= 2\,\alpha_1, \enskip X_3 =  \frac12(M + \barM).  
\end{equation*}
	Equality of two of the eigenvalues implies that $\alpha_1 = -\dfrac12 \tau_1 \pm \dfrac12 \kappa_1$
	or $\kappa_1 = 0$. In each case the isometry algebra is 5-dimensional. Therefore
	$\alpha_1 \neq  -\dfrac12 \tau_1 \pm \dfrac12 \kappa_1$ and $\kappa_1 \neq 0$, the eigenvectors 
	are distinct, and the residual group is generated by the discrete Lorentz transformations 
	$\{$$\mathcal R$, $\mathcal T$, $\mathcal V$, $\mathcal Y$, $\mathcal Z$$\}$. This case is given in 2.\ref{IIIdAbelianLZ}.1.
\par
\noindent
	{\bf Case II.} In the rotated basis $\epsilon_1 = \gamma_1 = 0$,  
	$\nu_1 = -\kappa_1$, $s_1= -\tau_1$, $s_2 = \kappa_1$, and $s_3 = 2\alpha_1$.
	The eigenvalues of $A$ (relative to $\tilde \eta$) are then 
\begin{equation*}
	\xi_1 = - \tau_1 - i\kappa_1, \enskip  X_1 =  -K +i\,L; \quad
	\xi_2 = - \tau_1 + i\kappa_1, \enskip X_2 = - K -i\,L;\quad
	\xi_3 = 2\,\alpha_1,  \enskip X_3 = \dfrac{i}2(M - \barM).
\end{equation*}
	With $\kappa_1 \neq 0$, the residual group is the discrete group with generators 
	$\{\mathcal R,\ \mathcal T,\ \mathcal Y,\ \mathcal Z\}$. This case is given in 2.\ref{IIIdAbelianLZ}.2.
	
\par
\noindent
	{\bf Case III.}  This implies that in the rotated basis 
	$ \epsilon_1 = \gamma_1 = \kappa_1 =0$, $s_1 = -\tau_1$, $s_2 = \nu_1$
	and $s_3 = 2\alpha_1$. 
	The eigenvalues of $A$ (relative to $\tilde \eta$) are then
\begin{equation*}
	\xi_1 =  \xi_2 = - \tau_1,\enskip  X_1 = K, \enskip \mathcal K = \langle K,\, L \rangle; \quad 
	\xi_3 = 2\,\alpha_1,\enskip  X_3 = \frac{1}2(M + \barM).
\end{equation*} 
	For $\xi_1 \neq \xi_3$, the residual group is the  group generated by the $KL$ boosts and
	$ \mathcal R$, $\mathcal Y$, $\mathcal T$. This case is given in 2.\ref{IIIdAbelianLZ}.3. Since the isometry 
	group is 5-dimensional if $\nu_1  =0$, the continuous residual group can
	be gauge-fixed, {\it e.g.,} by setting $\nu_1 = \pm 1$.

	If $\xi_1 = \xi_3$, then $ \tau_1 = -2\,\alpha_1$
	in which case the isometry algebra is 5-dimensional. Hence
	$ \tau_1 \neq - 2\,\alpha_1$.  

\par
\noindent
	{\bf Case IV.}  This implies that in the rotated basis 
	$\epsilon_1 = \kappa_1 = \nu_1 =0$, $\tau_1 =-2\alpha_1$, $s_1 =2\alpha_1$, 
	$s_2 = \gamma_1$. The eigenvalue/eigenvectors are 
\begin{equation*}
	\xi_1 = \xi_2 = \xi_3 =  2\,\alpha_1,\enskip   
	X_1 = K, \enskip  
	\mathcal K = \langle\, K, \frac12(M + \barM)\,\rangle.
\end{equation*}
	The residual group is again the  group generated by the $KL$ boosts and
	$ \mathcal R$, $\mathcal Y$, $\mathcal T$.  This case is given in 2.\ref{IIIdAbelianLZ}.4.
 If $\gamma_1 = 0$, the isometry algebra 
	is 10-dimensional, consequently the residual group (boosts) can be gauge-fixed, {\it e.g.,}
	by setting $\gamma_1 = 1$. 

\par
\medskip
\noindent
{\bf \ref{IIIdAbelianNS}. \abelthreeNS.\ } According to Table \ref{LeviTable}, the sub-group of the Lorentz group which fixes the flag 
	$\langle\, M + \barM \,\rangle \subset \langle\, K, M\, , \barM \, \rangle$ 
	includes the $KL$ boosts and the null rotations \eqref{Null} with $\varphi = i v$ ($v$ real). 
	Since $\kappa = \sigma = \rho = 0$, the  spin coefficients  $\lambda_0$, $\gamma_0$, $\mu_0$ 
	transform under the null rotations as	
\begin{equation*}
	\gamma_0' = \gamma_0 -4v\beta_1   \quad  \lambda_0' = \lambda_0 +   4v\beta_1 \quad 
	\mu_0' = \mu_0 -  4v\beta_1 .
\end{equation*}
	Since $\beta_1 \neq 0$, the null rotations can be gauge-fixed by setting any one of these spin coefficients to zero.
	We take $\gamma_0 = 0$, which defines the gauge used in 2.\ref{IIIdAbelianNS}.1.	
	Under the $KL$ boosts, the remaining independent spin coefficients transform as 
\begin{equation*}
	\lambda' = \boostparm^{-1} \lambda,\quad
	\mu_0' = \boostparm^{-1} \mu_0,\quad
	\nu'  = \boostparm^{-2} \nu, \quad
	\beta_1' = \beta_1.				
\end{equation*}
	Since the derived algebra is 1-dimensional when $\lambda = \mu_0 = \nu = 0$, one of these
	spin coefficients must be non-zero and the boosts can be gauge-fixed. Having gauge-fixed the null rotations 
	and the boosts the residual group is discrete.

\subsection{Spacetime Groups with  Simple Derived Algebra}

{\bf \ref{simpT}. \LevitwoT.} 
	The subgroup of the Lorentz group which fixes the time-like center 
	$\liez = \langle K + L \rangle$ is $\text{O}(3)\times \mathcal T $. 
	This group acts on the quotient algebra $\lieg/\liez$, 
	preserving the induced metric $\tilde \eta$  on $\lieg/\liez$, and may therefore be used to 
	diagonalize the Killing form $B$ for $\lieg/\liez$. With respect to the basis for $\lieg/\liez$
\begin{equation*}
	e_1 =\dfrac12(K-L) + \liez \ , \enskip
	e_2 = \dfrac12(M + \barM) + \liez , \enskip
	e_3 = \dfrac{i}{2}(M- \barM) + \liez ,
\end{equation*}
	the structure equations for $\lieg/\liez$ are
\begin{equation*}
	[\, e_1,\, e_2]  = 2\,c_4\, e_1 + \lambda_0\, e_2+  c_1\, e_3, \quad  
	[\, e_1,\, e_3]  = 2\, c_5\, e_1+ c_2\, e_2-\lambda_0\, e_3, \quad
        [\, e_2,\, e_3]  = - c_3\, e_1-2\, c_5\, e_2+2\ c_4\, e_3,
\end{equation*}
where
\begin{equation*}
	c_1 = \lambda_1-2\,\gamma_1+\mu_1,\enskip
	c_2 = \lambda_1+2\,\gamma_1-\mu_1, \enskip
	c_3 = -2\,\mu_1+2\,\epsilon_1+2\,\gamma_1, \enskip
	c_4 =  \nu_0-2\,\beta_0, \enskip
	c_5 = \nu_1+2\,\beta_1 .
\end{equation*}
	The components of the Killing form are
\begin{equation*}
	B =
\begin{bmatrix} 
	2\, c_1\, c_2+2\,\lambda_0^ 2 &-4\, c_1\, c_5-4\, c_4\,\lambda_0&-4\, c_2\, c_4+4 \,\lambda_0\, c_5
\\[2\jot] 
	-4\, c_1\, c_5-4\, c_4\,\lambda_0&2\, c_1\, c_3+8\, c_4^2&-2\,\lambda_0\, c_3+8\, c_4\, c_5
\\[2\jot]
	-4\, c_2\, c_4+4\,\lambda_0\, c_5&-2\,\lambda_0\, c_3+8c_4\, c_5 &
	-2\, c_2\, c_3+8\, c_5^2
\end{bmatrix}.
\end{equation*}
	We solve the equations $B_{12} =B_{13} = B_{23} = 0$, 
	subject to the condition $\det(B) \neq 0$, to deduce that, in the basis which diagonalizes $B$, the spin coefficients satisfy $\lambda_0 = 0$ and $c_4 = c_5 = 0$,
	that is, 
\begin{equation*}
	\lambda_0=0, \quad 
	\nu_0 =2\beta_0,\quad 
	\nu_1=-2\, \beta_1. 
\end{equation*}
	The eigenvalues  $\xi_i$ and eigenvectors  $X_i$ of $B$ are
\begin{equation*} 
	\xi_1 =  2\,c_1\,c_2, \enskip X_1 =  \frac12\,(K-L); \quad 
	\xi_2 =  2\,c_1\,c_3, \enskip X_2 =  \frac12\,(M + \barM); \quad
	\xi_3 = -2\,c_2\,c_3, \enskip X_3 =  \frac i2\,(M - \barM).
\end{equation*} 
	Since the Killing form must be non-degenerate, none of the $c_i$ can vanish. If the eigenvalues 
	are distinct, then the residual group is the discrete group generated
	by $\{$$\mathcal R$, $\mathcal T$, $\mathcal U$,  $\mathcal V$, $\mathcal Y$, $\mathcal Z$$\}$.  See 2.\ref{simpT}.1.

	If two of the eigenvalues coincide then, 
	by a permutation of the basis elements,  we may suppose that $\xi_2 = \xi_3$ so that $c_1 = -c_2$ and 
	$\lambda_1=0$. The residual group then becomes  the group generated by $M\barM$ rotations and the discrete transformations 
	$\{$$\mathcal T$, $\mathcal Y$, $\mathcal Z$$\}$. See 2.\ref{simpT}.2.
	The rotations  transform the independent
	spin coefficient as 
\begin{equation*}
	\beta' = e^{i\theta}\beta, \enskip
	\gamma_1' = \gamma_1, \enskip
	\mu_1' = \mu_1.
\end{equation*}
	If $\beta =0$, the isometry group has dimension 5.  With  $\beta \neq 0$, the residual group can be
	gauge-fixed, for example, by choosing $\beta$ to be real and positive.
	Finally, if all three eigenvalues coincide, then $c_1 = - c_2 = c_3$. This is equivalent to  $\lambda_1=0$ and 
	$\mu_1 = 2\,\epsilon_1$ and the isometry dimension is 5. 
	Thus, in all cases, the residual group can be gauge-fixed to a discrete group.	

\noindent
{\bf \ref{simpS}. \LevitwoS.}  In this case the residual group fixes $\liez =\langle K - L\rangle$ and consists of ${\rm O}(2,1) \times \mathcal Z$.
	The  O(2,1) subgroup  can be used to transform the Killing form
	of the quotient algebra  to one of the four normal forms in \eqref{Segre}. 
	With respect to the  basis for $\lieg/\liez$
\begin{equation*}
	\bfk = \dfrac12 (K+L + M + \barM) +\liez, \enskip 
	\bfell= \dfrac12 (K+L - M - \barM) +\liez , \enskip
	\bfm =  \dfrac{i}2\, ( M - \barM) + \liez,
\end{equation*}
	the structure equations for the quotient algebra  are
\begin{equation*}
	[\, \bfk ,\, \bfell]  = c_1\, \bfk + c_2\,\bfell + c_3\, \bfm,	
	\quad
	[\, \bfk,\, \bfm]  = c_4\,\bfk -  c_5\,\bfell- c_2\, \bfm
	\quad
	[\,\bfell,\, \bfm]  = -c_6\,\bfk  -  c_4\,\bfell +  c_1\,\bfm,
\end{equation*}
where
\begin{alignat*}{3}
	c_1 &= 4\,\beta_0 + \lambda_0 + 2\,\nu_0,
&\quad
	c_2 &= 4\,\beta_0-\lambda_0+2\,\nu_0,
&\quad
	c_3 &= -4\,\gamma_1 + 2\,\lambda_1 + 2\,\mu_1,
\\
	c_4 &=  \epsilon_1-2\,\gamma_1 - \frac12\lambda_1 + \frac32\,\mu_1,
&\quad
	c_5 &= -4\,\beta_1 -\epsilon_1 -\frac12\lambda_1   - \frac12 \mu_1   + 2\,\nu_1,
&\quad
	c_6 &= -4\,\beta_1  +\epsilon_1 +  \frac12 \lambda_1 +  2\,\nu_1+ \frac12 \mu_1.
\end{alignat*}
	The covariant components of the Killing form are  computed to be
\begin{equation*}
	B = 
\begin{bmatrix} 
	2\,c_2^2-2\,c_3\,c_5
&	-2\,c_1\,c_2-2\,c_3\,c_4
&	2\,c_1\,c_5+2\,c_2\,c_4
\\[2\jot]	
	-2\,c_1\,c_2-2\, c_3\,c_4
&	2\,c_1^2+2\,c_3\,c_6
&	2\,c_1\,c_4-2\,c_2\,c_6
\\[2\jot] 
	2\,c_1\,c_5+2\,c_2\,c_4
&	2\,c_1\,c_4-2\,c_2\,c_6
&	2\,c_4^2+2\,c_5\,c_6
\end{bmatrix}.
\end{equation*}

	The normalizations are then
\begin{alignat*}{2}
\text{I}:	
&\  	B_{11} =  B_{22},\ B_{13} = B_{23} =0 
&\quad \Longrightarrow \quad & c_1=0,  \enskip c_2 =0, \enskip  c_5 = -c_6, 
	\enskip s_1 = 2\, c_3c_4,\enskip s_2 = 2\,c_3c_6, \enskip s_3 = 4\,(c_4^2 - c_6^2)
\\
&& \quad \Longrightarrow \quad &	\nu_0 = -2\,\beta_0, \enskip \nu_1= 2\,\beta_1, \enskip  \lambda_0 = 0 .
\\[2\jot]
\text{II}:&\	B_{11} = - B_{22},\ B_{13} = B_{23} =0 
&\quad \Longrightarrow \quad &  
	c_1=0,  \enskip c_2 =0, \enskip  c_5 = c_6, \enskip 
	s_1 = 2\,c_3c_4, \enskip s_2 = -2\,c_3c_6, \enskip s_3 = 4\,(c_4^2 + c_6^2)
\\[1\jot]
&& \quad \Longrightarrow \quad &	\nu_0 = -2\,\beta_0, \enskip  \lambda_0 = 0, \enskip 
	\lambda_1 = -2\,\epsilon_1 - \mu_1 .
\\[2\jot]
\text{III}:&\  B_{11} =  B_{13} = B_{23}  =0 
&\quad \Longrightarrow \quad & 
	c_1 =0, \enskip c_2 =0, \enskip c_5 = 0, \enskip
	s_1= 2\,c_3 c_4, \enskip s_2 = 2\,c_3 c_6, \enskip s_3 = 4\,c_4^2
\\[1\jot]
&& \quad \Longrightarrow \quad &	\nu_0 = -2\,\beta_0, \enskip  \lambda_0 = 0, \enskip 
	\lambda_1=  -8\,\beta_1 -2\,\epsilon_1 - \mu_1 +4\,\nu_1.
\\[2\jot]
\text{IV}:&\ B_{11}= B_{22} = B_{13} =0,
&\quad \Longrightarrow \quad &  c_2=0, \enskip c_5 = 0, \enskip c_3 = 2\,c_4, \enskip c_1^2 = -2\,c_4c_6, \enskip
	s_1 = 4\,c_4^2, \enskip s_2 = 2\,c_1 c_4
\\[1\jot]
& \enskip  B_{33} = - \frac12 B_{12} & \quad \Longrightarrow \quad & 
	\nu_0=-2\,\beta_0 + \dfrac12\lambda_0,\enskip
	\nu_1= 2\,\beta_1 + \lambda_1, \enskip
	\epsilon_1 = \frac{3}{2}\lambda_1 - \frac12 \mu_1, 	
\\[1\jot]
&&&
	\lambda_0^{2}= -2\lambda_1\left(\lambda_1 - 2\gamma_1 + \,\mu_1\right). 
\end{alignat*}

	It remains to show how the residual group can be gauge-fixed in each case. We remark that, in all cases, the 
	eigenvalues are invariants of the residual group.

\smallskip
\noindent
{\bf Case I.}
	We find that the eigenvalues and eigenvectors of $B$, relative to $\tilde \eta$, are
\begin{equation*}
	\xi_1 = 2\,c_3(c_4 +c_6), \enskip X_1 =  \frac12(M + \barM); \quad
	\xi_2 = 2\,c_3(c_4 - c_6),\enskip X_2 = K + L;\quad
	\xi_3 = 4(c_4^2 -c_6^2),\enskip X_3 = \frac{i}{2}(M - \barM). 
\end{equation*}
	None of the factors appearing in the eigenvalues can vanish -- otherwise $\det(B) = 0$. 
	If the eigenvalues are distinct then the residual group is generated by $\mathcal Z$ along with the discrete group of permutations and reflections of the eigenvectors generated by $\{\mathcal T,\ \mathcal U,\ \mathcal Y\}$. See 2.\ref{simpS}.1. If the eigenvalues 
	are all equal then $c_6 = 0$ and $c_3 = 2 c_4$ so that  $\lambda_1 = 0$ 
	and $\mu_1 = -2\epsilon_1$ and the isometry
	algebra is 5-dimensional.  It remains to examine the cases where two of the eigenvalues coincide.

\noindent 
\smallskip
{\bf Case Ia.} If $\xi_1 = \xi_2$, then $c_6 =0$ so that $\lambda_1 = -\mu_1 - 2\epsilon_1$.
	The sub-group of O(2,1)  which fixes $\langle X_3\rangle$  is 
	induced from the subgroup of $\rm O(3,1)$  fixing $\langle i\,(M - \barM)\rangle$ and the center $\langle K-L\rangle$.
	This is the group generated by the boosts in the  $K+L,  M + \barM$ plane 
	and the discrete transformations  $\mathcal R\circ T$, $\mathcal Y$, and $\mathcal Z$.  The boosts, which we denote by $B_{K+L,M+\barMs}$,
 are given explicitly by
\begin{equation}
\begin{aligned}	
	K' &=\frac {(\boostparm +1)^2}{4\boostparm}\, K+ \frac{(\boostparm-1)^2}{4\boostparm}\, L 
	+\frac {{\boostparm}^{2}-1}{4\boostparm}\, M + \frac {{\boostparm}^{2}-1}{4\boostparm}\, \barM,
\\
	L' &= \frac {(\boostparm -1)^2}{4\boostparm}\, K+ \frac{(\boostparm + 1)^2}{4\boostparm}\, L 
	+\frac {{\boostparm}^{2}-1}{4\boostparm}\, M + \frac {{\boostparm}^{2}-1}{4\boostparm}\, \barM,
\\
	M' &= \frac{\boostparm^2-1}{4\boostparm}\,K + \frac{\boostparm^2-1}{4\boostparm}\, L  
	+\frac{(\boostparm + 1)^2}{4\boostparm}\, M + \frac{(\boostparm - 1)^2}{4\boostparm}\,\barM,
\\
	\barM' &= \frac{\boostparm^2-1}{4\boostparm}\,K + \frac{\boostparm^2-1}{4\boostparm}\, L 
	+ \frac{(\boostparm - 1)^2}{4\boostparm}\, M + \frac{(\boostparm + 1)^2}{4\boostparm}\,\barM,
\end{aligned}
\label{bigBoost}
\end{equation}
where $\boostparm>0$.  We can extend $\boostparm$ to include $\boostparm<0$, and we denote the resulting group by $B^*_{K+L,M+\barMs}$. This group is generated by the boosts (\ref{bigBoost}) with $\boostparm>0$ and the transformation (\ref{bigBoost}) with $\boostparm=-1$, where this latter transformation is equal to $\cal R\circ T\circ Y\circ Z$.  See 2.\ref{simpS}.2.

	The transformation of the spin coefficients implies
\begin{equation*}
	(\,\beta_1 - \dfrac12(\epsilon_1 -\gamma_1))' = 
	\,\boostparm\,(\,\beta_1 - \dfrac12(\epsilon_1 -\gamma_1)),
	\quad
	(\,\beta_1 + \dfrac12(\epsilon_1 -\gamma_1))' = 
	\,\boostparm^{-1}\,(\,\beta_1 + \dfrac12(\epsilon_1 -\gamma_1))	.
\end{equation*}
	If $\beta_1 =0$ and $\gamma_1 = \epsilon_1$, then the isometry algebra has dimension 5. Otherwise, the 	
	residual group can be gauge-fixed, {\it e.g.,}  by  setting 
	$\beta_1 \pm \dfrac12(\epsilon_1 -\gamma_1) =1$.

\noindent 
\smallskip
{\bf Case Ib.} 
	If $\xi_1 = \xi_3$ then $c_3 = 2(c_4 - c_6)$ so that $\lambda_1 =0$. 
	The group which fixes $\langle X_2\rangle$ is 
	induced from the group fixing $\langle K-L\rangle $ and $\langle K+L\rangle$. This is the group generated by $\mathcal T$, $\mathcal Y$, $\mathcal Z$, and $M\barM$ rotations, for which 
	$\beta' = e^{i\theta}\beta$. See 2.\ref{simpS}.3. If $\beta =0$ the isometry group has dimension 5; therefore a
	 gauge condition such as $\beta_1 =0$ and $\beta_0 >0$ reduces the residual group to a discrete group. 

\noindent 
\medskip	
{\bf Case II.}   The eigenvalues/eigenvectors are 
\vspace{-6pt}
\begin{align*}
	\xi_1 &= 2\,c_3(c_4 - i\,c_6), \enskip X_1 = (K + L + M + \barM) + i\,(K + L - M - \barM);
\\
	\xi_2 &= 2\,c_3(c_4 + i\,c_6), \enskip X_2 = (K + L + M + \barM) - i\,(K + L - M - \barM);
\\
	\xi_3 &= 4\,(c_4^2 + c_6^2), \enskip X_3 = \frac{i}{2}(M - \barM).
\end{align*}
	The residual group  fixes the eigenspaces and is generated by $\{\mathcal R,\ \mathcal T,\ \mathcal Y,\ \mathcal Z\}$.  See 2.\ref{simpS}.4.

\noindent 
\medskip
{\bf Case III.} The eigenvalues/eigenvectors are 
\vspace{-6pt}
\begin{equation*}
	\xi_1 = \xi_2 =  2\,c_3c_4, \enskip X_1 = \frac12( K + L + M + \barM), \enskip 
	\mathcal K = \langle X_1 ,X_2= \frac12( K + L -  M - \barM) \rangle; \quad \xi_3 =  4c_4^2, \enskip  
	X_3 =  \frac{i}{2}(M - \barM). 
\end{equation*}
	The residual group is  the group generated by boosts in the $X_1$, $X_2$ plane, given by  \eqref{bigBoost} with $\boostparm \neq0$, and the discrete transformations $\mathcal  Y$, $\mathcal Z$, $\mathcal R \circ \mathcal T$. See 2.\ref{simpS}.5.
	The independent spin coefficients are $\{\beta_0, \beta_1, \epsilon_1, \gamma_1, \mu_1 \}$
	and the action of the boosts on the spin coefficients implies
\begin{align*}
	&(\nu_1 -2\beta_1)' = \boostparm^{-2}(\nu_1 -2\beta_1), \quad
	(2\nu_1 - 2\beta_1 -\epsilon_1 + \gamma_1)' = \boostparm(\,2\nu_1 - 2\beta_1 -\epsilon_1 + \gamma_1),\
\\
	&(2\nu_1 - 2\beta_1 +\epsilon_1 - \gamma_1)' = \boostparm^{-1}(2\nu_1 - 2\beta_1 +\epsilon_1 - \gamma_1).
\end{align*}	
	The spin coefficient combinations $c_3$ and $c_4$ are invariant. 
	The boosts can therefore be gauge-fixed unless  $\nu_1 =0$, $\beta_1 =0$, $\epsilon_1 = \gamma_1$,
	but then the isometry algebra is 5-dimensional.  Thus the residual group can be reduced to a discrete group by normalization.

	If all the eigenvalues coincide then $c_3 = 2\, c_4$ so that  
	$\nu_1=2\,\beta_1+\dfrac23\,\epsilon_1 + \dfrac13\mu_1$. In this case, the residual group
	fixes  $\langle K-L\rangle$ and the eigenvector 2-plane $\langle K + L +M + \barM, i(M- \barM)\rangle$ and 
	therefore consists of the foregoing boosts \eqref{bigBoost} with $\boostparm\neq0$,
	 the 1-parameter family of null rotations fixing $K-L$ and  $K + L +M + \barM$, and the discrete transformations  from the previous case. 
	The null rotations  are given by
\begin{equation*}
\begin{aligned}	
	K' &= ({u}^{2}+1)\, K+{u}^{2}\, L+({u}^{2}+iu)\, M+({u}^{2}-iu)\, \barM,
\\
	L' &= {u}^{2}\, K+({u}^{2}+1)\, L+({u}^{2}+iu)\, M+({u}^{2}-iu)\, \barM,
\\
	M' &= (-{u}^{2}-iu)\, K+(-{u}^{2}-iu)\, L+(1-{u}^{2}-2\,iu)\, M-{u}^{2}\, \barM,
\\
	\barM' &= (-{u}^{2}+iu)\, K+(-{u}^{2}+iu)\, L-{u}^{2}\, M+(1-{u}^{2}+2\,iu)\, \barM.
\end{aligned}
\end{equation*}
	The action of these null rotations on the independent spin coefficients is
\begin{align*}
	\beta_0'&=2\,\Upsilon\,u+\beta_0,\quad
	\beta_1'=-2\,\Upsilon\,{u}^{2}-2\,\beta_0\,u +\beta_1, \quad
	 \epsilon_1'= -2\,\Upsilon\,{u}^{2}-2\,\beta_0\,u+\epsilon_1,
\\
	 \gamma_1' &=2\,\Upsilon\,{u}^{2}+2\,\beta_0\,u+\gamma_1, \quad
	\mu_1' =4\,\Upsilon\,{u}^{2}+4\,\beta_0\,u +\mu_1,
	\enskip\text{where}\enskip \Upsilon = \beta_1 +\frac16\epsilon_1+\frac12 \gamma_1 + \frac13 \mu_1.
\end{align*}
	We remark that $\Upsilon$ is invariant under the null rotations. To gauge fix this null rotation subgroup of the residual group we must consider two 	cases.  If $\Upsilon\neq0$ we may gauge fix via the normalization $\beta_0=0$.  This yields 2.14.6.  The boosts can be gauge-fixed as described in the case of distinct eigenvalues, and the resulting residual group is then discrete.

If $\Upsilon=0$, then $\beta_0\neq0$ or the isometry algebra is 5-dimensional. We can therefore use the null rotations to normalize $\beta_1=0$.  This reduces the gauge group to the discrete group  $\{\mathcal R \circ \mathcal T,\  \mathcal Y\}$.
All together, these conditions lead to 2.14.7. 

\smallskip
\noindent
{\bf Case IV.} The eigenvalue/eigenvectors are 
\begin{equation*}
	\xi_1 = \xi_2 = \xi_3 = 4\,c_4^2, \enskip X_1 =K + L + M + \barM,
	\enskip \mathcal K  = \langle(X_1, i\,(M- \barM)\rangle.
\end{equation*}
	The residual group is generated by  boosts in the  $K + L$, $M + \barM$ plane and the discrete transformations
	$\mathcal  Y$, $\mathcal Z$, $\mathcal R \circ \mathcal T$.  The boosts are given in \eqref{bigBoost} with $\boostparm\neq0$
	and transform $\lambda_0$ and $\lambda_1$ by 
\begin{equation*}
	\lambda_0' = \boostparm^{-1} \lambda_0 \quad\text{and}\quad \lambda_1' = \boostparm^{-2}\lambda_1.
\end{equation*}
	One also checks that $c_4 = \lambda_1 -2\gamma_ 1 +\mu_1\neq 0$ is invariant.  Recall that, in this case, $\lambda_0^{2}= -2\lambda_1\left(\lambda_1 - 2\gamma_1 + \,\mu_1\right)$.  Consequently,  
	if $\lambda_1 = 0$ then $\lambda_0 = 0$, and conversely.  If $\lambda_0=\lambda_1=0$
	the isometry algebra is 5-dimensional.  Therefore  the boosts can always be  gauge-fixed.  We choose the gauge $\lambda_0=\lambda_1$ for this case.
	The discrete residual group is  then generated by  $\mathcal R \circ \mathcal T$ and $\mathcal Z$. See 2.14.8. 
\smallskip

\noindent
{\bf \ref{simpN}. \LevitwoN.} The subgroup of the Lorentz group which fixes the  null center 
	$\liez = \langle K \rangle$ is the 4-dimensional group generated by $M\barM$ rotations, 
	$KL$ boosts,  the two parameter family of null rotations which fix $K$ 
	(see \eqref{Rotation}--\eqref{Null}),  and the discrete transformations \{$\mathcal T$, $\mathcal Y$\}.
	This group acts on the quotient algebra $\lieg/\liez$ and may therefore be used to 
	transform its Killing form $B$ to a normal form.  We begin by deriving the possible normal forms.

	Set
\begin{equation}
	e_1 = L + \liez, \enskip e_2 = (M + \barM) + \liez, \enskip e_3 =  i\,(M- \barM) + \liez,	
\DGlabel{Qbasis}
\end{equation}
	and let  $\{ \omega_1, \, \omega_2, \, \omega_3\}$ be the associated dual basis.
	The induced action of the boosts, real null rotations, imaginary null rotations, and 
	spatial rotations are given, respectively, by  
\begin{alignat*}{2}
	\varphi_1(t)\,[\,e_1,\, e_2,\, e_3\,] &= [\,t\, e_1, \, e_2,\, e_3\,],
	&\quad 
	\varphi_2(t)\,[\,e_1,\, e_2,\, e_3\,] &= [\,e_1 + t\, e_2,\, e_2 ,\, e_3\,],
\\[1\jot]
	\varphi_3(t)\,[\,e_1,\, e_2,\, e_3\,] &= [\,e_1 + t\, e_3,\, e_2, \,  e_3\,],
	&\quad
	\varphi_4(t)\,[\,e_1,\, e_2,\, e_3\,] &= [\,e_1,\,  \cos(t)\,e_2 - \sin(t)\,e_3,\, \sin(t)\, e_2 + \cos(t)\, e_3\,].
\end{alignat*}
	Note that 
\begin{equation*}
	\varphi_2^*(t)\,[\,\omega_1,\, \omega_2,\, \omega_3\,] = 
	[\,\omega_1,\, t\,\omega_1 +\omega_2,\, \omega_3\,], \quad\text{and}\quad
	\varphi_3^*(t)\,[\,\omega_1,\, \omega_2,\, \omega_3\,] = 
	[\,\omega_1,\, \omega_2,\,  t\,\omega_1 +\omega_3\,].
\end{equation*}

	The task at hand is to transform a non-degenerate quadratic  form,
	\begin{equation*}
	S = s_1 \,\omega_1\odot \omega_1 + 2\, s_2\, \omega_1\odot \omega_2 + 2\,s_3\, \omega_1\odot \omega_3
	 + s_4\,\omega_2\odot \omega_2  + 2\,s_5\,\omega_2\odot \omega_3 + s_6\, \omega_3\odot \omega_3,
	\end{equation*} to normal form. The rotation $\varphi_4$ acts  on $S$ by 
	conjugation of the symmetric matrix 
	$\begin{bmatrix} s_4 & s_5 \\ s_5 & s_6 \end{bmatrix}$ and accordingly we may use $\varphi_4$ to
	set $s_5 =0$.  The non-degeneracy of $S$ then implies that $s_4$ and $s_6$ cannot both vanish;
	if necessary, by a further rotation of $90^\circ$ we  may assume $s_4 \neq 0$. The transformation 
	$\varphi_2$ is then used to set $s_2=0$.

	Two cases are now considered. If $s_6 \neq 0$ (Case V), then  the 
	transformation $\varphi_3$ can be used to set 
	$s_3=0$.  If  $s_6 = 0$ (Case VI), then  we must have $s_3 \neq 0$ and  $
	\varphi_3$ can be used to set $s_1=0$.
	The normal forms for $S$ are therefore represented by the symmetric matrices 
\begin{equation}
V = \begin{bmatrix}
	s_1 &0 &0\\ 0 & s_4 &0 \\ 0 &0 & s_6
\end{bmatrix}
\quad\text{and}\quad 
VI = 	
\begin{bmatrix}
	0 & 0 & s_3 
\\	0&  s_4 &0 
\\ 	s_3 &0 & 0
\end{bmatrix}.
\end{equation}

	In Case V the residual group is generated by the 1-parameter scaling group $\varphi_1(t)$, corresponding to $B^*_{K,L}$, along with $\mathcal U$, 		$\mathcal Y$. In the special case $s_4= s_6$, the residual group also includes the rotations $\varphi_4(t)$, corresponding to $R_{M,\barMs}$. For Case VI, the residual group is generated by
	$\varphi_1(t)$, corresponding to $B^*_{K,L}$, along with $\mathcal R$, $\mathcal Y$.  The determinant of the restriction of $S$  to the Riemannian 2-plane  $\langle e_2, e_3\rangle$ 
	is a relative invariant of the residual group, which is non-zero 
	in case V and is zero in case VI.

	We now return to the original problem of gauge fixing the residual group for {\bf simpCN}.\
	With respect to the basis \eqref{Qbasis}, the structure equations for the quotient are
\begin{equation*}
	[\, e_1,\, e_2\,]  = -8\,\beta_0\, e_1-\lambda_0\, e_2 -  c_1\, e_3,  \quad 
	[\, e_1,\, e_3\,]  = 8\,\beta_1\, e_1- c_2\, e_2+\lambda_0\, e_3, \quad 
	[\, e_2,\, e_3\,]  = 8\,\epsilon_1\, e_1-8\,\beta_1\, e_2-8\,\beta_0\, e_3,
\end{equation*}
	where  $c_1=\lambda_1-2\,\gamma_1+\mu_1$ and $c_2 = \lambda_1+2 \gamma_1-\mu_1$. The components of the Killing form are 	
\begin{equation*}
B=
\begin{bmatrix}
	2\, c_1\, c_2+2\,\lambda_0^2
&	-16\,\beta_0\lambda_0+16\, c_1\,\beta_1
&	-16\,\beta_0 c_2-16\,\beta_1\,\lambda_0
\\[1\jot]
	-16\,\beta_0\lambda_0+16\, c_1\,\beta_1
&	128\,\beta_0^2+16\, c_1\,\epsilon_1
&	-128\,\beta_0\beta_1-16\,\lambda_0\,\epsilon_1
\\[1\jot]
	-16\,\beta_0 c_2-16\,\beta_1\,\lambda_0
&	-128\,\beta_0\beta_1-16\,\lambda_0\,\epsilon_1
&	-16\, c_2\,\epsilon_1+128\,\beta_1^2
\end{bmatrix}.
\end{equation*}

	For Case V, we solve the equations 
	$B_{12} = B_{13} = B_{23} = 0$ to find $ \beta_0 = \beta_1 = \lambda_0 =0$.  See 2.15.1.
	The $KL$ boosts transform the independent spin coefficients 
	$\left\{\epsilon_1, \gamma_1, \lambda_1, \mu_1, \nu  \right\} $
	 by
\begin{equation*}
	\epsilon_1' = \boostparm\, \epsilon_1, \enskip
	\gamma_1' = \boostparm^{-1}\, \gamma_1, \enskip
	\lambda_1' = \boostparm^{-1}\lambda_1, \enskip
	\mu_1'= \boostparm^{-1}\mu_1, \enskip
	\nu' = \boostparm^{-2}\nu.
\end{equation*}
	The non-degeneracy of the Killing  form requires 
	$c_1 \neq 0$ and $c_2 \neq 0$, therefore
	the residual boosts  can be gauge-fixed, {\it e.g.,} by setting $c_1 =1$ or $c_2 = 1$.
	In the special case  where $c_1 = -c_2$, that is, $\lambda_1=0$, one must have $\nu \neq 0$ 
	(or the isometry  algebra is 5-dimensional).   See 2.15.2 for this case. 
	Since $\nu\neq 0$, the residual spatial rotations can be gauge-fixed {\it e.g.},  by setting $\nu$  real and positive.

	For Case VI, we solve  $B_{11} = B_{12} = B_{23}= B_{33}=0$ to arrive at 
	$\beta_1 = \epsilon_1 =  \lambda_0 =0$ and $c_1=0$, that is, $\lambda_1 = 2 \gamma_1 - \mu_1$.
	See 2.15.3.
	The independent spin coefficients 
	$\left\{\beta_0,\gamma_1, \mu_1,\nu  \right\}$ transform under the $KL$ boosts as above.
	The non-degeneracy of the Killing form requires $2\gamma_1 - \mu_1 \neq0 $, 
	so  the boosts can be gauge-fixed {\it e.g.,} by setting  $2\,\gamma_1 - \mu_1  = 1$.


\subsection{Spacetime Groups with   2-Dimensional Derived Algebras}

	For spacetime Lie algebras with 2-dimensional derived algebras,  
	we have in each case reduced the residual group to a subgroup of the 
	group generated by $KL$ boosts, $M\barM$ rotations, and various discrete transformations. For each such  spacetime Lie algebra 
	it is easy to see that there are enough non-zero spin coefficients to ensure that the residual group 
	can be reduced to a finite discrete group by gauge fixing. The details are as follows.

\par
\smallskip
\noindent
{\bf \ref{IIdAbelR1}. \abeltwoRone.\ } 
	The sub-group of the Lorentz group which stabilizes $\lieg' = \langle\, M,\, \barM \,\rangle$ 
	is generated by the 2-dimensional group of $KL$ boosts, $M\barM$ rotations, and the discrete transformations $\mathcal Y$, $\mathcal Z$ (see Table \ref{LeviTable}).
	As shown in section \ref{TwoDDerived}, the normalizations for  \abeltwoRone,
	namely $\lambda$, $\sigma$ real  (with $\lambda = q \sigma$), reduce the residual group to a discrete group.
\par
\smallskip
\noindent
{\bf \ref{IIdAbelR2}. \abeltwoRtwo.\ }
	The residual group includes  $KL$ boosts and  $M\barM$
	rotations. Since $\sigma \neq 0$, these transformations can be gauge-fixed, {\it e.g.} by setting $\sigma = 1$.
\par
\smallskip
\noindent
{\bf \ref{IIdAbelR3}. \abeltwoRthree.\ }
	The residual group again includes boosts and rotations, which  act on the independent 
	spin coefficients by
\begin{equation*}
	\beta' = e^{i\theta}\,\beta,\enskip
	\epsilon_1' = \boostparm\, \epsilon_1,\enskip
	\gamma_1' = \boostparm^{-1} \gamma_1,\enskip
	\mu_1' = \boostparm^{-1} \mu_1, \enskip
	\rho_0' = \boostparm\, \rho_0.
\end{equation*}
	If $\beta =0$,  the isometry algebra is 7 dimensional. If 
	$\epsilon_1 = \gamma_1 = \mu_0 = \rho_0 =0 $, then 
	the isometry algebra is 5-dimensional.
	Thus both the boosts and rotations can always be gauge-fixed by normalization of  independent spin coefficients.
\par
\smallskip
\noindent
{\bf \ref{IIdAbelL1}. \abeltwoLone.\ }
	 The continuous part of the residual group includes boosts, which act on the independent spin coefficients by
\begin{equation*}
	\epsilon_1' = \boostparm\,\epsilon_1\,\enskip
	\gamma_1' = \boostparm^{-1}\gamma_1,\enskip
	\kappa_0' = \boostparm^{2} \kappa_0 ,\enskip
	\nu_0' = \boostparm^{-2}{\nu_0}, \enskip
	\beta_0' = \beta_0,\enskip
	\tau' = \tau.
\end{equation*}
	The isometry algebra has dimension 7 when 
	$\epsilon_1= \gamma_1 = \kappa_0 = \nu_0 = 0$ 
	and therefore the boosts can always be gauge-fixed.
\par
\smallskip
\noindent
{\bf \ref{IIdAbelL2}. \abeltwoLtwo.\ }
	The continuous part of the residual group includes boosts and rotations, which act on the
	independent spin coefficients by
\begin{equation*}
	\tau' = e^{i\, \theta}\tau, \enskip
	\epsilon_1' = \boostparm\,\epsilon_1, \enskip
	\gamma_1' = \boostparm^{-1}\,\gamma_1.
\end{equation*}
	If either 
	$\tau =0$ or $\epsilon_1=\gamma_1=0 $, then
	the dimension of the isometry algebra increases to 5 or 10, respectively. Consequently the boosts and rotations can be gauge-fixed.
\par
\smallskip
\noindent
{\bf \ref{IIAbelN1}. \abeltwoNone.\ }
	Here  the residual group includes  $KL$ boosts. Since $\epsilon_1 \neq 0$, these transformations are easily gauge-fixed.  
\par
\smallskip
\noindent
{\bf \ref{IIAbelN2}. \abeltwoNtwo.\ }
	Again, the  residual group includes  $KL$ boosts, which act on the independent spin coefficients by
\begin{equation*}
	\nu' = \boostparm^{-2}\nu, \enskip	
	\gamma_0 ' = \boostparm^{-1} \gamma_0, \enskip
	\beta_1' = \beta_1, \enskip
	\tau_0' = \tau_0.
\end{equation*}
	Since the isometry algebra is 5-dimensional when $\nu = \gamma_0 =0$, 
	the boosts can always be gauge-fixed.
\par
\smallskip
\noindent
{\bf \ref{IIAbelN3}. \abeltwoNthree.\ }
	The  residual group includes  $KL$ boosts, which act on the independent spin coefficients by
\begin{equation*}
	\gamma_1' = \boostparm^{-1} \gamma_1,\enskip
	\mu_0' = \boostparm^{-1} \mu_0,\enskip
	\nu_1' = \boostparm^{-2} \nu_1,\enskip
	\beta_1' = \beta_1,\enskip
	\tau_1' = \tau_1.
\end{equation*}
	Since the isometry algebra is 5-dimensional when $\gamma = \mu_0 = \nu_1 =0$, 
	the  boosts can always be gauge-fixed.
\par
\smallskip
\noindent
{\bf \ref{IIAbelN4}. \abeltwoNfour.\ }
	The  residual group includes  $KL$ boosts, which act on the independent spin coefficients by
\begin{equation*}
	\nu' = \boostparm^{-2} \nu, \enskip
	\mu_1' = \boostparm^{-1}\mu_1,  \enskip 
	\beta_1' = \beta_1.
\end{equation*}
	Since the isometry algebra is 5-dimensional when $\nu = \mu_1 = 0$, 
	the boosts can always be gauge-fixed.
\par
\smallskip
\noindent
{\bf \ref{IIAbelN5}. \abeltwoNfive.\ }
	The  residual group includes  $KL$ boosts, which act on the independent spin coefficients by
\begin{equation*}
	\nu' = \boostparm^{-2} \nu, \enskip
	\gamma_0' = \boostparm^{-1}\gamma_0,  \enskip 
	\beta_1' = \beta_1.
\end{equation*}
	Since the isometry algebra is 5-dimensional when $\nu = \gamma_0 =0$, 
	the boosts are easily gauge-fixed.

\section{Software Implementation}\label{software}
\newcommand{\DGio}[2]{\begin{footnotesize}\tt > #1  \begin{equation*} #2 \end{equation*}\vskip -4pt \end{footnotesize}}

\newcommand{\DGin}[1]{{\footnotesize\tt > #1}}
\newcommand{\DGhead}[1]{ \vskip 5pt \noindent \kern -10pt {\sc$\blacktriangleright$ #1}}
\parindent = 0pt
The extensive computations required to obtain the results of this paper were performed using the {\sc Maple} package {\sc DifferentialGeometry} and the sub-package,  
{\sc SpacetimeGroups},  which provides a comprehensive toolbox for verifying and applying the results of this paper. Current versions of both packages are available at \cite{Anderson-Torre}.

In the following sub-sections we provide illustrative examples of the use of this software to support the results of the previous sections.  We have edited some of the input and output  to clarify the exposition.

\medskip
%

\DGhead{1. SpaceTimeLieAlgebra}

The command {\tt SpaceTimeLieAlgebra} is used to initialize any one of the  spacetime Lie algebras defined in this paper.
With no arguments the command returns the admissible equation numbers which can be passed to SpaceTimeLieAlgebra.

\newcommand\qt{\text{"}}
\DGio{Eqlist := SpaceTimeLieAlgebra()}
{Eqlist := [\qt1.4\qt,\, \qt2.1,\, \qt2.10\qt,\, \qt2.10.1\qt,\, \qt2.10.2\qt,\, \ldots, \qt5.38\qt,\, \qt5.39\qt,\, \qt5.40\qt,\, \qt5.41\qt,\, \qt5.7\qt,\, \qt5.9\qt]}

%

The starting point for all of our analysis is the set of Newman-Penrose structure equations (1.4). These structure equations may be initialized with 

\DGio{SpaceTimeLieAlgebra("1.4");}{NPalg}

The vectors $K, L, M, \barM$ and their Lie brackets are now known to {\sc Maple} 
(compare with \eqref{masterNP}), for example:

\DGio{LieBracket(K, L)}{-(\gamma + \bar\gamma)\,K - (\epsilon + \bar\epsilon)\,L + 
(\pi + \bar\tau)\,M + (\bar\pi + \tau)\, \barM }

\DGhead{2. Verifying the Classification of Spacetime Lie algebras}

{\it One can use the {\sl DifferentialGeometry} software to check every step of our classification proof}.  We illustrate this for the spacetime Lie algebra {\bf abel3LT}.  The following computations follow the arguments  presented in Section 5.4.

Our starting point is the set of structure equations (5.9). These are the structure equations for any spacetime Lie algebra with a 3-dimensional derived algebra which is Lorentzian.  The following command initializes the vector space and defines the structure equations; these data are named {\tt LDerived1}.

\DGio{SpaceTimeLieAlgebra("5.9"):}
{\text{LDerived1}}
The structure equations can be obtained with the command {\tt LieAlgebraData}.  For brevity, we suppress the output (see (5.9)).

\DGin{LD1 := LieAlgebraData(LDerived1):}

\par
Calculate a basis for the derived algebra.

\DGio{DerAlg := DerivedAlgebra();}{ \text{DerAlg} := [ K+L, M, \barM]}

For the derived algebra to be abelian the following Lie brackets must vanish.

\DGio{Eq1 := LieBracket(K + L, M);}
{\text{Eq1} := (-\kappa+\bar\nu)\, K+(-\kappa+\bar\nu)\, L+(\frac32\,\bar\rho+\epsilon-\bar\epsilon-\mu_0- \frac12\rho+2\,i\gamma_1)\, M+(\sigma-\bar\lambda)\, \barM}

\DGio{Eq2 := LieBracket(M, barM);}
{\text{Eq2} :=(\rho-\bar\rho)\, K+(\rho-\bar\rho)\, L+(-\nu-\bar\tau+2\,\bar\beta)\, M+(\bar\nu+\tau-2\,\beta)\, \barM}

The components of these Lie brackets can be extracted with {\tt DGinformation/"CoefficientSet"} and the solution obtained via the {\sl Maple} algebraic solver.  The result is
 

\DGin{Soln1 := $[\
\epsilon = \epsilon_0-i\gamma_1,\enskip
\bar\epsilon =  \epsilon_0+i\gamma_1,\enskip
\mu_0=\rho_0,\enskip
\lambda=\bar\sigma,\enskip
\bar\lambda=\sigma,\enskip
\nu=\bar\kappa,\enskip
\bar\nu=\kappa,\enskip
\rho=\rho_0,\enskip
\bar\rho=\rho_0,\enskip\\
\phantom{xxxxxxxxxxx}
\tau=-\kappa+2\,\beta,\enskip
\bar\tau=-\bar\kappa+2\,\bar\beta \ ]$ }

Substitute this solution back into the structure equations.

\DGio{LD2 := DGsimplify(subs(Soln1,  LD1));}
{LD2 :=[K, L] = -2\,\epsilon_0\,K -2\,\epsilon_0,L + 2\,\bar\kappa\, M +2\,\kappa\ \barM, \enskip
[K, M] = -\kappa K - \kappa L +(\rho_0 -2\,I\, \gamma_1)\,M + \sigma\, \barM,}
\DGin{
$\phantom{xxxxxxxxxxxx}
[L, M] = \kappa K + \kappa L +(\rho_0 + 2\, I \gamma_1)M - \sigma  \bar M$
}

We use the command {\tt DGequal} to check that these structure equations match (5.21). 

\DGio{SpaceTimeLieAlgebra("5.21"):}{abel3LTB}

\DGio{DGequal(LD2, abel3LTB);}{true}

The Jacobi identities are satisfied at this point.

\DGio{Query("Jacobi");}{true}

To complete our analysis of the family {\bf abel3LT}, we calculate the vector defined by equation (1.6). 
The calling sequence is {\tt SkewAdjointDirection(X, N, G)}, where {\tt X} is a vector complementary to the derived subalgebra, 
{\tt N} is a vector normal to the derived algebra, and {\tt G} is the metric tensor on the spacetime Lie algebra.

\DGio{zeta :=SkewAdjointDirection(K - L, K - L, eta);}
{\zeta := -4\,\gamma_1\, K-4\,\gamma_1\, L+(4\,i\bar\beta-4\,i\bar\kappa)\, M+(-4\,i\beta+4\,i\kappa)\, \barM}

For {\bf abel3LT} we  assume that $\zeta$ is timelike. We can then use a Lorentz transformation, fixing the derived algebra, to move $\zeta$ to a multiple 
of  $K + L$.  In this tetrad $\beta = \kappa$; we define

\DGin{Soln2 := $\{\beta = \kappa, \bar \beta = \bar \kappa\}$}

Substitute these values into the structure equations (5.21).
 
\DGin{LD2 := DGsimplify(subs(Soln2,  LieAlgebraData(abel3LTB))):}

The result matches (2.8) and our analysis is complete.

\DGio{SpaceTimeLieAlgebra("2.8");}{abel3LT0}

\DGio{DGequal(LD2, abel3LT0);}{true}

%
%

\DGhead{3. The Residual Group and the Equivalence Problem for Spacetime Lie Algebras}

The command {\tt NPResidualGroup} calculates the continuous subgroup  
of the Lorentz group which stabilizes a given 
	flag of subspaces in a spacetime Lie algebra.  This subgroup can then be used to normalize the spin coefficients.

In this section we show how the software is used to obtain the residual group and normalizations presented in Section 6.2.4.

First initialize the spacetime Lie algebra {\bf heisLN.}

\DGin{SpaceTimeLieAlgebra("2.4", coefficients = "real"):}

The derived algebra is the Lorentzian  subspace $\langle K, L, M + \barM \rangle$
and the second derived algebra is the null subspace $\langle K\rangle$.

The 2-parameter subgroup of the Lorentz group which stabilizes these subspaces is:
  
\DGin{chi := NPResidualGroup(heisLN, [[K], [K, L, M + Mb]]):}

The connected component of the residual group therefore consists of boosts

\DGio{chi[1];}{K \to e^u\, K , L\to e^{-u}L, M\to M, \bar M \to \bar M}

and null rotations with a real parameter

\DGio{chi[2];}{K\to K, L\to u^2 K + L + u M + u \barM , M\to uK +M, \bar M\to uK + \bar M}

The free  spin coefficients for {\bf heisLN} are:

\DGio{SpinCoefficientsInStructureEquations(heisLN0);}{\left\{ \nu_1,\tau_1,\beta_1,\epsilon_1,\gamma_1,\mu_1, \nu_0
 \right\}}

We wish to use the null rotations to normalize one of these spin coefficients to 0. 

The command {\tt TransformNPSpinCoefficients} gives us the transformation rules for the free   spin coefficients.  For the null rotation these are:

\DGio{TransformNPSpinCoefficients(chi[2]);}
{ \{\ldots,  
	\tilde \tau_1 =2\,\epsilon_1\,u+ \tau_1,\enskip
	\tilde \beta_1 =\epsilon_1\,u+\beta_1,	\enskip
	\tilde \mu_1 =2\,\epsilon_1\,{u}^{2}+\tau_1\,u+2\,\beta_1\,u+\mu_1,
	\tilde \nu_0=\nu_0,  \ldots\}
}

This proves, as stated in Section 6.1.4,  that if $\epsilon_1 \neq0$ then we may transform $\tau_1 \to 0$.  If $\epsilon_1 = 0$, we then see that we can set $\mu_1=0$ if $\tau_1 + 2\beta_1 \neq 0$. 
 We check that this first normalization gives the structure equations (2.4.1).

\DGin{LD1 := subs(It = 0, LieAlgebraData(heisLN0)):}

\DGin{SpaceTimeLieAlgebra("2.4.1", coefficients = "real"):}

\DGio{DGequal(LD1, heisLN1);}{true}

\DGhead{4. Classifying  Spacetime Lie algebras I. Identification}

In this section and the two that follow we demonstrate the 3 commands in the 
{\tt SpaceTimeGroups} package 
which can be used to  solve the equivalence problem for spacetime Lie algebras.

For the purposes of illustration, we consider the inheriting Einstein-Maxwell solution
of \cite{Ozsvath:1965a} (see 3.17).

First we retrieve the metric and a null tetrad from the 
{\sc DifferentialGeometry} database of exact solutions.

\DGio{h, nt0  := Library:-Retrieve("ExactSolutions","GR", ["Ozsvath1965a", 7, 14, 1], manifoldname = M, \\ output = ["Metric", "NullTetrad", "seq"]):}{
h := 
- \frac{{\rm e}^{-2\, x^3}}{a^2}\, dx^0 \odot dx^0
+ 4\,\sqrt{2}\, \frac{{\rm e}^{-3\,x^3}}{a^2}\,dx^0\odot dx^1
- 7\,\frac{{\rm e}^{-4\, x^3}}{a^2}\,dx^1\odot dx^1
+ \frac {{\rm e}^{4\, x^3}}{a^2}\,dx^2 \odot dx^2
+ \frac{1}{a^2}\,dx^3\odot dx^3,
}
\begin{footnotesize}
\begin{align*}
nt0[1] &:= \frac{\sqrt {2}}{2}\,a\,{\rm e}^{ x^3}\, \partial_{x^0} + 
\frac{\sqrt {2}}{2}\,a\, \partial_{x^3},
\quad
nt[2] := \frac{\sqrt {2}}{2}a\,{\rm e}^{ x^3}\, \partial_{x^0}
	-\frac{\sqrt {2}}{2}a\, \partial_{x^3}, 
\\
nt0[3] &:=2\,a\,{\rm e}^{ x^3}\, \partial_{x^0} 
	+ \frac{\sqrt {2}}{2}\,a\,{\rm e}^{2\, x^3}\, \partial_{x^1} 
	+ i\, \frac{\sqrt {2}}{2}a\,{{\rm e}^{-2\,{\it x^3}}}\, \partial_{x^2},
\quad
\\
nt0[4] &:= 
	2\,a\,{\rm e}^{ x^3}\, \partial_{x^0} 
      + \frac{\sqrt{2}}{2}\,a\,{\rm e}^{2\, x^3}\, \partial_{x^1}
     -i \frac{\sqrt {2}}{2}a\,{{\rm e}^{-2\,{\it x^3}}}\, \partial_{x^2}
\end{align*}
\end{footnotesize}

\noindent
We use {\tt nt0} to create a Lie algebra with basis {\tt nt}. We then use the {\tt infolevel} environment variable to track the 
command {\tt ClassifySpacetimeLieAlgebra} as it
performs the steps required for classification.

\DGin{DGEnvironment[LieAlgebra](nt0, alg, vectorlabels = '[K, L, M, Mb]'):}

\DGin{nt := [K, L, M, Mb]
}

\DGin{infolevel[ClassifySpacetimeLieAlgebra] := 2:}

\DGin{ClassifySpacetimeLieAlgebra(nt, sideconditions = \{${a > 0 }$\});}

\begin{footnotesize}
{\it 
$\bullet$ The spacetime Lie algebra (STLA) is of type: abel3

\qquad $\bullet$   The determinant of the induced metric on the derived algebra is -1
  
\qquad $\bullet$   The skew-adjoint direction is [0, 0, 0, $2\,a$]

\qquad $\bullet$    The norm${}^2$ of the skew-adjoint vector is $4\,a^2$

$\bullet$ The induced metric on the derived algebra is Lorentzian

$\bullet$ The STLA is of type: abel3L

$\bullet$ The skew-adjoint vector is spacelike

$\bullet$ The STLA is of type: abel3LS}
\end{footnotesize}
\begin{equation*}
	\text{\tt "}abel3LS\text{\tt "}
\end{equation*}

We thus find that the Ozsvath solution is a spacetime Lie group of type  
{\bf abel3LS.}

\DGhead{5. Classifying  Spacetime Lie algebras II. Finding an Adapted Null Tetrad}

Now that we know  the Ozsvath spacetime is of type {\bf abel3LS}, we  find a null tetrad $\{K_1, L_1, M_1, \barM_1\}$ such that, 
in accordance with equations (2.9), 
the derived algebra is $\langle K_1, L_1, M_1 + \barM_1 \rangle$ and the skew-adjoint line is along  $M_1 + \barM_1$. The structure equations are then aligned with (2.9). 
All this is accomplished with the command {\tt STLAAdaptedNullTetrad}.

\DGin{nt1 := STLAAdaptedNullTetrad[abel3LS](nt) assuming $a >0$;}
\begin{footnotesize}
\begin{alignat*}{2}
nt1[1] &:= \dfrac12 \, (K + L + M + \barM),
&\quad 
nt1[2] &:= \dfrac12\,(K +L - M - \barM),
\\
nt1[3] &:= \dfrac12\,(-K+L-M-\barM),
&\quad 
nt1[4] &:= \dfrac12\,(K-L-M+\barM)
\end{alignat*}
\end{footnotesize}
Check that these vector fields define a null tetrad.

\begin{footnotesize}
\DGio{TensorInnerProduct(h, nt1, nt1)}{\left[ \begin {array}{cccc} 0&-1&0&0\\ \noalign{\medskip}-1&0&0&0
\\ \noalign{\medskip}0&0&0&1\\ \noalign{\medskip}0&0&1&0\end {array}
 \right] }
\end{footnotesize}
Next, we initialize a spacetime Lie algebra  with the null tetrad {\tt nt1}.  
This is still the Ozsvath spacetime but now with respect to a null tetrad adapted to our classification 
scheme.  The null tetrad will be denoted by $(K_1, L_1, M_1, \barM_1)$, with dual basis $(\omega_1, \omega_2, \omega_3, \omega_4)$.

\DGin{DGEnvironment[LieAlgebra](nt1, alg1, vectorlabels = [K1, L1, M1, M1b], 

\quad formlabels = [omega1, omega2, omega3, omega4]):}

We check that the derived algebra and skew-adjoint line are aligned with those of (2.9).

\DGio{DerivedAlgebra();}{[K_1, L_1, M_1 + \barM_1]}

\medskip
\DGio{eta1 := evalDG(-2*omega1 \&s omega2 + 2*omega3 \&s omega4);}{\eta := -2   \omega_1\, \odot\, \omega_2+ 2\omega_3\, \odot \, \omega_4}

\DGio{SkewAdjointDirection(M1 - Mb1, M1 - Mb1, eta1);}{2\,a\, M_1 + 2\, a\, \barM_1}

\DGhead{6. Classifying  Spacetime Lie algebras III. Matching the Newman-Penrose Spin Coefficients}

Finally, to complete our solution to the equivalence problem, we need to determine the relationship between the spin coefficients which appear in (2.9) and the structure constants in the Ozsvath spacetime
 ({\tt alg1} -- initialized above).
 
\DGio{SpaceTimeLieAlgebra("2.9", coefficients = "real");}{abel3LS0}

 \DGio{vars := SpinCoefficientsInStructureEquations(abel3LS0);}{\{\nu_1,\, \tau_1,\, \alpha_1,\, \beta_1,\, \epsilon_1,\, \kappa_1,\, \mu_1\}}

\DGin{Eq := MatchNPSpinCoefficients(abel3LS0, alg1, vars);}
\begin{footnotesize}
\begin{align*}
 \nu_1   &= \dfrac14\,( \sqrt {2}-4)\, a, \enskip
 \tau_1   = - \dfrac34\,\sqrt {2}\, a, \enskip
 \alpha_1 = - \dfrac12\,( 1+\sqrt {2})\, a,  \enskip
 \beta_1  = -\dfrac12\,(\sqrt {2}-1)\, a, \enskip
\\
 \epsilon_1 & = 0, \enskip
 \kappa_1 =  \dfrac14\,(\sqrt {2}+4)\, a, \enskip
 \mu_1 = 0.
\end{align*}
\end{footnotesize}

\noindent With these values for the spin coefficients the structure equations for the spacetime 
Lie algebra {\bf abel3LS} and the structure equations for the Ozsvath electrovac spacetime coincide.

\DGio{DGequal(subs(Eq, LieAlgebraData(abel3LS0)), alg1);}{true}

%
\DGhead{7. Equivalence of 2 Spacetime Lie algebras using the residual group}

Here  we illustrate another aspect of the equivalence problem for spacetime Lie algebras.
Consider  two spacetime Lie algebras, defined in terms of the spin coefficients by 

\DGin{NP1 := $\{\alpha=i,\enskip
       \beta=2\,i,\enskip
       \gamma=i,\enskip
       \epsilon=0,\enskip
       \kappa=-i/2,\enskip
       \lambda=i,\enskip
       \mu=i,\enskip
	\nu=i/2,\enskip
       \pi=-i,\enskip
       \rho=0,\enskip
       \sigma=0,\enskip
       \tau=-i\}$
}

\DGin{NP2 := $\{ \alpha=-2\,i,\enskip
       \beta=-i,\enskip
       \gamma=0,\enskip
       \epsilon=2\,i,\enskip
       \kappa=-2\,i,\enskip
       \lambda=0,\enskip
       \mu=0,\enskip
       \nu=i/8,\enskip
       \pi=i,\enskip
       \rho=2\,i,\enskip
       \sigma=2\,i,\enskip
       \tau=i\}$
}

In this section we show that   these spacetime Lie algebras are equivalent.  We begin by showing they are both of type 
{\bf abel3LS}. Therefore, if these algebras are equivalent, they must be related by the residual group of Lorentz transformations for {\bf abel3LS}.  
 We use {\tt MatchNPSpinCoefficients} to find such a transformation.

Start with the general Newman-Penrose structure equations.

\DGio{SpaceTimeLieAlgebra("1.4", coefficients = "real");}{NPalg}

Substitute the  spin coefficients {\tt NP1} into  these structure equations and initialize as {\tt alg1}, with basis given by $(x_1, x_2, x_3, x_4)$.

\DGin{LD1 := subs(NPComplexToReal(NP1), LieAlgebraData(NPalg, alg1)):}

\DGio{DGEnvironment[LieAlgebra](LD1, vectorlabels = [x]);}{alg1}

Similarly, substitute the second set of spin coefficients and initialize as {\tt alg2},  with basis given by $(y_1, y_2, y_3, y_4)$.

\DGin{LD2 := subs(NPComplexToReal(NP2), LieAlgebraData(NPalg, alg2)):}

\DGio{LD2 := DGEnvironment[LieAlgebra](LD2, vectorlabels = [y]);}{alg2}

Check that both spacetimes are of type {\bf abel3LS}.

\DGio{ClassifySpacetimeLieAlgebra([x1, x2, x3, x4]);}{\text{"abel3LS"}}

\DGio{ClassifySpacetimeLieAlgebra([y1, y2, y3, y4]);}{\text{"abel3LS"}}

The command {\tt STLAResidualGroup} retrieves the stored values for the residual group, including the discrete Lorentz transformation, for a given spacetime Lie algebra class. Here are the generators (given as  matrices defining Lorentz transformations of the null tetrad):

\DGio{STLAResidualGroup(abel3LS, output = "generators");}
{
[ \left[ \begin {array}{cccc} 1&0&0&0\\ \noalign{\medskip}0&1&0&0
\\ \noalign{\medskip}0&0&-1&0\\ \noalign{\medskip}0&0&0&-1\end {array}
 \right] , \left[ \begin {array}{cccc} 1&0&0&0\\ \noalign{\medskip}0&1
&0&0\\ \noalign{\medskip}0&0&0&1\\ \noalign{\medskip}0&0&1&0
\end {array} \right] , \left[ \begin {array}{cccc} 0&1&0&0
\\ \noalign{\medskip}1&0&0&0\\ \noalign{\medskip}0&0&1&0
\\ \noalign{\medskip}0&0&0&1\end {array} \right] 
\left[ \begin {array}{cccc} x&0&0&0\\ \noalign{\medskip}0&{x}^{-1}&0
&0\\ \noalign{\medskip}0&0&1&0\\ \noalign{\medskip}0&0&0&1\end {array}
 \right] 
]
}

These correspond to the transformations  $\mathcal R$, $\mathcal Y$, $\mathcal Z$, $B_{KL}^*$ introduced in Section 5.1. See also the residual group listed in 2.9.

The full residual group is generated by the 4 matrices above. It is retrieved as follows; we suppress the output which consists of 8 matrices.

\DGin{RG := STLAResidualGroup(abel3LS):}

The command {\tt MatchNPSpinCoefficients} searches through the list of all 8 matrices comprising the residual group to check for equivalence, {\it i.e.,} whether the two Lie algebras are related by an element of the group.  If the two algebras are equivalent, the output is a list of vectors which indicates the change of basis needed to identify the two spacetime Lie algebras.

\DGio{Equivalence := MatchNPSpinCoefficients(alg1, alg2, \{\ \}, RG)[1]:}
{[-2\,y_2, -\frac12\,y_1, y_4, y_3]}

Verify that this change of basis aligns the two spacetime Lie algebras.

\DGio{DGequal(LieAlgebraData(Equivalence), alg1);}{true}

\DGhead{8. Solutions to the Einstein Field Equations}

In this section we will check that the spin coefficients defined by  3.8 do indeed define a perfect fluid solution to the  Einstein field equations.

First, initialize the Lie algebra 3.8 and, at the same time, obtain the metric tensor and the matter field variables.

\DGio{eta, F := SpaceTimeLieAlgebra("3.8", coefficients = "real", output = "Fields");}{
\begin{aligned}
& eta, F := 
\\
&-2\,\Theta_1 \odot \Theta_2 + 2\Theta_3 \odot\Theta_4, \
\text{table([``U" = $\dfrac{\sqrt{2}}{2}(K +L)$, ``phi$\wedge$2" = $\dfrac{-2s^4+ 5s^2 -2}{a^2}$, ``psi$\wedge$2" = $\dfrac{-s^2 + 2}{a^2}$])}
\end{aligned}
}

Now calculate the Einstein tensor for the metric {\tt eta}:

\DGio{Ein := simplify(map(expand, EinsteinTensor(eta)));}
{Ein := 
-\,{\frac {2\,{s}^{4}-5\,{s}^{2}+2}{{2\,a}^{2}}}(\,K\otimes K + \,L\otimes L)
-{\frac {{s}^{4}-3\,{s}^{2}+2}{{a}^{2}}}(\,K\otimes L +\,L\otimes K)
-\,{\frac {{s}^{2}-2}{{2\,a}^{2}}}(M\otimes \bar M + \,\bar M\otimes M)}

The energy momentum tensor is 

\DGin{T := evalDG(phi2*U \&t U + psi2*InverseMetric(eta)):}{}

and the field equations hold:

\DGio{DGsimplify(Ein \&minus T);}{0\, K \otimes K}

\DGhead{9. Algebraically Special Spacetime Lie Groups}

In this section we show that  the  only spacetime of class {\bf abel3LZ3} and Petrov type D is the Lie  algebra  4.3.3. 
Begin by initializing the Lie algebra and defining the metric tensor.

\DGin{eta := SpaceTimeLieAlgebra("2.11.3",  coefficients = "real", output = "Metric");}

We note that the Petrov type is generically type II.

\DGio{PetrovType([K, L, M, Mb])}{\text{\tt "}II \text{\tt"}}

We pass the keyword argument  {\tt output =  "D"} to  
{\tt PetrovType} to obtain the conditions on the spin coefficients for the spacetime Lie algebra to be of type D.

\DGio{factor(PetrovType([K, L, M, Mb], output = "D"));}{[0,0,0,0,-16/3\,\alpha_1^{3} \left( \tau_1+2\,\alpha_1 \right) ^{3}
\nu_1\, \left( \tau_1+\alpha_1 \right) ]}

One of these factors must vanish for the spacetime to be of type D. If $\alpha_1 =0$, the derived algebra is 2-dimensional. If $\nu_1 = 0$ or $ \tau_1 + 2\alpha_1 =0$, the isometry algebra is 5-dimensional. We conclude that 
$\tau_1 = - \alpha_1$ and this gives 4.3.3.

\DGhead{10. Jump in Isometry Algebra Dimension}

In this section we give a simple example which illustrates the
jump in the dimension of the isometry algebra that occurs when 
various spin coefficients vanish.  Begin by initializing the algebra 2.1 and defining the metric tensor.

\DGin{eta := SpaceTimeLieAlgebra("2.1", manifoldname = N, coefficients = "real",  output = "Metric", frame = "OT"):}

Use {\tt IsometryAlgebraData} to calculate the dimension of the Lie algebra of Killing vectors.

\DGio{IsometryAlgebraData(eta, output = ["Dimension"]);}{4}

According to the information in Section 2 for the spacetime Lie algebra 
2.1, the isometry algebra is 5-dimensional when $\kappa = \sigma = 0$.

We use {\tt InstantiateFrame} to set  $\kappa = \sigma = 0$ in the structure equations for 2.1. The original structure equations are 
stored in a backup frame and can be restored with {\tt RestoreFrame}.

\DGin{InstantiateFrame(N, $\{\kappa_0 = 0, \enskip \kappa_1 = 0,\enskip  \sigma_0 = 0, \enskip \sigma_1=0 \}$):}

With these values for   $\kappa$ and $\sigma$, the dimension of the isometry algebra increases:

\DGio{IsometryAlgebraData(eta, output = ["Dimension"]);}{5}

Therefore, when $\kappa = \sigma = 0$ the structure equations 2.1
define a  homogeneous space with a multiply transitive group and so, by our definition, they do not define 
a spacetime Lie algebra.

\DGhead{11. Lie's Third Theorem and Local Group Coordinates}

In this section we show how to obtain the local coordinate expression for the metric on the spacetime group defined by the solvable spacetime Lie algebra 3.12.

\DGio{eta := SpaceTimeLieAlgebra("3.12", output = "Metric", manifoldname = P, frame = "OT");}
{\eta := 2\left(-\omega1\otimes \omega1 + \omega2\otimes \omega2 + \omega3\otimes \omega3 + \omega4\otimes \omega4\right)}

Create group coordinates:

\DGin{DGEnvironment[Coordinate]([x, y, z, w], G);}

For solvable Lie algebras, the command {\tt LiesThirdTheorem} (in the {\sc GroupActions} package) calculates a set of vector fields 
whose commutators yield the same structure equations as the given abstract Lie algebra. 
These vector fields may be viewed as the left invariant vector fields on the Lie group.

\DGio{Gamma := GroupActions:-LiesThirdTheorem(P, G);}
{
[\partial_x, \enskip  \partial_y, \enskip  \partial_z, \enskip
(2\,\alpha_1\,x+2\,\kappa_1\,y)\, \partial_x+(2\,\alpha_1\,y-2\,\kappa_1\,x)\, \partial_y-4\,\alpha_1\,z\, \partial_z+\, \partial_w ]  
}

Check that the structure equations for the vector  fields $\Gamma$  match those of the Lie algebra  3.12.

\DGio{DGequal(LieAlgebraData(Gamma), P);}{true}

Calculate the dual basis to the left invariant vector fields. These are the (left invariant) Maurer-Cartan forms on $G$.

\DGio{Omega := DualBasis(Gamma);}{
[\, dx+(-2\,\alpha_1\,x-2\,\kappa_1\,y)\, dw, \enskip  \, dy+(-2\,\alpha_1\,y+2\,\kappa_1\,x)\, dw, \enskip \, dz+4\,\alpha_1\,z\, dw,\,\enskip  dw]}

The coordinate form of the metric for the spacetime Lie algebra 3.12 is therefore

\DGin
{
h := evalDG( 2*(-Omega[1]\,\&t\,Omega[1] + Omega[2]\,\&t\,Omega[2] + Omega[3]\,\&t\,Omega[3] + Omega[4]\,\&t\,Omega[4]));
}

The result is the metric displayed in the introduction to Section 3.

The Killing vectors for {\tt h}  are given by the right invariant vector fields,  that is, the Lie algebra of vector fields  each of which commutes with $\Gamma$. 
These are often referred to as the reciprocal vector fields.

\DGin{KV := GroupActions:-ReciprocalVectorFieldSystem(Gamma, [x = 0, y = 0, z = 0, w = 0]);
}
\begin{align*}
KV := [\ &
{{\rm e}^{2\,\alpha_1\,w}}\cos \left( 2\,\kappa_1\,w \right) \, \partial_x-{{\rm e}^{2\,\alpha_1\,w}}\sin \left( 2\,\kappa_1\,w \right) \, \partial_y
,\enskip
{{\rm e}^{2\,\alpha_1\,w}}\sin \left( 2\,\kappa_1\,w \right) \, \partial_x+{{\rm e}^{2\,\alpha_1\,w}}\cos \left( 2\,\kappa_1\,w \right) \, \partial_y
,
\\
&
{{\rm e}^{-4\,\alpha_1\,w}}\, \enskip \partial_z,\, \partial_w \ ]
\end{align*}

Check that these are indeed Killing vectors for the metric {\tt h}:

\DGio{LieDerivative(KV, h);}
{[0\,dx\otimes dx, \enskip 0\,dx\otimes dx, \enskip 0\,dx\otimes dx, \enskip 0\,dx\otimes dx]}

\section*{Acknowledgment}

We thank R. McLenaghan for providing us with a copy of G. Fee's thesis.
This work was supported in part by National Science Foundation grant ACI-1642404.



\appendix
\section{Isometries of Spacetime Groups}
\setlength{\parindent}{12pt}
	Our  classification of spacetime groups rests heavily on the 
	fact that the dimension of the isometry algebra 
	(and, indeed, the full structure equations of the 
	isometry algebra) can be computed directly  
	by purely algebraic means from the spacetime Lie algebra structure equations. 
	The details of this result seem not so readily available in the 
	 literature and, therefore,  it seems useful to summarize them here.

	Let $(M, g)$ be an $n$-dimensional  pseudo-Riemannian manifold with Levi-Civita connection $\nabla$. 
	If $X$ is a Killing vector field for the metric $g$,
	then it is well known that
\begin{equation}
	F_{ij}  \equiv \nabla_iX_j =  \nabla_{[i}X_{j]} 
	\quad \text{satisfies}\quad 
	\nabla_i F_{jk} =X_\ell R^\ell{}_{ijk}.
\label{KV}%
 \immediate\write5{"\getrefnumber{KV}" = "KV"}
\end{equation}
	Already, this shows that an (analytic) Killing vector  field $X$ 
	is uniquely determined  by its ``Killing data'' --- the values of the tensors $X$ and $F$
	at a given point --- and  the set  of all Killing vector fields
	is a finite-dimensional vector space of dimension 
	$\leq \dfrac{n(n+1)}{2}$. 
	A simple argument, which we now present,  sharpens this result considerably. 
	
	To this end, we introduce the vector bundle 
	$\mathbb K = T^*M \oplus \Lambda^2(M) $. 
	 Motivated by \eqref{KV},  we define 
	a  linear connection $\tilde \nabla$ on  $\mathbb K$ by the formula
\begin{equation*}
	\widetilde \nabla_i
	\begin{bmatrix} \mathbb X _j \\  \mathbb F_{kl} \end{bmatrix} 
	= \begin{bmatrix} \delta^a_j \nabla_i & -\delta^a_i\delta^b_j\\ -R^a{}_{ikl}& \delta^a_k\delta^b_l\nabla_i \end{bmatrix}  
	\begin{bmatrix} \mathbb X _a \\  \mathbb F_{ab} \end{bmatrix},	
\end{equation*}
	where $\begin{bmatrix} \mathbb X _j \\  \mathbb F_{kl} \end{bmatrix}$ is  a local section of $\mathbb K$.
	It then follows, rather easily, that there is 
	{a one-to-one correspondence between the Killing vectors for the 
	metric $g$ 
	and the parallel sections of the 
	connection $\widetilde \nabla$ on $\mathbb K$.}  The bundle $\mathbb K$ is called the 
	{\it  tractor bundle} for the Killing equations and $\widetilde \nabla$ the {\it tractor connection} (see \cite{Michel} and references therein). 
	The tractor bundle $\mathbb K$ is equipped with a bracket operation which is compatible with the Lie bracket of vector fields on $M$ 				\cite{AshtekarAshtekar}, namely,
\begin{equation}
	\label{AshtekarBracket}
	[\begin{bmatrix} \mathbb X_{(1)}\\  \mathbb F_{(1)} \end{bmatrix}, \begin{bmatrix} \mathbb X_{(2)} \\  \mathbb F_{(2)} \end{bmatrix}] = 
	\begin{bmatrix} \mathbb X_{(1)}^{k} F_{(2)\, km} - X_{(2)}^{k} F_{(1)\, km} \\  \mathbb 
	F_{(1)m}^{k} F_{(2)kn} - F_{(2)m}^{k} F_{(1)kn} - R_{rsmn}X_{(1)}^r X_{(2)}^s
		 \end{bmatrix}.
\end{equation}
	
	From this viewpoint, the calculation of Killing vectors  translates into the  general problem of calculating  parallel 
	sections  of a vector bundle  $\pi: E \to M$, endowed with a linear connection $\mathcal D$ \cite{atkins:2011a}. This problem 
	is best understood in terms of the holonomy of $\mathcal D$
	(see Besse's Fundamental Principle \cite{besse:1986}, paragraph 10.19). 
	Fix an  auxiliary linear connection ${\mathcal D}_0$ on $M$ and 
	let $\hat{\mathcal D}$ be the induced linear connection on $T^r_s(M) \oplus E$.  	
        Let $\mathcal R$ be the curvature tensor for $\mathcal D$ and define the infinitesimal holonomy of $\mathcal D$ 
	at order $k$ to be the set of all linear endomorphisms of $E_x$ given by
\begin{equation*}	
	\text{hol}^k_x({\mathcal D}) = \{( \hat{\mathcal D})^k \mathcal R)(X_1, X_2, \ldots, X_k) \text{\ for all \ }
	X_i \in T_x(M).
\end{equation*}
	Set 
\begin{equation*}
	 P_{k, x} = \{ v \in E_x  \,|\, L(v) = 0  \text{\ for all\ }  L \in \text{hol}^k_x({\mathcal D}) \}.
\end{equation*}
	One can check that these constructions are independent of the choice of ${\mathcal D}_0$.
	The infinitesimal holonomy of ${\mathcal D}$ at $x$ is 
	$\text{hol}_x({\mathcal D})  = \cup_{k =1}^{\infty}\text{hol}^k_x({\mathcal D})$. 
	In this setting, the Ambrose-Singer theorem asserts that the infinitesimal holonomy  
	is the Lie algebra of the holonomy group of ${\mathcal D}$.
	Moreover, the vector sub-bundle  $\mathbb P =  \cup_{x \in M} P_x$, where  
	$P_x = \cap_{k =1}^{\infty}\text{P}_{k, x}({\mathcal D})$, is an integrable sub-bundle of $E$. 
	Every parallel section of $E$ factors through $\mathbb P$ and and {\it  the rank of $\mathbb P$ equals  the
	dimension of the vector space of parallel sections.} 

	We may apply these results to the tractor connection for the Killing equations to determine the dimension
	of the space of Killing vectors.  However, from a computational point of view, 
	it more efficient to calculate the parallel section bundle  $\mathbb P$ directly as the set of  
	$[\mathbb X_i, \mathbb F_{jk}]$ satisfying 
\begin{equation}
\begin{aligned}
& \mathbb X^{\ell} \nabla_{\ell}R_{ijhk } + (\mathbb F \cdot R)_{ijhk} = 0
\\[1\jot]
& 
\mathbb X^{\ell} \nabla^2_{\ell p}  R_{ijhk } + (\mathbb F \cdot \nabla R)_{pijhk} = 0
\\[1\jot]
&\mathbb X^{\ell} \nabla^3_{\ell p q} R_{ijhk } + (\mathbb F \cdot \nabla^2 R)_{pqijhk} = 0, \quad \ldots
\end{aligned}
\label{tractorhol}
\end{equation}	
	where $(F \cdot R)_{ijhk} = F_i^\ell R_{\ell jhk} + F_j^\ell R_{i \ell hk} + \cdots	$.   The first of these equations is equivalent to the definition of $P_{0,x}$, with subsequent 
	equations being the differential consequences. 
	
	If $M$ is a homogeneous space and $h$ is a $G$-invariant metric on $M$, then the 
	parallel section bundle  $\mathbb P \subset \mathbb K$ is $G$-invariant, that is,  $a\cdot P_{x} = P_{a\cdot x}$ for $a\in G$,
	and calculations can be done at a single point. Finally, we also remark that there are algebraic formulas for 
	the curvature and its covariant derivatives in terms of the structure equations for the Lie algebra of $G$.  With all these results in hand, it is possible to calculate the isometry algebra of a spacetime group directly from the structure equations of its Lie algebra.

	The {\sc DifferentialGeometry} command {\tt IsometryAlgebraData} calculates the vector space 
	of solutions to the linear 
	systems \eqref{tractorhol} order by order, stopping when the solutions define a Lie algebra, that is, define an integrable distribution according to the bracket (\ref{AshtekarBracket}).
	The following {\sc DifferentialGeometry} worksheet illustrates the process. We emphasize 
	that the method works for any  metric, without the need for local coordinates 
	and without integrating the Killing equations. 

\DGhead{Killing Data for the Isometry Algebra }

\DGin{with(DifferentialGeometry): with(LieAlgebras): with(Tensor): with(DGApplications:-SpacetimeGroups)}

We use the spacetime Lie algebra 3.11 to illustrate the key ideas. 

\DGio{eta := SpaceTimeLieAlgebra("3.11", output = "Metric", frame = "OT", coefficients = "real")}
{\eta = - \omega_1 \otimes  \omega_1  + \omega_2 \otimes \omega_2 + \omega_3 \otimes \omega_3 + \omega_4 \otimes \omega_4}

Set the infolevel to 2 in order to get detailed information about the computations.
 
\DGin{infolevel[IsometryAlgebraData] := 2:}

\DGin{LD := IsometryAlgebraData(eta, alg);}

\begin{footnotesize}
\it 

$\bullet$ Computing the curvature

$\bullet$ Computing the derivatives of curvature at order 1

$\bullet$ Solving the equations $X\cdot \nabla R + F \cdot R= 0$ 

$\bullet$ The upper bound on the dimension of the isometry algebra is 5

$\bullet$ The stopping criterion is the integrability of the Killing data  (X, F)

\quad $\bullet$ The Killing data (X, F) are not integrable

$\bullet$ Computing the derivatives of curvature at order 2


$\bullet$ Solving the equations $X\cdot \nabla^2 R + F \cdot \nabla R = 0$ 

$\bullet$ The upper bound on the dimension of the isometry algebra is 4.

$\bullet$ The stopping criterion is the integrability of the Killing data  (X, F)

\quad  $\bullet$ The Killing data (X, F) are integrable
 
$\bullet$ Calculating the Lie algebra data from the Killing data

\end{footnotesize}
\begin{equation*} 
LD = [e_1, e_2] = 0, \enskip  [e_1, e_3] = -2\lambda_0 e_3 -(2\lambda_1 + \mu_1)e_4, \ \ldots
\end{equation*}

Let's look more closely at the Killing data calculated in the first step.  We use the keyword arguments {\tt minisometrydim = 5}, which causes the program to terminate once the Killing data is of dimension 5 or less.

\DGin{KD := IsometryAlgebraData(eta, alg1, minisometrydim = 5, output = ["KillingData"]):}

\medskip\noindent
The list {\tt KD} is a basis for the 5-dimensional space of Killing data. Each element in the list {\it KD} is a 2 element list consisting of a vector $\mathbb X^i$ and type (1,1) tensor $\mathbb F^i{}_j$. For example, the fifth basis element in {\tt KD} is 
\begin{footnotesize}
\begin{equation*}
KD[5] := [0\, E_1,\enskip \dfrac12 E_3 \otimes \omega_4 -  \dfrac12 E_4 \otimes \omega_3] .
\end{equation*}
\end{footnotesize}

We check that this pair satisfies  the first equation in \eqref{tractorhol},

\DGin{R := CurvatureTensor(eta):}

\DGio{InducedDerivationOnTensors(KD[5][2], R):}
{0\, E_1\, \omega^1 \otimes \omega^1 \otimes \omega^3,}

but {\it not} the next equation:

\DGin{R1 := CovariantDerivative(R, Christoffel(eta)):}

\DGio{InducedDerivationOnTensors(KD[5][2], R1);}{
-16\lambda_1(\lambda_0^2 + \lambda_1^2 -\mu_1^2) E_1 \otimes \omega^3 \otimes \omega^3 \otimes \omega^4\otimes \omega^3 + \cdots}

At the next step in calculating the parallel section bundle $\mathbb P$, the 5th solution  at the first step, KD[5], is eliminated and the bound on the rank of $\mathbb P$ (and hence the bound on the number of Killing vectors) decreases to 4.

\setlength{\textwidth}{7in}
\section {4-dimensional Lie algebras}

Here we provide a classification of 4-dimensional real Lie algebras based upon the results of \cite{Patera-Sharp-Winternitz-Zassenhaus:1976} (for indecomposable Lie algebras).  For those Lie algebras depending upon parameters, we have case-split according to  the dimension of the algebra of derivations, which is denoted by $d$. 

\begin{multicols}{2}
\small
\setlength{\abovedisplayskip}{6pt}
\setlength{\belowdisplayskip}{4pt}
\setlength{\baselineskip}{10pt}
\parindent = 0pt
\columnseprule = .5pt
\parskip = 4pt
\setlength{\tabcolsep}{2 pt}
\noindent
{\bf 4-Dimensional Decomposable Lie Algebras}

\medskip
\vbox{
{\bf[4, 0]}
\medskip

\hbox{ 

$\begin{tabular}[t]{L| C  C  C  C } \
&\ e_1  &e_2   &e_3   &e_4 \\ \hline
e_{{1}} & . & . & . & .\\
e_{{2}} &  & . & . & .\\
e_{{3}} &  &  & . & .\\
e_{{4}} &  &  &  & .\\
\end{tabular}$

}
\begin{tabular}[t] {K K}
$\qquad d=16$ \\
\end{tabular}

}


\bigskip
\vbox{
{\bf[4, -1]}
\medskip

\hbox{ 

$\begin{tabular}[t]{L| C  C  C  C } \
&\ e_1  &e_2   &e_3   &e_4 \\ \hline
e_{{1}} & . & e_{{1}} & . & .\\
e_{{2}} &  & . & . & .\\
e_{{3}} &  &  & . & .\\
e_{{4}} &  &  &  & .\\
\end{tabular}$

\qquad
}
\begin{tabular}[t] {K K}
$\qquad d=8$ \\
\end{tabular}

}

%

\medskip
\vbox{
{\bf[4, -2]}
\medskip

\hbox{ 

$\begin{tabular}[t]{L| C  C  C  C } \
&\ e_1  &e_2   &e_3   &e_4 \\ \hline
e_{{1}} & . & e_{{1}} & . & .\\
e_{{2}} &  & . & . & .\\
e_{{3}} &  &  & . & e_{{3}}\\
e_{{4}} &  &  &  & .\\
\end{tabular}$

}
\begin{tabular}[t] {K K}
$\qquad d=4$ \\
\end{tabular}
}
\medskip

%
\vbox{
{\bf[4, -3]}
\medskip

\hbox{ 

$\begin{tabular}[t]{L| C  C  C  C } \
&\ e_1  &e_2   &e_3   &e_4 \\ \hline
e_{{1}} & . & . & . & .\\
e_{{2}} &  & . & e_{{1}} & .\\
e_{{3}} &  &  & . & .\\
e_{{4}} &  &  &  & .\\
\end{tabular}$

}
\begin{tabular}[t] {K K}
$\qquad d=10$ \\
\end{tabular}
\medskip
}


\medskip
\vbox{
{\bf[4, -4]}
\medskip

\hbox{ 

$\begin{tabular}[t]{L| C  C  C  C } \
&\ e_1  &\ e_2  &e_3  &e_4 \\ \hline
e_{{1}} & . & . & e_{{1}} & .\\
e_{{2}} &  & . & {\it a1}\,e_{{2}} & .\\
e_{{3}} &  &  & . & .\\
e_{{4}} &  &  &  & .\\
\end{tabular}$

}
\begin{tabular}[t] {K K}
parameters:  $ [-1\leq {\it a1}< 1,{\it a1}\neq 0,{\rm \ and\ } {\it a1}=1]$ \\
$d = 6$\ if\ $a1\neq1$; $d=8$\ if\  $a1=1$
\end{tabular}
\medskip
}

\medskip

%
%

\medskip
\vbox{
{\bf[4, -5]}
\medskip

\hbox{ 

$\begin{tabular}[t]{L| C  C  C  C } \
&\ e_1  &e_2   &e_3   &e_4 \\ \hline
e_{{1}} & . & . & e_{{1}} & .\\
e_{{2}} &  & . & e_{{1}}+e_{{2}} & .\\
e_{{3}} &  &  & . & .\\
e_{{4}} &  &  &  & .\\
\end{tabular}$

}
\begin{tabular}[t] {K K}
$\qquad d=6$ \\
\end{tabular}
\medskip
}

\medskip


\medskip
\vbox{
{\bf[4, -6]}
\medskip

\hbox{ 

$\begin{tabular}[t]{L| C  C  C  C } \
&\ e_1  &e_2   &e_3   &e_4 \\ \hline
e_{{1}} & . & . & {\it a1}\,e_{{1}}-e_{{2}} & .\\
e_{{2}} &  & . & e_{{1}}+{\it a1}\,e_{{2}} & .\\
e_{{3}} &  &  & . & .\\
e_{{4}} &  &  &  & .\\
\end{tabular}$

}
\begin{tabular}[t] {K K}
parameters: & $ [0\leq {\it a1}]$, $d=6$ \\
\end{tabular}
\medskip
}

\medskip


\medskip
\vbox{
{\bf[4, -7]}
\medskip

\hbox{ 

$\begin{tabular}[t]{L| C  C  C  C } \
&\ e_1  &e_2   &e_3   &e_4 \\ \hline
e_{{1}} & . & e_{{1}} & -2\,e_{{2}} & .\\
e_{{2}} &  & . & e_{{3}} & .\\
e_{{3}} &  &  & . & .\\
e_{{4}} &  &  &  & .\\
\end{tabular}$

}
\begin{tabular}[t] {K K}
$\qquad d=4$ \\
\end{tabular}
\medskip
}


\medskip
\vbox{
{\bf[4, -8]}
\medskip

\hbox{ 

$\begin{tabular}[t]{L| C  C  C  C } \
&\ e_1  &e_2   &e_3   &e_4 \\ \hline
e_{{1}} & . & e_{{3}} & -e_{{2}} & .\\
e_{{2}} &  & . & e_{{1}} & .\\
e_{{3}} &  &  & . & .\\
e_{{4}} &  &  &  & .\\
\end{tabular}$

}
\begin{tabular}[t] {K K}
$\qquad d=4$ \\
\end{tabular}
\medskip
}

\medskip

\medskip

\noindent
{\bf 4-Dimensional In-decomposable Lie Algebras}

\medskip

\vbox{
{\bf[4, 1]}
\smallskip

\hbox{ 

$\begin{tabular}[t]{L| C  C  C  C } \
&\ e_1  &e_2   &e_3   &e_4 \\ \hline
e_{{1}} & . & . & . & .\\
e_{{2}} &  & . & . & e_{{1}}\\
e_{{3}} &  &  & . & e_{{2}}\\
e_{{4}} &  &  &  & .\\
\end{tabular}$

%
}
\begin{tabular}[t] {K K}
$\qquad d=7$ \\
\end{tabular}
\medskip
}

\medskip


\medskip
\vbox{
{\bf[4, 2]}
\medskip

\hbox{ 

$\begin{tabular}[t]{L| C  C  C  C } \
&\ e_1  &e_2   &e_3   &e_4 \\ \hline
e_{{1}} & . & . & . & e_{{1}}\\
e_{{2}} &  & . & . & ae_{{2}}\\
e_{{3}} &  &  & . & be_{{3}}\\
e_{{4}} &  &  &  & .\\
\end{tabular}$

\qquad
}
\begin{tabular}[t] {K K}
parameters:  $ [a\neq 0,b\neq 0,a\leq 1,b\leq a,-1\leq b]$ \\
$d=  6$\ if $a\neq b$;\enskip $d=8\ {\rm if}\ a=b\neq1;$
$d=12\  {\rm if}\ a=b=1$
\end{tabular}
}

%
%
\medskip
\medskip
\vbox{
{\bf[4, 3]}
\medskip

\hbox{ 

$\begin{tabular}[t]{L| C  C  C  C } \
&\ e_1  &e_2   &e_3   &e_4 \\ \hline
e_{{1}} & . & . & . & e_{{1}}\\
e_{{2}} &  & . & . & e_{{1}}+e_{{2}}\\
e_{{3}} &  &  & . & ae_{{3}}\\
e_{{4}} &  &  &  & .\\
\end{tabular}$

}
\begin{tabular}[t] {K K}
parameters:  $ [a\neq 0]$ \\
$d=6$,\ if\ $a\neq1$; $d=8$\ if\ $a=1$ 
\end{tabular}
}



\medskip\medskip
\vbox{
{\bf[4, 4]}
\medskip

\hbox{ 

$\begin{tabular}[t]{L| C  C  C  C } \
&\ e_1  &e_2   &e_3   &e_4 \\ \hline
e_{{1}} & . & . & . & .\\
e_{{2}} &  & . & . & e_{{1}}\\
e_{{3}} &  &  & . & e_{{3}}\\
e_{{4}} &  &  &  & .\\
\end{tabular}$

}
\begin{tabular}[t] {K K}
$\qquad d=6$ \\
\end{tabular}
\medskip
}

\medskip\medskip


\medskip
\vbox{
{\bf[4, 5]}
\medskip

\hbox{ 

$\begin{tabular}[t]{L| C  C  C  C } \
&\ e_1  &e_2   &e_3   &e_4 \\ \hline
e_{{1}} & . & . & . & e_{{1}}\\
e_{{2}} &  & . & . & e_{{1}}+e_{{2}}\\
e_{{3}} &  &  & . & e_{{2}}+e_{{3}}\\
e_{{4}} &  &  &  & .\\
\end{tabular}$

}
\begin{tabular}[t] {K K}
$\qquad d=6$ \\
\end{tabular}
\medskip
}

\medskip


\medskip
\vbox{
{\bf[4, 6]}
\medskip

\hbox{ 

$\begin{tabular}[t]{L| C  C  C  C } \
&\ e_1  &e_2   &e_3   &e_4 \\ \hline
e_{{1}} & . & . & . & ae_{{1}}\\
e_{{2}} &  & . & . & be_{{2}}-e_{{3}}\\
e_{{3}} &  &  & . & e_{{2}}+be_{{3}}\\
e_{{4}} &  &  &  & .\\
\end{tabular}$

}
\begin{tabular}[t] {K K}
parameters:  $ [a\neq 0,0\leq b]$, 
$d=6$
\end{tabular}
}

\vbox{
{\bf[4, 7]}
\medskip

\hbox{ 

$\begin{tabular}[t]{L| C  C  C  C } \
&\ e_1  &e_2   &e_3   &e_4 \\ \hline
e_{{1}} & . & . & . & (a+1)e_{1}\\
e_{{2}} &  & . & e_{{1}} & e_{{2}}\\
e_{{3}} &  &  & . & -e_{{3}}\\
e_{{4}} &  &  &  & .\\
\end{tabular}$

\qquad

}
\begin{tabular}[t] {K K}
parameters:  $ [-1\leq a\leq 1]$ \\
$d=5$\ if\ $a\neq 1$;\ $d=7$\ if\ $a=1$
\end{tabular}
\medskip
}

\vbox{
{\bf[4, 8]}
\medskip

\hbox{ 
$\begin{tabular}[t]{L| C  C  C  C } \
&\ e_1  &e_2   &e_3   &e_4 \\ \hline
e_{{1}} & . & . & . & .\\
e_{{2}} &  & . & e_{{1}} & -e_{{3}}\\
e_{{3}} &  &  & . & e_{{2}}\\
e_{{4}} &  &  &  & .\\
\end{tabular}$

}
\begin{tabular}[t] {K K}
$\qquad d=5$
\end{tabular}
}

\bigskip

\vbox{
{\bf[4, 9]}
\medskip

\hbox{ 

$\begin{tabular}[t]{L| C  C  C  C } \
&\ e_1  &e_2   &e_3   &e_4 \\ \hline
e_{{1}} & . & . & . & \left (a+1\right )e_{{1}}\\
e_{{2}} &  & . & e_{{1}} & e_{{2}}\\
e_{{3}} &  &  & . & ae_{{3}}\\
e_{{4}} &  &  &  & .\\
\end{tabular}$

\qquad
}
\begin{tabular}[t] {K K}
parameters:  $ [-1<a\leq 1, a\neq1;\ a=1]$ \\
$d=5$\ if $a\neq 1$,\ $d=7$ if $a=1$.
\end{tabular}
}
%
%

\medskip
\vbox{
{\bf[4, 10]}
\medskip

\hbox{ 
$\begin{tabular}[t]{L| C  C  C  C } \
&\ e_1  &e_2   &e_3   &e_4 \\ \hline
e_{{1}} & . & . & . & 2\,e_{{1}}\\
e_{{2}} &  & . & e_{{1}} & e_{{2}}\\
e_{{3}} &  &  & . & e_{{2}}+e_{{3}}\\
e_{{4}} &  &  &  & .\\
\end{tabular}$

}
\begin{tabular}[t] {K K} \\[-10pt]
$\qquad d=5$
\end{tabular}
}


\medskip
\vbox{
{\bf[4, 11]}
\medskip

\hbox{ 

$\begin{tabular}[t]{L| C  C  C  C } \
&\ e_1  &e_2   &e_3   &e_4 \\ \hline
e_{{1}} & . & . & . & 2\,ae_{{1}}\\
e_{{2}} &  & . & e_{{1}} & ae_{{2}}-e_{{3}}\\
e_{{3}} &  &  & . & e_{{2}}+ae_{{3}}\\
e_{{4}} &  &  &  & .\\
\end{tabular}$

}
\begin{tabular}[t] {K K}
parameters:  $ [0<a]$,\ $d=5$ \\
\end{tabular}
}
\smallskip



\vbox{
{\bf[4, 12]}
\medskip

\hbox{ 

$\begin{tabular}[t]{L| C  C  C  C } \
&\ e_1  &e_2   &e_3   &e_4 \\ \hline
e_{{1}} & . & . & . & e_{{1}}\\
e_{{2}} &  & . & e_{{1}} & e_{{2}}\\
e_{{3}} &  &  & . & .\\
e_{{4}} &  &  &  & .\\
\end{tabular}$

}
\begin{tabular}[t] {K K} \\[-10pt]
$d=5$
\medskip

\end{tabular}
}

\vbox{
{\bf[4, 13]}
\medskip

\hbox{ 

$\begin{tabular}[t]{L| C  C  C  C } \
&\ e_1  &e_2   &e_3   &e_4 \\ \hline
e_{{1}} & . & . & e_{{1}} & -e_{{2}}\\
e_{{2}} &  & . & e_{{2}} & e_{{1}}\\
e_{{3}} &  &  & . & .\\
e_{{4}} &  &  &  & .\\
\end{tabular}$

}
\begin{tabular}[t] {K K} \\[-10pt]
$d=4$
\end{tabular}
}


\end{multicols}
\bigskip
\section{Symbols and Notation}

\begin{tabular}{ll}

$\{K, L, M, \overline M\}, \{\Theta_K, \Theta_L, \Theta_M, \Theta_{\barMs}\}$  &$G$-invariant null tetrad and dual basis\\
\\
$\alpha,\beta, \gamma, \epsilon, \kappa, \lambda,
	\mu, \nu, \pi, \rho, \sigma,\tau$ & spin coefficients\\
	$\alpha = \alpha_0 + i \alpha_1$ & real and imaginary parts\\
	\\
$\{E_1, E_2, E_3, E_4\}, \{\omega^1, \omega^2, \omega^3, \omega^4\}$  &$G$-invariant tetrad and dual basis\\
\\
$\langle A, B, C, \dots\rangle$  & span of  $A, B, C, \dots$\\ 
\\
$\mathfrak{g},\ \mathfrak{g}^\prime,\ \mathfrak{g}^{\prime\prime}$  & spacetime Lie algebra, first derived algebra, second derived algebra\\
\\
$\zeta^a$  &skew-adjoint line (see (\ref{defzeta}))\\
\\
${\rm O}(\eta)$  & Lorentz group which fixes  $\eta$\\
\\
$R_{M\barMs}$  &1-parameter group of rotations in the $M$-$\barM$ plane (see \S \ref{Lorentz})\\
\\ 
$B_{KL}$  &1-parameter group of boosts in the $K$-$L$ plane (see \S  \ref{Lorentz})\\
\\
$N_K$ & 2-parameter group of null rotations fixing $K$ (see \S  \ref{Lorentz})\\
\\
$N_{K,u}$  & 1-parameter group of null rotations fixing $K$ and $i(M-\barM)$ (see \S  \ref{Lorentz})\\
\\
$N_{K,iv}$  & 1-parameter group of null rotations fixing $K$ and $M+\barM$ (see \S  \ref{Lorentz})\\
\\
${\cal R}, {\cal Y}, {\cal T}, {\cal Z}, {\cal U},  {\cal V}$ &discrete Lorentz transformations (see \S  \ref{Lorentz})\\
\\
$J^a_{bcd}$ & Jacobi tensor (see \S \ref{NPPrel})\\
\\
$\liez$ & one-dimensional center of spacetime Lie algebra
\end{tabular}

\immediate\write5{"eof" = EOF}

\end{document}